\documentclass[twocolumns]{aastex7}
\usepackage{color}
\usepackage{graphicx}
\usepackage[normalem]{ulem}
\usepackage{rotating}

\usepackage{lineno}
\usepackage{braket}
\usepackage{gensymb}
\linenumbers

\shorttitle{The Spatial Distribution of ICGCs in Fornax cluster}
\shortauthors{}

\date{\today}

\begin{document}

\title{The Spatial Distribution of Intra-Cluster Globular Clusters in the Fornax Cluster}

\correspondingauthor{Raffaele D'Abrusco}
\email{rdabrusc@cfa.harvard.edu}

\author[0000-0003-3073-0605]{Raffaele D'Abrusco}
\affil{Smithsonian Astrophysical Observatory, 60 Garden Street, Cambridge, MA 20138, USA}
\email{rdabrusc@cfa.harvard.edu}

\author[0009-0007-6055-3933]{Marco Mirabile}
\affil{INAF Osservatorio Astronomico d’Abruzzo, Via Maggini, 64100 Teramo, Italy}
\affil{Gran Sasso Science Institute, Viale Francesco Crispi 7, 67100 L’Aquila, Italy}
\email{marco.mirabile@inaf.it}

\author[0000-0003-2072-384X]{Michele Cantiello}
\affil{INAF Osservatorio Astronomico d’Abruzzo, Via Maggini, 64100 Teramo, Italy}
\email{michele.cantiello@inaf.it}

\author[0000-0003-4291-0005]{Enrica Iodice}
\affil{INAF Osservatorio Astronomico di Capodimonte, Salita Moiariello, 16, 80131 Napoli, Italy}
\email{enrichetta.iodice@inaf.it}

\author[0000-0002-4175-4728]{Avinash Chaturvedi}
\affil{Leibniz-Institut für Astrophysik Potsdam (AIP), An der Sternwarte 16, D-14482 Potsdam, Germany}
\email{avi.chaturvedi@aip.de}

\author[0000-0002-2363-5522]{Michael Hilker}
\affil{European Southern Observatory, Karl-Schwarzschild-Straße 2, 85748 Garching, Germany}
\email{mhilker@eso.org}

\author[0000-0002-3554-3318]{Giuseppina Fabbiano}
\affil{Smithsonian Astrophysical Observatory, 60 Garden Street, Cambridge, MA 20138, USA}
\email{gfabbiano@cfa.harvard.edu}

\author[0000-0002-6427-7039]{Marilena Spavone}
\affil{INAF Osservatorio Astronomico di Capodimonte, Salita Moiariello, 16, 80131 Napoli, Italy}
\email{marilena.spavone@inaf.it}

\author[0000-0003-4210-7693]{Maurizio Paolillo}
\affil{Dipartimento di Fisica ``Ettore Pancini'', Universit\`a di Napoli ``Federico II'', via Cinthia 9, 80126 Napoli, Italy}
\affil{INFN - Sezione di Napoli, via Cinthia 9, 80126 Napoli, Italy}
\affil{INAF Osservatorio Astronomico di Capodimonte, Salita Moiariello, 16, 80131 Napoli, Italy}
\email{paolillo@na.infn.it}

\begin{abstract}

We investigate the spatial distribution of the Intra-Cluster Globular
Clusters (ICGCs) detected in the core of the Fornax galaxy cluster. By separately modeling  
different components of the observed population of Globular Clusters (GCs), we confirm 
the existence of an abundant ICGCs over-density with a geometrically complex, 
elongated morphology roughly centered on the cluster dominant galaxy 
NGC\,1399 and stretching along the E-W direction. We identify several areas 
in the ICGCs distribution that deviate from a simple elliptical model and feature large 
density enhancements. These regions are characterized based on their 
statistical significance, GCs excess number, position, size and location relative 
to the galaxies in their surroundings. The relations between the spatial distribution
and features of the ICGCs structures, mostly populated by blue GCs, and properties of
the intra-cluster light and dwarf galaxies detected in the core of the 
Fornax cluster are described and discussed. The line-of-sight velocity distribution 
of spectroscopically confirmed GCs within the ICGCs structures is compatible with the 
systemic velocities of nearby bright galaxies in the Fornax cluster, suggesting that the 
ICGCs population is at least partially composed by GCs stripped from their hosts. 
We argue that the findings here presented suggest that, on sub-cluster scales,
different mechanisms contribute to the growth of the ICGC. The western region of Fornax is 
likely associated with
old merging events that predate 
the assembly of the Fornax cluster. The eastern side instead points to a mix of tidal disruption 
of dwarf galaxies and stripping from the GCSs of massive hosts, more
typical of relaxed, high-density cluster environments. 

\end{abstract}

\section{Introduction}
\label{sec:introduction}

Clusters of galaxies present a unique opportunity to investigate how frequent, violent interactions
between galaxies under the effect of a strong gravitational potential, shape the observable properties of the 
visible matter in their surroundings, on scales and magnitudes that are not accessible in other, less dense environments. 
One of the most interesting observational trait of galaxy clusters is the 
existence of Globular Clusters (GCs) at large distances from any galaxy and, for this reason, unlikely to belong
to the GCs system (GCS) of any potential host galaxy. These GCs are usually called Intra-Cluster Globular Clusters (ICGCs).

The ICGCs were first postulated by~\cite{white1987}, and 
used by~\cite{west1995} to explain the high specific frequency S$_{N}$, 
defined as the number of GCs per unity galaxy luminosity~\citep{harris1981,harris1991}, of bright elliptical galaxies 
in the center of clusters.~\cite{west1995} suggested that the origin of the ICGCs population could be 
attributed to the combined effect of tidal stripping of portions of the 
GCSs of interacting galaxies and {\it in situ} formation. After the detection
of small samples of ICGCs in the Virgo~\citep{williams2007} 
and the Fornax clusters~\citep{bassino2003,bergond2007,schuberth2008} 
follow-up studies detected large numbers
of ICGCs in Virgo~\citep{lee2010,durrell2014,powalka2018}, Fornax~\citep{dabrusco2016,cantiello2020}. 
Coma~\citep{peng2011,madrid2018}, 
Abell\,1689~\citep{alamomartinez2013}, Abell\,1185~\citep{west2011} and Perseus~\citep{harris2020,kluge2024}.
Recent cosmological simulations of the galaxy-cluster halos that follow the evolution of GCs seeds
in the galactic halos over cosmological times~\citep{ramosalmendares2018} have shown 
that in the 
local Universe $\sim60\%$ of the GCs initially bound to single galaxies are removed by galaxy-galaxy interactions 
from their original halo and dispersed into the intra-cluster space, generating 
the ICGCs population. New observational facilities with large field-of-view (f.o.v.) combined with high sensitivity and spatial 
resolution
are pushing forward our ability to observe and characterize 
ICGCs in observational regimes so far unexplored. 
Early results from JWST images of distant 
clusters~\citep{lee2022,harris2023} reported 
relatively young ICGC populations with complex morphology with spatial distributions consistent 
with that of the Intra-Cluster Light (ICL). These ICGCs distributions do not follow 
the geometry of the dark matter halo of the clusters as derived from weak lensing~\citep{diego2023,martis2024}.

The Fornax cluster, the second richest nearby cluster of galaxies after Virgo at 
d$_{\mathrm{Fornax}}\sim\!20$ Mpc~\citep{blakeslee2009}, has two sub-structures, 
one centered around NGC\,1399 and identified as the core of the cluster, 
and the other associated with NGC\,1316, which is located $\approx3\degree$ to the 
South-West (SW) of NGC\,1399. The SW component, called
the Fornax A sub-cluster, is bound to Fornax
and appears to infall towards the core~\citep{drinkwater2001}. The main galaxy over-density has a larger
fraction of ETGs than the Virgo cluster~\citep{jordan2007}, indicating a relatively advanced evolutionary 
stage. Using kinematics and structural properties derived from Integral 
Field Unit (IFU) spectroscopic data of a subset of Fornax galaxies~\cite{iodice2019a},~\cite{iodice2019b}
and~\cite{spavone2022}
discovered the existence of two distinct
classes of galaxies in the main sub-substructure of the cluster: a compact clump of early infallers, 
located NW of the center of the cluster, and intermediate and recent infallers which are located 
at larger cluster-centric distances. The core of the Fornax cluster has been the 
target of studies investigating 
the properties of its abundant ICGCs through large f.o.v., deep 
imaging data~\citep{bassino2003,bassino2006,dabrusco2016,cantiello2020} and 
spectroscopic campaigns~\citep{bergond2007,schuberth2008,schuberth2010,pota2018,chaturvedi2022}. Recent
results from the analysis of {\it Euclid} Early Release Observations~\citep{cuillandre2024} of 
Fornax (ERO-F)~\citep{saifollahi2024} confirmed the existence of a very large population 
of GCs in $\sim\!0.5(\degree)^{2}$ field located NW 
of the cluster, roughly centered around the position of NGC\,1387.

In this paper, we perform a detailed analysis of the spatial properties of the population of 
GCs detected in Fornax out to one virial radius~\citep[R$_{\mathrm{vir}}\!=\!0.7$ Mpc,][]{drinkwater2001} using
large coverage, deep data from the Fornax Deep Survey~\citep[FDS,][]{peletier2020}. In order to highlight 
the spatial properties of the population of GCs in the core of the cluster, we separately simulated 
distinct components that contribute to the observed distribution of GCs, and extracted a map of the 
ICGCs by taking advantage of 1) the
detailed models of the luminosity and radial distributions of the GCSs of bright galaxies in the literature, 
and 2) well-known correlations between the brightness of the hosts and the size and
radial profiles of the GCSs for galaxies for which GCS-specific models
are not available. 

This manuscript is organized as follows. In Section~\ref{sec:data} the data to extract the sample of 
GCs employed in this work are introduced; Section~\ref{sec:method} presents the statistical 
method that has produced the residual maps of the spatial distribution of the ICGCs. In 
Section~\ref{sec:results} and Section~\ref{sec:discussion} we describe our findings and compare
them to other observables that can constrain the nature of the GCs structures.
We use cgs units unless otherwise stated. Optical magnitudes used in this manuscript are in 
the Vega system and are not corrected for the Galactic extinction. Standard cosmological 
parameter values have been used for all calculations~\citep{bennett2014}.

\section{Data}
\label{sec:data}

FDS~\citep{peletier2020} includes a combination of guaranteed time 
observations 
from the FOrnax Cluster Ultra-deep Survey (FOCUS, P.I. R.~Peletier) and the VST Early-
type GAlaxy Survey (VEGAS, P.I. E.~Iodice), described by~\cite{spavone2017} 
and~\cite{iodice2021}. Both surveys used the 
Italian National Institute for Astrophysics (INAF) VLT Survey Telescope (VST) while
operated by the European Southern Observatory (ESO). This 2.6 m diameter optical survey telescope, 
located at Cerro Paranal, Chile~\citep{schipani2010}, has imaged the sky in the $u$, 
$g$, $r$ and $i$-bands using the large field of view ($1\times1$ degrees) camera 
OmegaCAM~\citep{kuijken2011}. FDS covers $\simeq$22$(\degree)^{2}$ around the core of 
cluster up to the virial radius of the cluster, plus additional $\simeq$6$(\degree)^{2}$ 
centered on the Fornax A sub-cluster. The depth of the FDS images, averaged over an area of 
1$(\arcsec)^{2}$, corresponds to surface brightness of 28.3, 28.4, 27.8, 27.2 
mag$/(\arcsec)^{2}$ in the $u$, $g$, $r$ and $i$ bands, respectively~\citep{peletier2020}.

\subsection{Fornax Deep Survey Globular Cluster Candidates}
\label{subsec:fdsgc}

We extracted the GC candidates from the VST FDS 
following~\cite{cantiello2020}. The large inter-field FWHM variations 
was corrected by combining
$g$, $r$ and $i$-band single exposures with median FWHM lower than $0\arcsec.9$
into co-added master-detection frames. The FWHM threshold was determined as a 
trade-off between the final resolution and depth of the coadded stakes. The effects 
of the independent calibrations of each field on the source detection were
accounted for through a two-steps procedure consisting in a first derivation of a 
standard calibration plan of VST frames, followed by the extraction
of a calibration matrix matching the full FDS source catalog to a unique 
reference. In this way, the offset between fields becomes 
negligible, and the re-calibrated photometry matches well published data and the 
predictions from stellar population 
synthesis models. This approach resulted in 
a homogeneous sample of sources across the large f.o.v.~covered by FDS images. 
The GC candidates have been selected using both photometric and 
morphometric selection criteria (the $g$ band magnitude, 
colors $g\!-\!i$, $g\!-\!r$, $u\!-\!r$ and their uncertainties, the concentration index $CI_{n}$, 
the stellarity index CLASS$\_$STAR, the FWHM of the detection, flux radius, the elongation 
and the sharpness) on the general
catalog of $\sim1.7$ million matched sources extracted from the $ugri$ images. 
The intervals of parameter values used for the selection of the FDS GC candidates 
are based on a reference catalog of spectroscopically confirmed 
GCs~\citep{schuberth2010,pota2018,chaturvedi2022} and {\it bona fide} GCs selected from high 
resolution images obtained with the Advanced Camera for
Surveys (ACS) on board the Hubble Space Telescope (HST) by the ACSFCS 
collaboration~\citep{jordan2009,jordan2015}. 

\begin{figure*}[ht]
    \centering
    \includegraphics[width=0.495\linewidth]{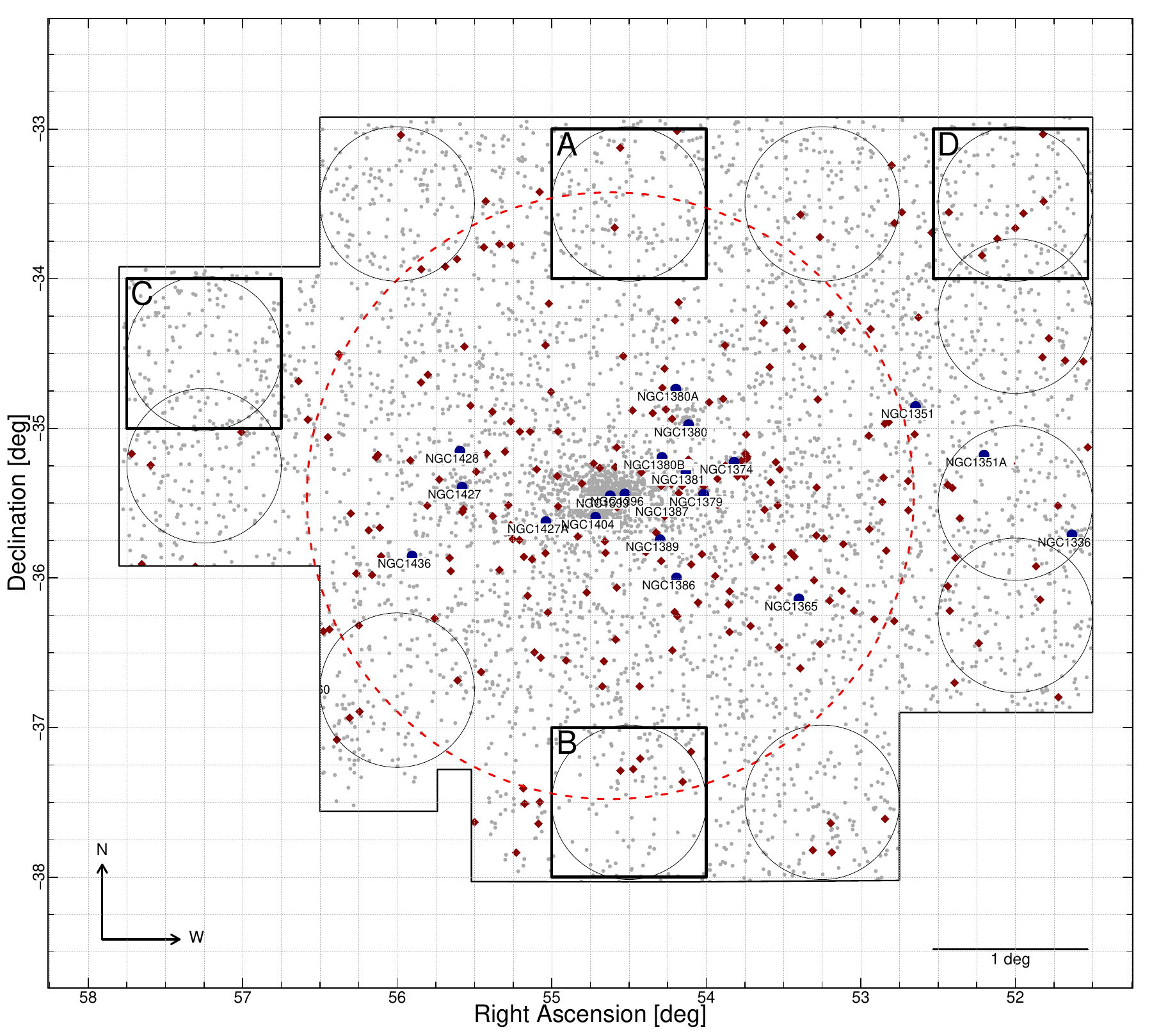}
    \includegraphics[width=0.495\linewidth]{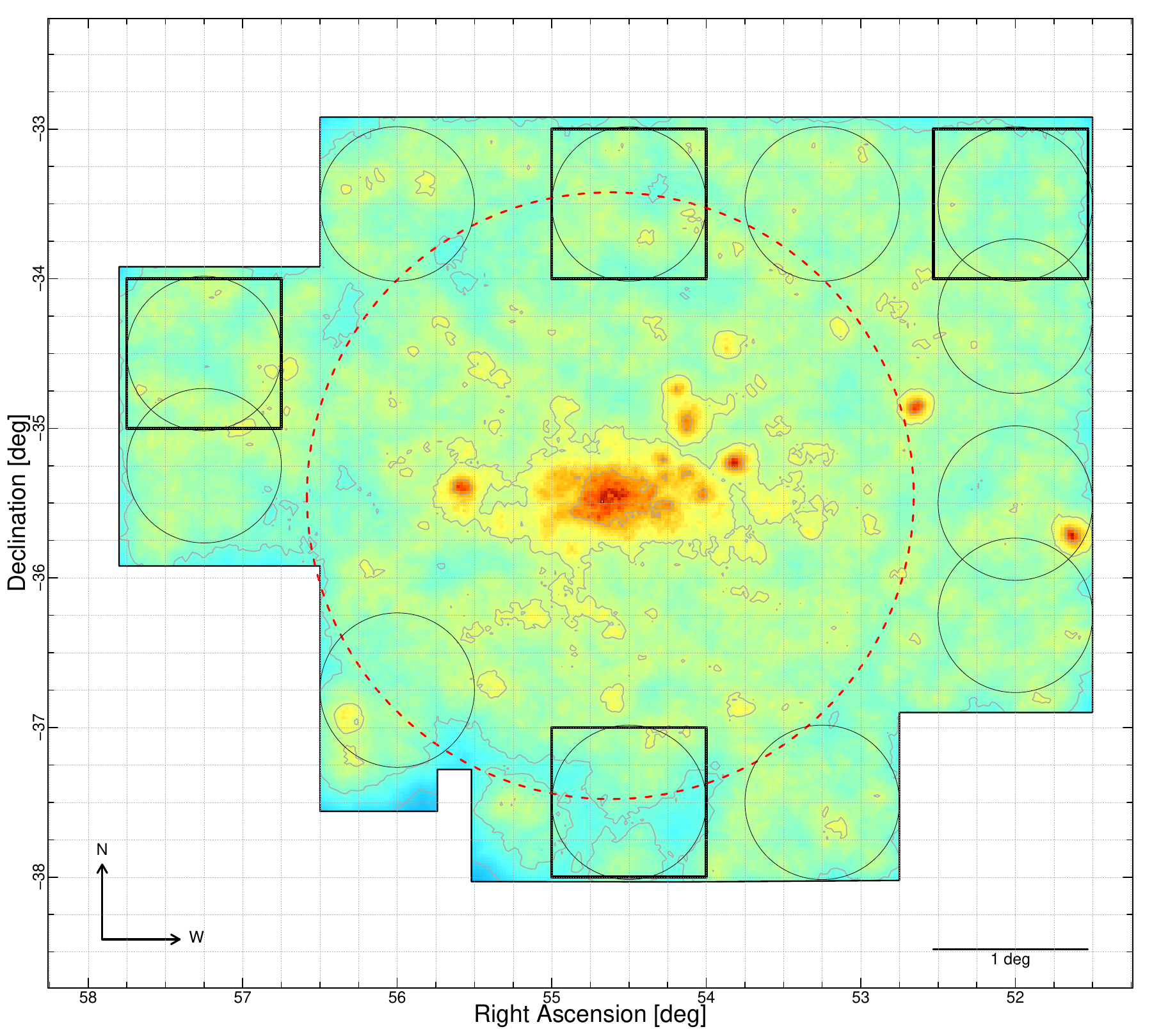}	
    \caption{Left: Position of FDS GC candidates. The red diamonds show 
    the positions of the galaxies from the Fornax Cluster Catalog~\citep[FCC,][]{ferguson1989};  
    the brightest galaxies in the field are labeled and highlighted with a blue symbol. 
    Right: Density map of FDS GC candidates obtained using the $K$-Nearest Neighbor (KNN) 
    method~\citep{dressler1980} with $K\!=\!3$, with low-density levels highlighted. In both plots, 
    the four black boxes display the regions used to estimate the density of the homogeneous, 
    component of the GC candidates 
    population according to the first method described in 
    Section~\ref{subsec:simulcomponents}, while the gray circles are employed to determined the 
    uncertainty of the density of the homogeneous component from candidate GCs located outside
    of the virial radius of the cluster (see Section~\ref{subsec:simulcomponents} for 
    details on both methods). The red dashed circles represent the virial radius of the Fornax 
    cluster~\citep{drinkwater2001}.}
    \label{fig:gc_position}
\end{figure*}

The resulting catalog of GC candidates contains 5653 
sources~(see Figure~\ref{fig:gc_position}) in an area of $\sim21$ square degrees 
roughly centered around the position of NGC\,1399 and extending out to the cluster virial radius 
of $\sim0.7$ Mpc~\citep{drinkwater2001}. The completeness 
of the candidate GCs catalog is determined by the $u$-band, which is shallowest among the four bands 
used to extract the FDS sample of photometric sources. The choice of a bright limiting $u$ magnitude for the GCs sample
($\sim\!24.1$ as reported by~\cite{cantiello2020}), approximately 1.4 magnitude brighter than the 
median turn-over magnitude of the $u$-bad GC luminosity function in Fornax galaxies~\citep{jordan2007}, 
is required to minimize the stellar contamination and ensure maximal homogeneity of the GC population 
over the large area covered by the FDS f.o.v.. 

\section{Method}
\label{sec:method}

The spatial distribution of the GCs detected in the Fornax cluster is investigated in this paper 
through a modification of the technique used in~\cite{dabrusco2013,dabrusco2015,dabrusco2022}, 
who derived the GCs residual maps by comparing the observed GC distribution to two-dimensional 
simulations based on the average radial profiles of the GCSs and homogeneous azimuthal
distributions. In particular,~\cite{dabrusco2022} also compared the observed GC distribution 
to a radial profile based on the stellar surface brightness of the host galaxy to emphasize 
deviations from the expected model.

In this work, the total distribution of GCs in the Fornax cluster is modeled as the 
superposition of three different, independent sub-populations with different spatial properties: 

\begin{itemize}
    \item Component A: a homogeneous, isotropic, flat distribution component over the entire observed area 
    that accounts for the combined contribution of background and foreground 
    contamination~(see Section~\ref{subsubsec:homocomponent}).
    \item Component B: a spatially complex component resulting from the superposition of multiple spatially compact, 
    resolved sub-components associated with the GCSs of the galaxies in the area covered by the FDS sample~(see 
    Section~\ref{subsubsec:galcomponent}).
    \item Component C: a spatially extended component including GCs associated with the GCs over-density observed in the 
    core of the Fornax cluster~\citep{bassino2003,bassino2006,dabrusco2016,cantiello2020}. The properties of this 
    component are described in Section~\ref{sec:results}. 
\end{itemize}

The total number of observed GC candidates can then be expressed as: 

\begin{equation}
    N_{\mathrm{GCs}}^{\mathrm{(obs)}}=N_{\mathrm{GCs}}^{\mathrm{(A)}}\!+N_{\mathrm{GCs}}^{\mathrm{(B)}}+N_{\mathrm{GCs}}^{\mathrm{(C)}}\!\equiv\!\\     
                                      N_{\mathrm{GCs}}^{\mathrm{(hom)}}\!+\sum_{i=1}^{N}N_{\mathrm{GCs}}^{(\mathrm{gal}_{i})}+
                                      N_{\mathrm{GCs}}^{\mathrm{(ext)}}
    \label{eq:numgc_observed}                                  
\end{equation}

\noindent where $N_{\mathrm{GCs}}^{\mathrm{(A)}}$, 
$N_{\mathrm{GCs}}^{\mathrm{(B)}}\!=\!\sum_{i=1}^{N}N_{\mathrm{GCs}}^{(\mathrm{gal}_{i})}$ 
and $N_{\mathrm{GCs}}^{\mathrm{(C)}}$
are the total number of observed GCs for components A, B and C respectively. 
The galaxies-based component B is the 
sum of the independent contributions of the GC systems for each of the $N$ galaxies in the 
area covered by the FDS GCs sample. Since our goal is to characterize the statistical 
significance, size and morphology of the projected distribution of the ICGCs, we only
simulate components A and B and use the residual maps to investigate 
the third component of the observed GC distribution (C). The total number 
of GCs in each simulated instance of the first two components can be then expressed as: 

\begin{equation}
    \widehat{N}_{\mathrm{GCs}}^{\mathrm{(sim)}}=\widehat{N}_{\mathrm{GCs}}^{\mathrm{(hom)}}\!+
    \sum_{j=1}^{M}\widehat{N}_{\mathrm{GCs}}^{(\mathrm{gal}_{j})}
    \label{eq:numgc_simulated}
\end{equation}

\noindent where $\widehat{N}_{\mathrm{GCs}}^{\mathrm{(hom)}}$ and $\sum_{j=1}^{M}\widehat{N}_{\mathrm{GCs}}^{(\mathrm{gal}_{j})}$
are the estimated total numbers of GCs from components A and B. The sum in the galaxies-specific component runs 
to $M\leq\!N$, where $M$ is the number of galaxies that contribute non-negligibly to the observed distribution of GCs, which by definition
is equal or smaller of the total number of galaxies $N$ observed in the field. In the following, we describe in details the method used to simulate the 
two GCs components (Section~\ref{subsec:simulcomponents}) and the estimation of the residual maps (Section~\ref{subsec:residuals}).

\subsection{Components of the GC populations}
\label{subsec:simulcomponents}

\subsubsection{Component A: Homogeneous component}
\label{subsubsec:homocomponent}


This component accounts for background and foreground 
contamination of the catalog of candidate GCs. We consider its surface density
constant over the footprint of the FDS observation. 

We applied two different methods to estimate the density of this component. First, we 
estimated the density of the homogeneous GCs component by averaging the densities 
of the GC candidates detected within four distinct $1\degree\!\times\!1\degree$ fields located in the outskirts 
of the imaged area of the cluster~(squares in Figure~\ref{fig:gc_position}). The four fields 
were chosen to be outside or partly overlapping with the virial radius of the Fornax cluster, 
in areas with no bright galaxies 
and no obvious over-density in the distribution of the FDS candidate GCs (see right panel in 
Figure~\ref{fig:gc_position}). These fields (Table~\ref{tab:controlfields}), show a large variance 
in the number of GC candidates included.
The density of the homogeneous component, obtained as the average of the densities in each of the 
four fields, is $d^{(1)}_{hom}\!=\!161.2\!\pm\!29.2$ GCs/$(^{\circ})^{2}$ ($\sim\!0.044$ GCs/$(\arcmin)^{2}$) so 
that $N_{\mathrm{GCs}}^{\mathrm{(hom,1)}}\!=\!3442.6\!\pm\!623.3$ over the entire FDS area, 
corresponding to $\sim\!60.8\!\pm\!10.9\%$ of the total number of FDS candidate GCs. 

The second approach employs
all candidate GCs located outside of the virial radius of the cluster~\citep{drinkwater2001}, excluding those 
within five effective radii ($5\times\!r_{\mathrm{eff}}$) from the positions of galaxies in 
the FCC~\citep{ferguson1989}. For galaxies without measured r$_{\mathrm{eff}}$ 
in the FCC, we 
used the average of the r$_{\mathrm{eff}}$ values for all galaxies in the background region, corresponding 
to $\sim\!6\arcsec$. 
This approach yields a density of the homogeneous component $d^{(2)}_{hom}\!=\!188.4\!\pm\!21.7$ GCs/$(^{\circ})^{2}$ 
($\sim0.05$ GCs/$(\arcmin)^{2}$), corresponding to a total number of sources in the homogeneous component 
$N_{\mathrm{GCs}}^{\mathrm{(hom,2)}}\!=\!4021.6\!\pm\!463.2$, i.e. $\sim\!71.1\!\pm\!8.2\%$ of the total number 
of FDS candidate GCs over the FDS f.o.v. The uncertainty on the density of the homogeneous component 
determined with this method is estimated as the standard deviation of the distribution of densities calculated 
inside the 12 circles of diameter 1$\degree$ displayed in both plots of Figure~\ref{fig:gc_position}. GC candidates 
located in the regions of the circles overlapping the virial radius and within five effective radii from the 
positions of FCC galaxies were not counted and the areas of the circle reduced accordingly.

Both methods yield densities
compatible with the stellar contamination estimated from a detailed map of the spatial 
distribution of bright stars in the FDS f.o.v. (see Figure~13 of~\citealt{cantiello2020}). 
According to these authors, the stellar surface density varies between 0.02 to 0.06 sources/$(\arcmin)^{2}$, i.e. from 
72 to 216 sources/$(^{\circ})^{2}$ over the area covered by the catalog of FDS candidate 
GCs. On the other hand, there are indications that some of the sources located outside 
the virial radius of the cluster might not be contaminants.
Using imaging data from the S-PLUS survey covering $\sim\!208\degree$ around NGC\,1399,~\cite{lomelinunez2025} 
discovered over-densities of GC candidates within 1 and 2 virial radii NE and SE of the cluster center, 
that seem to follow the spatial distribution of Ultra-Diffuse Galaxies (UDGs) in the outskirts of the cluster. 
These findings corroborate the 
hypothesis that a non-negligible population of {\it bona fide} GCs not associated with bright host galaxies
can be still found in the outskirts of the FDS f.o.v., where the density of the homogeneous component used
in this paper is measured. We discuss the effects of different levels of the homogeneous component 
on the results of our analysis in Section~\ref{sec:results}.

\begin{deluxetable}{cccc}
    \tablecaption{Properties of the control fields used to estimate the density of the isotropic, homogeneous 
    component of the simulated GC distribution in the core of the Fornax cluster.}
    \label{tab:controlfields}
    \tablehead{
	                & \colhead{R.A.\tablenotemark{a}}    & \colhead{Dec\tablenotemark{b}}   &  \colhead{GCs density\tablenotemark{c}}   
              }				
    \startdata
	A	            & 	  [54,55] 			&	[-33,-34]	&     173.9   \\\relax
	B	            & 	  [54,55]           &	[-38,-37]   &     124.8   \\\relax
	C	            & 	  [56.75, 57.75]    &	[-35,-34]	&     192.9   \\\relax
	D	            & 	  [51.53, 52.53]    &	[-34,-33]   &     153.5   \\
    \enddata
     \tablecomments{(a): Range of Right Ascension values spanned by the control field [deg];
                    (b): Range of Declination values spanned by the control field [deg];
                    (c): Density of candidate GCs in the control field [N$_{\mathrm{GCs}}$/deg$^{2}$].
                    }     
\end{deluxetable}

\subsubsection{Component B: The GC Systems of galaxies}
\label{subsubsec:galcomponent}

The GC systems of the galaxies located within the area of the sky imaged by FDS data are 
simulated by using, when available, the observational models of their luminosity 
and spatial distributions published in the literature and
obtained from deep, large-scale, high-quality photometric 
data. The GCSs of the brightest, better characterized galaxies (hereinafter the Main Galaxies Sample or MGS)
and of all the other galaxies from the~\cite{ferguson1989} catalog located within the FDS field-of-view 
(the Secondary Galaxy Sample or SGS) have been generated by following two different approaches. 
 
In the case of the MGS, the properties of the simulated GC systems of MGS galaxies have 
been determined using galaxy-specific luminosity functions and spatial models derived from 
high-quality, space and ground-based datasets. For the radial profiles, 
we employed modified Hubble models~\citep{binney1987} that describe the radial profiles of GCSs 
up to large galacto-centric distances thanks to the combined use of HST and ground-based
data at different radial distances. The azimuthal distribution of the simulated GCs is modeled as the elliptical 
geometry of the diffuse stellar light of the host galaxies. The brightness distribution of simulated MGS GCs are 
drawn from the best-fit Gaussian models of the Globular 
Cluster Luminosity Functions (GCLF) in the $g$ (HST equivalent) band determined by~\cite{villegas2010} 
using ACSFCS HST data. We use the 
$u$-band as it almost entirely determines the completeness of GC detections in the FDS $ugri$ matched catalog 
rather than the significantly deeper $g$, $r$ and $i$-band data~\citep[see][]{mirabile2024}.
The turnover magnitudes of the GCLF models
were converted to the FDS $u$ band by applying a correction equal to the median $u-g$ GC color from the general catalog
of sources extracted from the FDS images~\citep{cantiello2020}. The correction applied 
is $\Delta(u-g)_{\mathrm{corr}}^{\mathrm{FDS}}\!=\!1.4\!\pm\!0.3$, 
where the uncertainty is 1.5 times the Median Absolute Deviation (MAD) of the $u-g$ color distribution.
The list of galaxies in the MGS for which this approach is applicable and the relevant properties are displayed in 
Table~\ref{tab:main_galaxies}.\\

The observed number of GCs in the FDS data and their spatial distribution for MGS hosts was obtained by 
following these steps:

\begin{itemize}
    \item The radial positions of the simulated GCs are drawn from the adopted radial distribution models. 
    The modified Hubble profiles is parametrized as: 
    \begin{equation}
        \Sigma(r)=a\bigg[1+\bigg(\frac{r}{r_{0}}\bigg)^{2}\bigg]^{-b}
    \end{equation}
    where $r_{0}$ is the core radius below which the density in the core flattens, and $b$ is the exponent
    of the power-law profile outside of the core of the GCS. The best-fit parameters of the modified Hubble 
    radial profiles of ten out of the twelve galaxies in the MGS, the total predicted number of GCs in their 
    GCSs and the projected sizes of the GCSs of the MGS hosts~\citep{caso2019,debortoli2022,caso2024} 
    are shown in Table~\ref{tab:main_galaxies_radial}. The radial distances are simulated within an interval 
    whose upper limit is set to the projected size of the GCS, when available.
    \item The azimuthal distribution of the simulated GCSs of MGS galaxies is modeled according to 
    the elliptical geometry of the diffuse stellar light of their host galaxies as derived from FDS 
    data~\citep{iodice2016,iodice2019b} when available, or data of comparable depth and 
    field-of-view. We used the ellipticity and position angles of the elliptical isophotes measured at one effective radius 
    by interpolating the values reported by~\cite{iodice2019b} in the $u$ band for all galaxies in the MGS, 
    except for NGC\,1399~\citep{iodice2016} and NGC\,1336 and NGC\,1351~\citep{lauberts1989}.
    While~\cite{dabrusco2022} found significant azimuthal anisotropies at small galacto-centric radii in the spatial 
    distribution of GCs detected by HST in ten bright galaxies in the Fornax cluster, 
    the assumption of smooth, elliptical angular distribution is mostly valid for GCs detected in shallower data 
    which cannot probe the inner regions of galaxies, although some exceptions have been noted in the 
    literature~(see, for example,~\citealt{cantiello2018} for NGC\,3115,~\citealt{dabrusco2014b} for NGC\,4278,
    and~\citealt{blom2014} for NGC\,4365).
    \item The initial numbers of GCs of the GCSs of the MGS galaxies are assumed to be equal to the total number 
    of GCs derived from the best-fit modified Hubble radial profiles N$_{\mathrm{GCs}}^{\mathrm{(mH)}}$, with the 
    exception of NGC\,1399, whose large GCS, which by size dominates the contributions from other galaxies in the MGS, 
    accounts for N$_{\mathrm{GCs}}^{\mathrm{(mH)}}(\mathrm{NGC\,1399})\!=\!6450\!\pm\!700$~\citealt{dirsch2003} and requires particular 
    care. In their paper,~\cite{dirsch2003} extract 
    GC candidates in a 36$\arcmin$ x 36$\arcmin$ area centered on NGC\,1399 from
    images taken with the CTIO MOSAIC camera in the {\it C} and {\it T1} bands. By applying a {\it T1} magnitude and 
    a {\it C-T1} color cuts to the $\sim\!10450$ point-like sources detected in 
    the field,~\cite{dirsch2003} identify $\sim\!2600\!\pm\!50$ GC candidates in the $[2\arcmin,17\arcmin]$ galacto-centric 
    distance interval, and $\sim$700 within $2\arcmin$ from the center of NGC\,1399. They obtain the total size of the NGC\,1399 GCS
    by averaging two slightly different estimates obtained by extrapolating the best-fit modified
    Hubble radial profile of the GC candidates, after correcting the observed number for incompleteness and 
    contamination. While FDS and data used by~\cite{dirsch2003} have comparable 
    depths, the significant difference between the number of GC candidates detected in the same region ($\sim\!800$ FDS GCs 
    within $2\arcmin$ and $17\arcmin$ vs $\sim\!2600$ from~\citealt{dirsch2003}) can be attributed to the more restrictive 
    GC selection performed by~\cite{cantiello2020}, based on multiple color cuts and optimized to minimize contaminants. 
    We corrected for this difference by  
    rescaling the total size of the NGC\,1399 GCS derived from the modified Hubble radial profile by~\cite{dirsch2003} 
    by the ratio of observed FDS to CTIO GC candidates in the $[2\arcmin,17\arcmin]$ interval. The rescaled 
    N$_{\mathrm{GCs}}^{\mathrm{(mH,scaled)}}(\mathrm{NGC\,1399})\!=\!2750$ is used to calculate
    the observed number of FDS GC candidates as described below.
    \item Each of the N$_{\mathrm{GCs}}^{\mathrm{(mH)}}$ GCs is assigned a $u$ magnitude randomly drawn from 
    the best-fit Gaussian model of the GCLF for the specific host galaxy~\citep{villegas2010}. The number of ACSFCS
    GCs used to model the GCLF and the GCLF parameters of each galaxy are shown in Table~\ref{tab:main_galaxies}.
    The ACSFCS GCLF model is convolved with the $u$-band completeness function of FDS GCs. This 
    function describes the 
    fraction of detected sources as function of their $u$-band magnitude, and is estimated through numerical 
    simulations carried 
    out by $i)$ injecting fake sources in the image, $ii)$ detecting them using the same procedures adopted to 
    measure the photometry of sources adopted by~\cite{cantiello2020}, and then $iii)$ estimating the ratio of 
    detected to injected sources as a function of the injected magnitude. We adopted the completeness obtained 
    using the $u$-band observation of one of the VEGAS target, NGC\,3640, because the $u$-band observations of NGC\,3640 
    and of the FDS fields were acquired under the same conditions (same instrument, exposure time of 3 hours 
    and very similar image quality). Hence, the two fields are expected to be equivalent in terms of 
    photometric depth. For more details on how the completeness function was obtained, see~\cite{mirabile2024}.
    While the probability of a single GC of being selected depends on its color, the dependence of the 
    completeness function on the colors of the sources has not been estimated. For a discussion of how this may affect 
    the interpretation of the colors of GCs located within the boundaries of the spatial structures observed in the residual
    maps of the ICGCs distribution, see Section~\ref{subsec:colorsresidualstructures}. 
    Figure~\ref{fig:gclf_selectionfunction} shows the best-fit GCLFs from~\cite{villegas2010} 
    (after normalizing their peaks to unity) for all galaxies in the MGS with the $u$-band GCs completeness 
    function for FDS observations.
    \item The convoluted GCLF, describing the brightness distribution of the GCs candidates that would be 
    detected in FDS data, is used to calculate the expected number of observed FDS GCs for each galaxy 
    N$_{\mathrm{GCs}}^{\mathrm{FDS}}$
    and to randomly extract the simulated GCs that would be detected in FDS data
    from the general sample of simulated GCs whose radial, azimuthal and luminosity 
    distributions follow the models described above. The final numbers of
    GCs observed in FDS data N$_{\mathrm{GCs}}^{\mathrm{(mH,FDS)}}$
    are corrected for the Eddington bias~\citep{eddington1913}, by assuming an average uncertainty on the $u$ 
    magnitudes for 
    the simulated GCs equal to the mean error on the observed FDS $u$ magnitudes for all FDS sources 
    ($\delta_{u}\!=\!0.15$). After the correction, the N$_{\mathrm{GCs}}^{\mathrm{(mH,FDS)}}$ increase 
    on average by $\sim\!3\%$. 
    \item The detection efficiency of GCs in ground-based data in the central, high surface brightness 
    area in the core of the MGS 
    galaxies is degraded. We estimated the radial dependence of the detection efficiency of FDS GC candidates
    in the central regions of the galaxies by deriving 
    the projected surface density of FDS GC candidates in elliptical annuli centered on the nominal positions
    of the galaxies and with elliptical geometry from the diffuse stellar component of the host galaxy. 
    Since the residual maps of the spatial distribution of observed GCs at large distances from 
    the galaxy centers, where the ICGCs are located, 
    are not affected by the deficits in the cores of the MGS hosts, we do not correct the simulated 
    spatial distribution according to the detailed detection efficiency profiles. The effects 
    of the reduced number of observed number GCs at the smallest galacto-centric distances are mitigated 
    by using the radius (measured along the major axis of 
    the host galaxies) of the innermost detected FDS GC candidate for each galaxy in the 
    MGS sample (called avoidance radius) to discard all simulated GCs whose 
    galacto-centric distances and azimuthal coordinates fall within the elliptical avoidance 
    area of their MGS host galaxy. 
    The rejected simulated GCs are excluded from the sample used to generate the residual maps and from the 
    estimation of the observed excess number of GCs associated with the ICGCs population in the core of the Fornax 
    cluster. The possible loss of ICGCs whose projected positions fall in the centers of the bright galaxies 
    can be corrected for by estimating the number of ICGCs located within the avoidance regions as the result 
    of the product of the average density of observed ICGCs within the core of the Fornax cluster by the total 
    area of the avoidance regions. 
    The final number of simulated GCs N$_{\mathrm{GCs}}^{\mathrm{(mH,FDS,a)}}$ and 
    the avoidance major axes, used to define the inner avoidance areas, are reported in 
    Table~\ref{tab:main_galaxies}. 
\end{itemize}

\begin{figure*}[ht]
    \centering
    \includegraphics[width=0.495\linewidth]{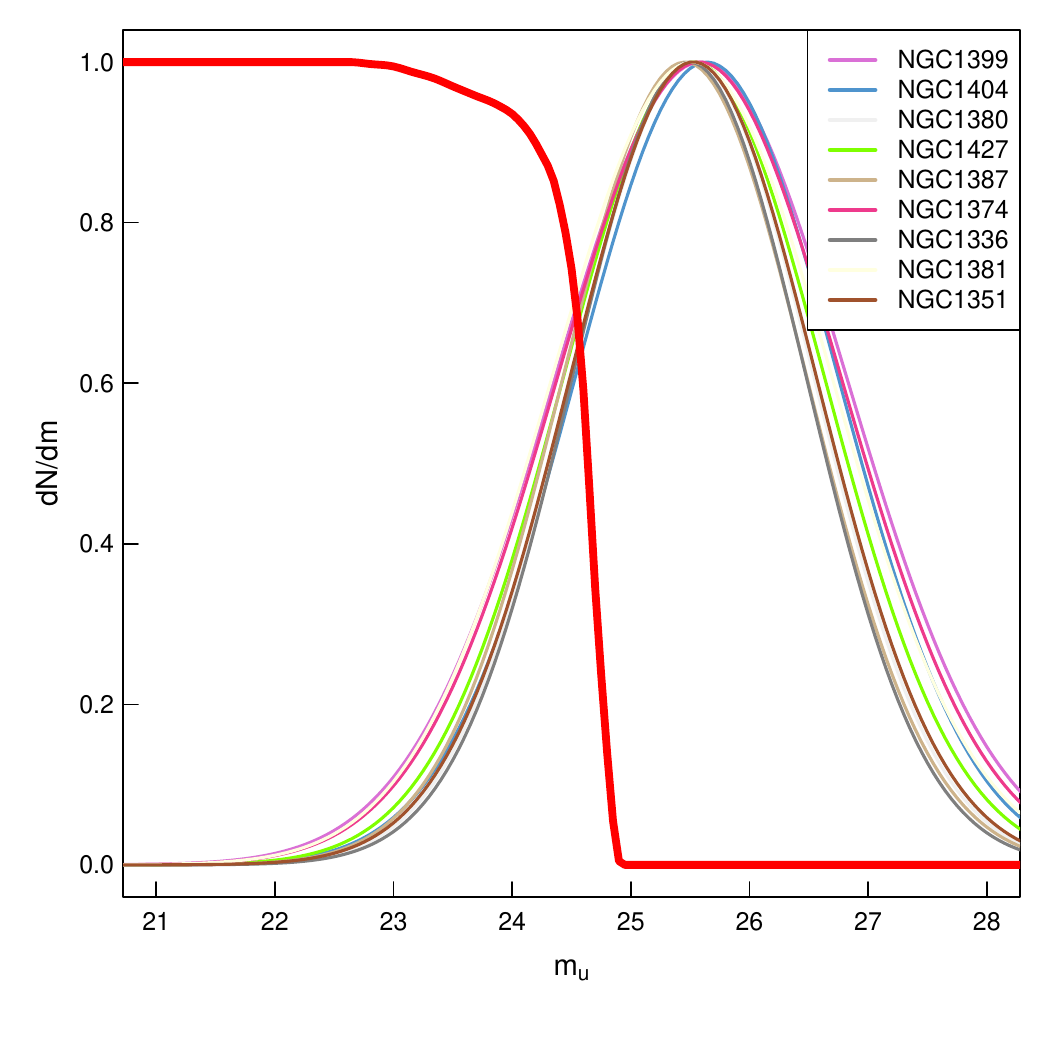}
    \caption{GCLFs for the MGS galaxies derived from ACSFCS 
    observations~\citep{villegas2010}. The peak of the Gaussian functions are 
    normalized to unity. The $u$-band completeness function of GCs detected in the
    FDS observations from~\cite{mirabile2024} is also shown.}
    \label{fig:gclf_selectionfunction}
\end{figure*}

The list of Fornax cluster galaxies in the MGS whose GC systems have been simulated by using the 
ACSFCS models of the GCLF, the modified Hubble models of their radial distributions and the elliptical 
geometry of the host galaxies are shown in Table~\ref{tab:main_galaxies}. 

\begin{deluxetable*}{cccccccccc}
 	\tablecaption{General properties of the MGS host galaxies and their GCSs, sorted by 
        decreasing brightness. The apparent magnitude in the $B$ band and the morphology of 
        the host galaxies are from~\cite{ferguson1989}, 
        while the effective radius along the major axis, the ellipticity and the position
        angle measured at the effective radius the $u$ band are from~\cite{iodice2019b} for 
        all galaxies except NGC\,1399~\citep{iodice2016}, NGC\,1336\ and NGC\,1351~\citep{lauberts1989}, 
        and NGC\,1396~\citep{hilker1999}.
        The GCLF Gaussian parameters have been converted to the 
        $u$ filter from their values in the ACSFCS $g$ band reported by~\cite{villegas2010}.}
 	\label{tab:main_galaxies}
 	\tablehead{
            \colhead{FCC\,ID\tablenotemark{a}} &
 	    \colhead{Name\tablenotemark{b}} &
 	    \colhead{BT$_{\mathrm{mag}}$\tablenotemark{c}} &   
            \colhead{Morph\tablenotemark{d}} &
            \colhead{R$_{e,\mathrm{gal}}^{(u)}$\tablenotemark{e}} &
            \colhead{e\tablenotemark{f}} &
            \colhead{PA\tablenotemark{g}} &   
 	    \colhead{N$^{\mathrm{(GCLF)}}_{\mathrm{GCs}}$\tablenotemark{h}} &
 	    \colhead{$\mu_{u}$\tablenotemark{i}} &
 	    \colhead{$\sigma_{u}$\tablenotemark{j}} 
            }
        \startdata
      213 & NGC\,1399 & 10.6 & E0 & 352.2$\pm$6.1 & 0.18$\pm$0.23 & 98.2$\pm$2.5 & 1074&                   25.59$\pm$0.05 & 1.231$\pm$0.038 \\
      219 & NGC\,1404 & 10.9 & E2 & 97.5$\pm$0.1 & 0.18$\pm$0.01 & 156.9$\pm$1.8 &  380&                 25.64$\pm$0.07 & 1.110$\pm$0.058  \\
      83 & NGC\,1351 & 10.9 & E5 & 32.5$\pm$0.5$^{\star\star}$ & 0.35$\pm$0.01 & 137$\pm$1 & 274&     25.53$\pm$0.07 & 1.040$\pm$0.058 \\
      167 & NGC\,1380 & 11.3 & S0a & 57.9$\pm$0.3& 0.47$\pm$0.01 & 5.0$\pm$0.3 &  424&                    25.52$\pm$0.06 & 1.022$\pm$0.046   \\
      276 & NGC\,1427 & 11.8 & E4 & 56.1$\pm$0.5 & 0.29$\pm$0.01 & 79.1$\pm$0.9 &  361&                   25.53$\pm$0.07 & 1.102$\pm$0.063 \\
      147 & NGC\,1374 & 11.9 & E0 & 38.3$\pm$0.1 & 0.1$\pm$0.01 & 118.2$\pm$1.9 &  320&                    25.58$\pm$0.09 & 1.197$\pm$0.067 \\
      184 & NGC\,1387 & 12.3 & SB0 & 32.3$\pm$0.1 & 0.09$\pm$0.01 & 107.5$\pm$1.8 &  306&                  25.46$\pm$0.07 & 1.029$\pm$0.054 \\ 
      170 & NGC\,1381 & 13.0 & S0  & 20.9$\pm$0.5$^{\star\star}$ & 0.17$\pm$0.01& 138.2$\pm$0.5 &  71&    25.52$\pm$0.19 & 1.182$\pm$0.167 \\
      47  & NGC\,1336 & 13.3 & SA0 & 20.4$\pm$0.5$^{\star}$& 0.32$\pm$0.01&  20$\pm$1&  276&             25.49$\pm$0.07 & 0.988$\pm$0.053 \\
      177 & NGC\,1380A& 13.3 & S0 & 31$\pm$1 & 0.75$\pm$0.01& 176.1$\pm$0.9 & 70 & 25.40$\pm$0.14  & 0.928$\pm$0.108   \\
      190& NGC\,1380B & 13.9 & S0 & 26.3$\pm$0.1 & 0.09$\pm$0.01 & 204$\pm$35 & 156 & 25.43$\pm$0.09  &0.932$\pm$0.090   \\
      202 & NGC\,1396 & 14.8 & SA0 & 10.7$\pm$0.1 & 0.5$\pm$0.1& 70$\pm$1&  232&                           25.49$\pm$0.08 & 1.101$\pm$0.068   \\
      \enddata
     \tablecomments{(a): FCC identifier;
                    (b): NGC name; 
                    (c): BT magnitude;
                    (d): Morphology;
                    (e): Effective radius in the $u$ band~[arcsec];
                    (f): Ellipticity;           
                    (g): Position angle~[degrees]; 
                    (h): Total number of GCs used to determine the GCLF~\citep{villegas2010};
                    (i): Turnover magnitude of the GCLF in the $u$ band~\citep{villegas2010};
                    (j): Dispersion of the GCLF in the $u$ band~\citep{villegas2010};
                    } 
\end{deluxetable*} 

\begin{deluxetable*}{cccccccccc}
 	\tablecaption{Properties of the models used to simulate the radial distribution of the MGS GCSs. 
        The best-fit parameters of the modified Hubble radial profiles and the 
        total number of GCs are from~\cite{caso2024} for NGC\,1399, NGC\,1427, 
        NGC\,1374; from~\cite{debortoli2022} for NGC\,1404, NGC\,1380, NGC\,1387, NGC\,1380A and from~\cite{caso2019}
        for NGC\,1336, NGC\,1351 and NGC\,1380B. The projected size of the GCSs are all from~\cite{debortoli2022} with    
        the exception of NGC\,1399~\citep{dirsch2003}. The nominal number of GC candidates for NGC\,1399 from~\cite{dirsch2003} 
        has been reduced as described in Section~\ref{subsubsec:galcomponent} to 
        N$_{\mathrm{GCs}}^{\mathrm{(mH,scaled)}}(\mathrm{NGC\,1399})\!=\!2750$.}
 	\label{tab:main_galaxies_radial}
 	\tablehead{
            \colhead{FCC\,ID\tablenotemark{a}} &
 	    \colhead{Name\tablenotemark{b}} &
            \colhead{N$_{\mathrm{GCs}}^{\mathrm{(mH)}}$\tablenotemark{c}} &    
            \colhead{a$^{\mathrm{(mH)}}$\tablenotemark{d}}&
            \colhead{r$_{0}^{\mathrm{(mH)}}$\tablenotemark{e}}&
            \colhead{b$^{\mathrm{(mH)}}$\tablenotemark{f}}&
            \colhead{r$_{L}^{\mathrm{(mH)}}$\tablenotemark{g}}&
            \colhead{N$_{\mathrm{GCs}}^{\mathrm{(mH,FDS)}}$\tablenotemark{h}}&
            \colhead{r$^{\mathrm{(a)}}$\tablenotemark{i}}&
            \colhead{N$_{\mathrm{GCs}}^{\mathrm{(mH,FDS,a)}}$\tablenotemark{l}}
 	    } 
        \startdata
        213 & NGC\,1399 & 6450$\pm$700$^{\star}$    & 2.26$\pm$0.01 & 0.83$\pm$0.02 & 0.81          & 15  & 521.1$\pm$148.1   & 0.8 & 478.1$\pm$135.7 \\
        219 & NGC\,1404 & 311$\pm$8                 & 2.13$\pm$0.02 & 0.50$\pm$0.05 & 0.94$\pm$0.07 & 16.3& 20.9$\pm$1.7      & 0.7 & 16.7$\pm$2.2 \\
        83  & NGC\,1351 & 370                       & 2.49$\pm$0.03 & 0.28$\pm$0.02 & 1.14$\pm$0.03 & 6.8 & 66.2$\pm$0.3      & 0.9 & 35.5$\pm$4.2  \\\
        167 & NGC\,1380 & 504$\pm$77                & 2.29$\pm$0.02 & 0.66$\pm$0.05 & 1.36$\pm$0.08 & 6.9 & 91.6$\pm$16.5     & 0.75& 61.4$\pm$11.8 \\
        276 & NGC\,1427 & 470$\pm$40                & 2.26$\pm$0.03 & 0.44$\pm$0.02 & 1.00          & 4.6 & 94.8$\pm$8.4      & 1   & 62.9$\pm$6.9 \\
        147 & NGC\,1374 & 360$\pm$17                & 2.29$\pm$0.04 & 0.45$\pm$0.02 & 1.15          & 2.7 & 75.5$\pm$3.6      & 0.75 & 34.4$\pm$4.6  \\
        184 & NGC\,1387 & 299$\pm$31                & 2.06$\pm$0.03 & 0.81$\pm$0.09 & 1.34$\pm$0.11 & 8.05& 55.2$\pm$6.9      & 0.75& 37.0$\pm$5.8 \\
        170 & NGC\,1381 & -                         & -             & -             & -             & -   & 13.8              & 0.85 & 13\\
        47  & NGC\,1336 & 355                       & 2.28$\pm$0.03 & 0.41$\pm$0.01 & 1.24$\pm$0.07 & 6.3 & 64.8$\pm$0.5      & 1   & 35.5$\pm$4.1  \\
        177 & NGC\,1380A& 67$\pm$9                  & 1.85$\pm$0.04 & 0.34$\pm$0.05 & 1.09$\pm$0.09 & 4.7 & 11.7$\pm$2.0      & 1   & 7.7$\pm$2.1\\
        190 & NGC\,1380B& 170                       & 2.31$\pm$0.04 & 0.82$\pm$0.03 & 1.37$\pm$0.09 & 3.9 & 32.1$\pm$0.1      & 1   & 10.2$\pm$2.7\\
        202 & NGC\,1396 & -                         & -             & -             & -             & -   & 1.9               & 0.75& 1  \\
        \enddata
     \tablecomments{(a): FCC identifier;
                    (b): NGC name; 
                    (c): Total number of GCs in the GCS of the host galaxy derived from modified Hubble radial model of the GCS;
                    (d): Central density of the modified Hubble radial profile;
                    (e): Flattening radius of the modified Hubble radial profile~[arcmin];
                    (f): Exponent of the power-law section of the modified Hubble profile;
                    (g): Size of the GCSs used to fit the modified Hubble profile~[arcmin];
                    (h): Total number of simulated FDS GCs assuming the modified Hubble model for the galaxy radial profile;
                    (i): Avoidance major axis~[arcmin];
                    (l): Average number of simulated FDS GCs located outside of the avoidance area;
                    } 
\end{deluxetable*} 

For galaxies in the SGS~(not included in Table~\ref{tab:main_galaxies}), a set of assumptions regarding 
the dependence of the properties of their GCSs on the host magnitude were made to estimate the expected total and 
observed number of GCs and simulate the GCS spatial distribution. 
For each galaxy: 

\begin{itemize}
    \item The total number of GCs, N$_{\mathrm{GCs}}^{\mathrm{(tot)}}$, is derived from the 
    correlation between the absolute magnitude M$_{V}$ of the host galaxy 
    and its specific frequency $S_{N}$~\citep[][Figure 11]{harris2013}: 
    \begin{equation}
        S_{N}=N_{\mathrm{GCs}}\cdot10^{0.4(\mathrm{M}_{V}+15)}
    \end{equation}
    Each realization of the specific frequency of the galaxy is randomly drawn from the observed range of $S_{N}$ values 
    as a function of the host visible absolute magnitude~\cite[Figure~11]{harris2013} obtained from the 
    BT$_{\mathrm{mag}}$ magnitude available in the FCC~\citep{ferguson1989} using conversion from the blue to
    visible band from~\cite{blair1982}, the average distance modulus of the Fornax
    cluster of galaxies $(m-M)_{0}\!=\!31.50\!\pm\!0.03$~\citep{blakeslee2009}. The $S_{N}$ are drawn from different 
    intervals based on the morphology of the host galaxy, i.e. Elliptical or dwarf Elliptical (\textit{E*} or~\textit{dE*}), 
    lenticular (\textit{S*} or \textit{dS*}) or spiral/irregular galaxy according to the classification available in 
    FCC~\citep{ferguson1989}. 
    The binning in absolute magnitude used for different morphological types and the intervals of $S_{N}$ where the 
    simulated values are drawn from are discussed in more details in the Appendix~\ref{sec:appendix1}.
    \item The observed number of GCs, N$_{\mathrm{GCs}}^{\mathrm{(FDS)}}$, is derived from the 
    total number of GCs N$_{\mathrm{GCs}}^{\mathrm{(tot)}}$ by assuming that the GCLF of the simulated GCSs 
    is described by a Gaussian whose $g$-band turnover magnitude $\mu_{g}$ and dispersion $\sigma_{g}$ are 
    determined 
    from the absolute magnitude M$_{B}$ of the host galaxy using the~\cite{jordan2007} relations derived for the Virgo 
    cluster galaxies:
    \begin{equation}
        \mu_{g}\!=\!-7.2\!\pm\!0.2
    \end{equation}
    \begin{equation}
        \sigma_{g}\!=\!(1.14\!\pm\!0.01)-(0.100\!\pm\!0.007)(M_{B}\!+\!20)
    \end{equation}
    The turnover magnitudes estimated from this relation are corrected using the~\cite{blakeslee2009} distance modulus of the Fornax
    cluster of galaxies $(m-M)_{0}\!=\!31.50\!\pm\!0.03$ and converted to the FDS $u$ filter using the 
    correction $(u-g)_{\mathrm{corr}}^{\mathrm{FDS}}\!=\!1.4\!\pm\!0.3$ discussed above. The N$_{\mathrm{GCs}}^{\mathrm{(tot)}}$ 
    GCs are distributed according to the GCLF shape described by the parameters thus determined. Then, the observed 
    N$_{\mathrm{GCs}}^{\mathrm{(FDS)}}$ is determined, as done for the main set of host galaxies, by convolving qthe 
    simulated magnitude distribution of simulated GCs with the FDS selection function.   
    \item The slopes of the GCSs radial profiles are correlated with the absolute magnitude of the hosts~\citep{harris1991}. As described
    in the Appendix~\ref{sec:appendix2}, the radial 
    profile of the simulated N$_{\mathrm{GCs}}^{\mathrm{(obs)}}$ GCs is assumed to follow  
    a power-law whose slope $\alpha_{\mathrm{GCs}}$ can be determined from the absolute magnitude M$_{V}$ of the host galaxy using 
    a linear relation derived from the data shown in Figure~5.2 of~\cite{ashman2008}: 
    \begin{equation}
        \alpha_{\mathrm{GCs}}\!=\!-8.079-0.296 M_{V}
    \end{equation}
    The local uncertainty on the regression has been estimated as the local dispersion of the linear fit
    (see discussion in Appendix~\ref{sec:appendix2}). 
\end{itemize}

Each simulation will produce a slightly different total number of GCs associated with each 
Fornax cluster galaxy because the parameters for a given simulation are allowed to vary within 
the observed uncertainties of the models used. An additional random noise component, randomly from the 
$[-10\%,+10\%]$ interval centered on the observed value of each model parameter, is added to the
parameters used in each simulation. The
results discussed in this paper have been observed by performing 10,000 simulations, where the 
homogeneous component and the contribution of each galaxy in the MGC and the SGC are independently
simulated. The average total number of simulated GCs of the galaxies' component is 
$N_{\mathrm{GCs}}^{\mathrm{(gal)}}\!=\sum_{i=1}^{N}N_{\mathrm{GCs}}^{(\mathrm{gal}_{i})}\!=\!1212.2\!\pm\!143.8$
($\sim\!21.4\!\pm\!2.5\%$ of the total number of FDS GC candidates), with the MGS and SGS accounting 
for $779.4\!\pm\!136.9$ and $433.1\!\pm\!43.9$ GCs ($\sim\!14.9\%$ and $\sim\!8.3\%$ of the total sample, 
respectively).

\subsection{Residual maps}
\label{subsec:residuals}

The residual maps of the GC distributions are determined according to the following steps:

\begin{itemize}
    \item Density maps of the observed spatial distribution of GCs are obtained with the $K$-Nearest Neighbor
    (KNN) method~\citep{dressler1980} on a fixed, regular rectangular grid that covers the full imaged area 
    of the cluster. The density value at the center of each cell is:
    \begin{equation}
        D_K\!=\!\frac{K}{A_D(d_{K})}
        \label{eq:density}
    \end{equation}
    where $K$ indicates the $K$-th closest GC (or {\it neighbor}) and $A_D(d_{K})\!=\!\pi\cdot\!d_{k}^{2}$ 
    is the area of the circle with radius equal to the distance of the $K$-th neighbor. 
    \item Multiple simulations of the observed GCs system are produced using a Monte Carlo 
    approach. The procedures used to generate the two components making up the simulated GCs spatial distribution, 
    i.e. the homogeneous and galaxies' components, are described in Section~\ref{subsec:simulcomponents}. 
    \item Density maps of all the distinct realizations of the simulated GCs 
    spatial distribution are generated on the same spatial grid and for each different value of the $K$ parameter
    considered in this analysis.
    \item The residual maps for each distinct $K$ values are calculated by subtracting, on a cell-by-cell basis,  
    the average density of all the simulated maps from the observed density map for the same $K$. Accordingly, 
    the residual value R$_{i}$ of the $i$-th cell of the grid is defined as: 
    \begin{equation}
        R_{i}\!=\!\frac{(O_{i}-<\!S\!>_{i})}{<\!S\!>_{i}}
        \label{eq:residual}
    \end{equation}
    where $O_{i}$ and $<\!S\!>_{i}$ are the observed density and the average of the density 
    from all the simulations in cell $i$. 
\end{itemize}

Since the simulated density values for each cell of the residual maps are well approximated by Gaussians, the statistical 
significance of the residual value in each cell (single-pixel significance) can be estimated as the deviation from the 
expected, mean value of the distribution of simulated densities. The spatial structures in the residual maps are defined 
as sets of adjacent cells with average single-cell significance larger 1$\sigma$. The cumulative significance of the spatial 
structures discussed in what follows, instead, is estimated by estimating the frequency of structures with comparable area 
and shape across residual maps obtained from single simulated distributions of GCs. The same procedure used for the detection 
of structures in the observed residual map is applied to each of the single-simulation residual maps. The total statistical 
significance of each observed residual structure is then expressed as the fraction of simulated density maps where mock residual 
structures with comparable features have been detected.

The free parameter $K$ of the KNN method measures the expected scale of the spatial structures emerging
from the residual maps and the density contrast of these structures over the average local density. 
Different $K$ values highlight spatial features at different scales: small $K$ values allow 
the exploration of small features, while larger values of $K$ bring out more extended structures. The loss 
of spatial resolution for large $K$ values is balanced by the smaller relative fractional error 
which is proportional to the inverse of the square root of $K$:

\begin{equation} 
    \frac{\sigma_{D(K)}}{D(K)}=\frac{1}{\sqrt{K}} 
    \label{eq:fracerror}
\end{equation}

Moreover, larger values of $K$ are more suitable to detect structures located in areas where the 
density of GCs is larger, because only high-contrast structures can be reliably detected over 
high-density background. Conversely, smaller $K$'s are more apt at detecting structures in regions 
where the total number of GCs is smaller, as in the outskirts of the cluster. The distributions of 
simulated density values for each cell of the final residual maps are well approximated by Gaussian 
distributions, thus simplifying the estimation of the statistical significance of each cell. 

\section{Residual structures and their characterization}
\label{sec:results}

In order to characterize the spatial distribution of the diffuse population of ICGCs,
we modeled the homogeneous and galaxies' components of the observed spatial distribution of FDS 
GC candidates as described in Section~\ref{subsubsec:homocomponent} and Section~\ref{subsubsec:galcomponent} 
respectively, by drawing 10,000 independent simulations. We derived the residual maps 
as described in Section~\ref{subsec:residuals}, using $K\!=\!(5, 10, 25, 50, 75, 100)$; the residual 
maps for all the $K$ values are shown and discussed in Appendix~\ref{sec:appendix4}. 
The residual maps are estimated on a regular grid of cells with equal angular
extents $\approx\!1.25\arcmin$ ($\sim\!0.021\degree$) along the Right Ascension and Declination 
axes, corresponding to physical 
linear size of $\approx$8 kpc assuming the distance of NGC\,1399 for all members of the
Fornax cluster. Each cell has area $\sim$1.56(\arcmin)$^{2}$ ($\sim4.3\cdot10^{-4}(^{\circ})^{2}$) 
and, on average over the whole FDS field-of-view, is occupied by $\sim$0.1 GC candidates. 
This choice guarantees that the average occupancies of cells located in different regions of the 
cluster are comparable: in particular, this value yields $\sim$0.103 GC candidates/cell in the area outside of 
the virial radius of the cluster and $\sim$ 0.109 GC candidates/cell inside the virial radius. Significantly larger or 
smaller cell sizes would produce large differences in the average cell occupancy as a function
of their cluster-centric distance that would make difficult to meaningfully evaluate the statistical 
significance of spatial structures and compare them.

In order to probe the effect of relatively small differences in cell size on the final residual 
maps of the spatial distribution of candidate GCs observed, following a similar approach 
to~\cite{dabrusco2015,dabrusco2022}, we derived the residual maps using 10 different, linearly 
spaced cell sizes (along both axes) within a $\pm$30\% interval centered on the adopted cell size 
for each $K$ value used in this paper and for the same set of simulations parameters employed
for the experiment discussed in the manuscript. The direct comparison of the different residual 
maps~(Figure~\ref{fig:residuals_cell_sizes}) rules out significant systematic effects on the spatial 
features observed. Further discussion of the tests can be found in Appendix~\ref{sec:appendix4}.

\begin{figure*}[ht]
    \centering
    \includegraphics[width=\linewidth]{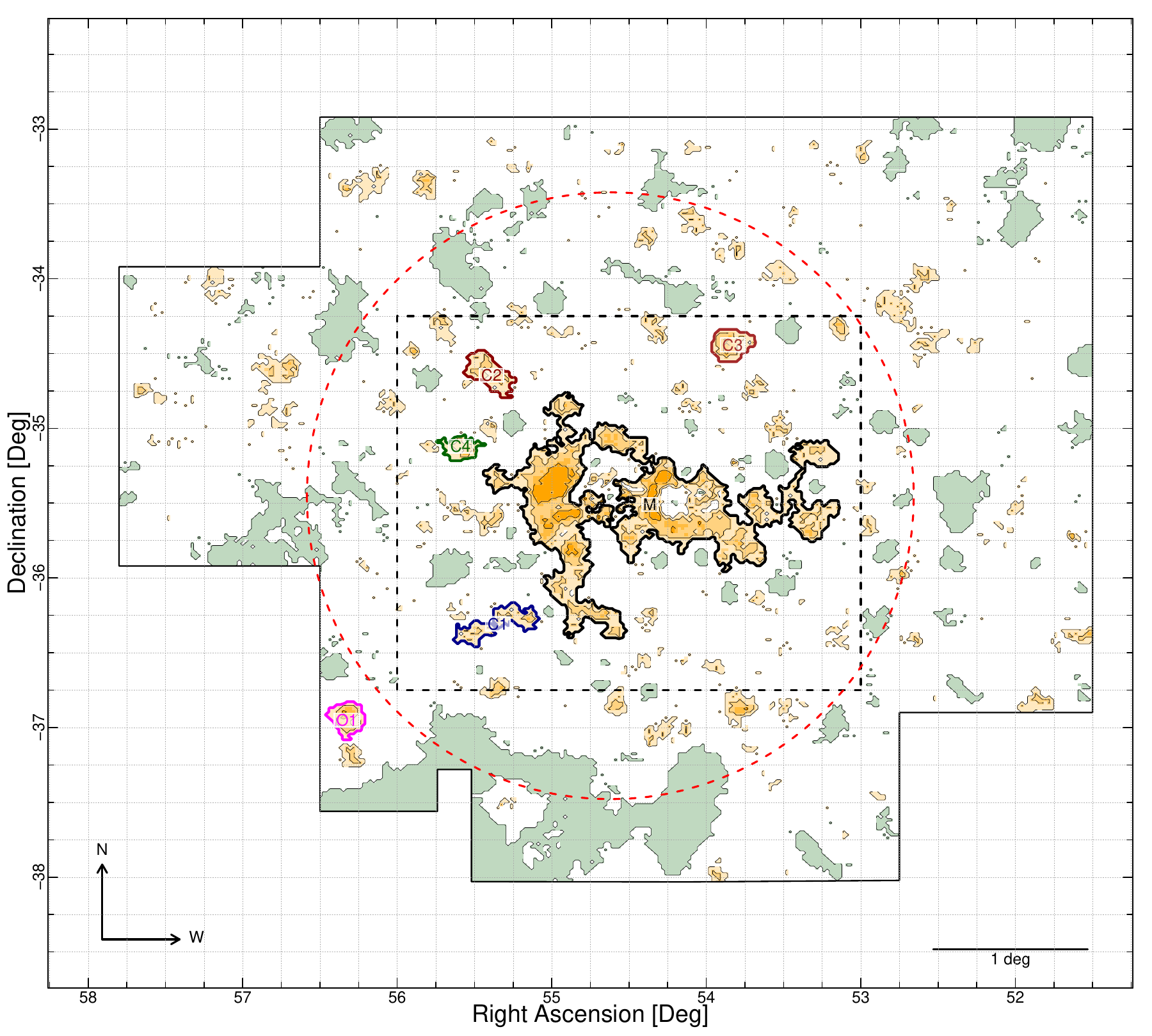}
    \caption{Residual map of the spatial distribution of GC candidates obtained with 
    KNN density maps with $K\!=\!10$. Orange and green pixels indicate over-dense and 
    under-dense regions. Iso-residual contours (thin black lines) and green/orange shading
    indicate pixels with single-pixel 
    statistical significance $-3\sigma,-2\sigma,-1\sigma,1\sigma,2\sigma,3\sigma$ (no
    residual pixels with single-pixel residual $<\!-1\sigma$ are present in the map). The 
    thick black line shows the contours of the dominant over-density in the core 
    region of the Fornax cluster (M), bounded by the continuous 1$\sigma$ single-pixel
    significance iso-residual contour. The red dashed circle represents the virial radius 
    of the Fornax cluster~\citep{drinkwater2001}, and the black dashed rectangle shows
    the region of the core of the cluster displayed in Figure~\ref{fig:core_residual_k10}.}
    \label{fig:residual_k10}
\end{figure*}

In what follows we will focus on the analysis of the residual map~(Figure~\ref{fig:residual_k10}) of 
the observed distribution of FDS GC candidates obtained with the following parameters: $K\!=\!10$, 
cell size $\approx\!1.25\arcmin$, density of the homogeneous
component derived from the area outside of the virial radius of the cluster (excluding sources within 
5r$_{\mathrm{eff}}$ from the position of FCC galaxies), and radial and azimuthal distributions of
simulated GCs of MGS galaxies following the best-fit modified Hubble profiles and elliptical geometry
of the host galaxies, respectively. We will describe the
spatial features whose boundaries are defined by the iso-residual contours associated with adjacent groups 
of $N_{\mathrm{pix}}\!\geq\!30$ pixels whose average single-pixel significance is $\geq\!1\sigma$ and
total statistical significance $\geq\!3\sigma$~(see Section~\ref{subsec:residuals} for details). 
Specific, smaller areas of large, enhanced positive residuals located within larger spatial structures 
will also be described.
As discussed in Section~\ref{subsec:residuals}, different values of the parameter $K$
highlight different spatial scales and contrast levels between a background population and 
the excess of candidate GCs. The best $K$ value for the reconstruction of the residual
map of a given distribution of points (and reference model) cannot be determined from first
principle, but it can be determined based on additional, independent information. In this case, 
we choose $K\!=\!10$ to characterize the residual maps of the GC candidates
because it produces structures that have sizes similar to the sizes of the spatial structures
observed in the density maps derived from the same dataset~\citep{dabrusco2016,cantiello2020}.
Table~\ref{tab:residual_maps_features} displays the main properties of the residual structures 
and positive residual enhancement regions described below.  

\subsection{The spatial structure of ICGCs in the core of the Fornax cluster}
\label{subsec:icgcstructure}

Based on the average numbers of simulated GCs associated with the spatially homogeneous 
component~($4021.6\!\pm\!463.2$, see Section~\ref{subsubsec:homocomponent})
and the galaxies' component~($1212.2\!\pm\!143.8$, Section~\ref{subsubsec:galcomponent}), both estimated
using the standard set of simulations parameters, the total average number of simulated GCs is 
$5230.1\!\pm\!493.8$ ($\sim\!92.3\!\pm\!8.2$\% of the observed GC candidates), 
corresponding to an average excess number of observed FDS GC candidates $422.9\!\pm\!409.6$ 
($\sim\!7.5\%$ of the total FDS sample) over the whole imaged field, where the uncertainty on the 
excess number is reported as the standard deviation of the distribution of excess numbers estimated 
from each of the 10,000 simulated distributions of GC candidates
generated as described in Section~\ref{sec:method}. The large uncertainties on numbers of simulated
GCs and the excess of observed GC candidates result from the conservative
approach to the estimation of the variance of the contribution to the global simulated GC populations of 
GCSs~(Section~\ref{subsubsec:galcomponent}), derived from the uncertainty on the model parameters 
for the single host, and the homogeneous component~(Section~\ref{subsubsec:homocomponent}), which 
takes into account cross-field variations of its density due to statistical fluctuations and potential presence
of spatial structures across the FDS f.o.v. of the population of contaminants. Moreover, the source 
density for the homogeneous component is potentially over-estimated, as discussed
in Section~\ref{subsubsec:homocomponent}, since the regions outside or partly overlapping with 
the cluster virial radius that have been used to estimate its
density may contain
members of the diffuse ICGC populations or GCs associated with faint and low surface brightness 
hosts (like UDGs, see~\citealt{lomelinunez2025}) not included in the 
FCC and, for this reason, not excluded from the density estimation\footnote{The 
excess of observed GC candidates increases to $998.1\!\pm\!639.7$ if the density of the homogeneous component 
from the control field density for the homogeneous component $d^{(1)}_{hom}\!=\!161.2\!\pm\!29.2$ GCs/$(^{\circ})^{2}$ 
~(see Section~\ref{subsubsec:homocomponent} for details) is used.}. 

The dominant feature of the GCs residual map is a large, geometrically complex structure located in 
the core of the Fornax cluster (M) and roughly centered on the position of NGC\,1399, whose position
and shape match those of the GC over-density attributed to a rich population of 
ICGCs~\citep{dabrusco2016,cantiello2020}. 
This structure is defined by the longest continuous, bounded iso-residual contour corresponding to a 
$1\sigma$ single-pixel 
significance in the residual map for $K\!=\!10$. The observed GCs excess number within the boundaries of 
M (Table~\ref{tab:residual_maps_features}) is N$_{\mathrm{GCs}}^{(\mathrm{obs,exc})}\!=\!627.1\!\pm\!74.3$
($989.5\!\pm\!138.3$ if the lower density 
for the homogeneous component from the four control fields described in Section~\ref{subsubsec:homocomponent} 
is used), larger than the GC excess number over the full area covered by FDS observation {\bf($422.9\!\pm\!409.6$)}. 
This discrepancy can be explained by taking into account the effect of the presence of areas of negative residuals in 
the outskirts of the cluster 
(see Figure~\ref{fig:residual_k10}):
in such regions the number of simulated GCs is smaller than the number of observed GC candidates (mostly
associated with the homogeneous component), and their contributions reduce the
total GC excess number but do not affect the local excess in the core of the cluster and within 
the boundaries of structure M. The total number of ICGCs contained in M can be calculated 
by assuming that the observed GC excess number in M is entirely associated with the intra-cluster GCs 
population: by correcting for the incompleteness using a Gaussian GCLF whose turn-over magnitude and dispersion 
are fixed to the mean values of the best-fit parameters from the ACSFCS GCLFs of all MGS galaxies~\cite{villegas2010}
($\bar{\mu}_{u}\!=\!25.53\!\pm\!0.09$, $\bar{\sigma}_{u}\!=\!1.10\!\pm\!0.08$), and for the geometric effect
due to the removal of the avoidance regions around the center of the brightest galaxies in the FDS f.o.v.~(see 
Section~\ref{subsubsec:galcomponent})
we find that the ICGCs account for N$_{\mathrm{ICGCs}}\!=\!6750\!\pm\!554$ (N$_{\mathrm{ICGCs}}\!=\!10650\!\pm\!1090$ 
for the alternative estimate of the homogeneous component density).

\begin{figure*}[ht]
    \centering
    \includegraphics[width=\linewidth]{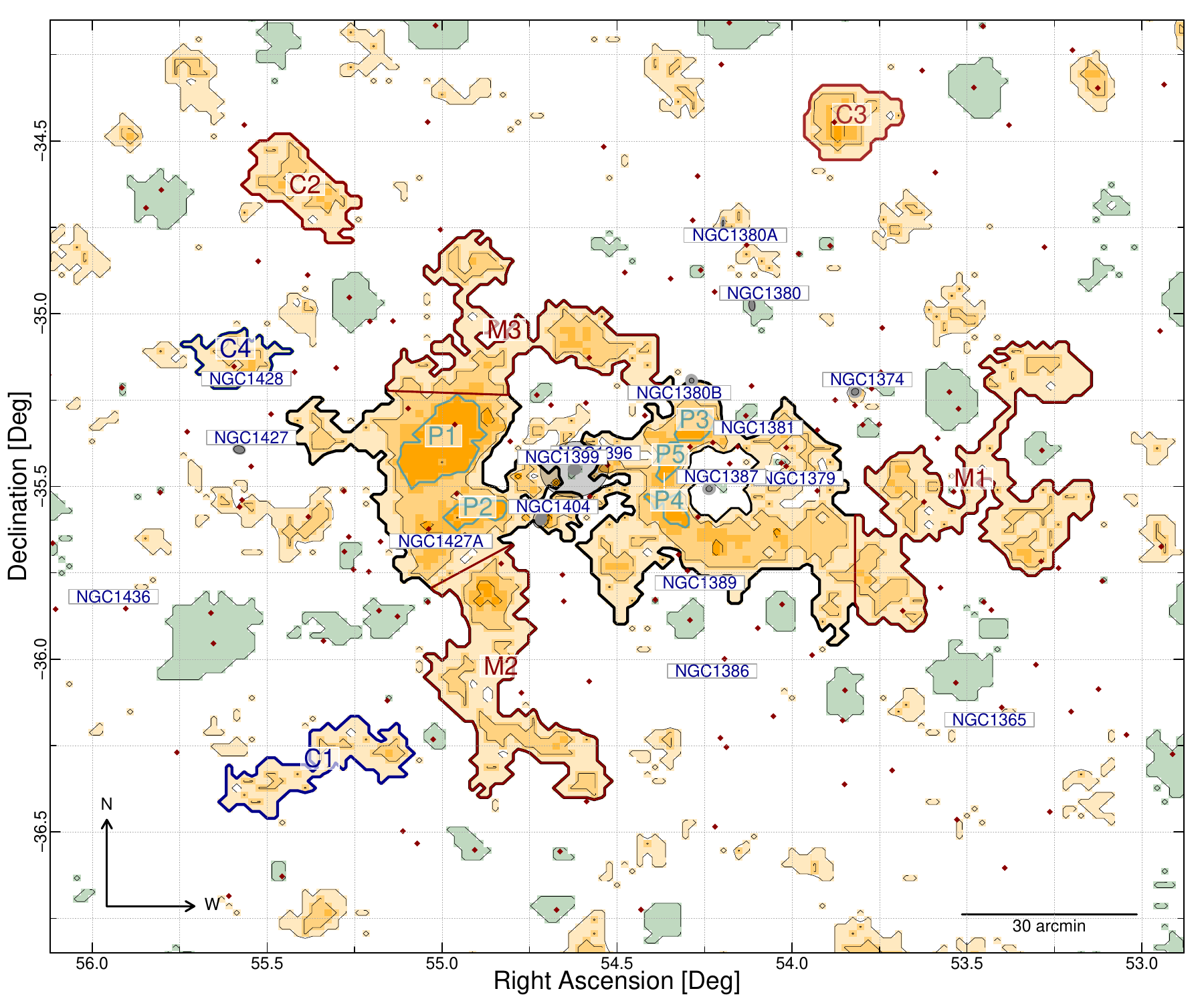}
    \caption{Residual map of the spatial distribution of all FDS GCs in the core of the Fornax 
    cluster, obtained with with $K\!=\!10$. The color lines show the boundaries of the spatial features
    of the residual map described in the text, that have been selected by 
    identifying spatial substructure that significantly deviate from the elliptical model of the 
    main over-density M (see Section~\ref{sec:results} for details). The red diamonds show the 
    positions of the FCC galaxies in the SGS, while the annotated larger blue symbols indicate MGS galaxies. 
    The gray-shaded circles represent five effective radii in the $g$ band for the main galaxies
    in the plot~\citep{iodice2019b}. Color coding of the residual maps and isoresidual contours 
    are defined as in Figure~\ref{fig:residual_k10}.}
    \label{fig:core_residual_k10}
\end{figure*}

The structure M (thick black line in 
Figures~\ref{fig:residual_k10} and~\ref{fig:core_residual_k10}) has an area of $\sim0.9(^{\circ})^{2}$ and
an elongated shape that can be modeled as an ellipse with major axis aligned 
along the W-E direction, a small $\sim4.2\degree$ tilt towards S in the S-W quadrant, and ellipticity 
$\epsilon_{\mathrm{M}}=\!1-b/a\!\sim\!0.77$. The length of M, measured as the major axis of the ellipse, 
is $\sim\!2.26\degree\!\pm\!0.08\degree$, and the minor axis of the 
best-fit ellipse, is $\sim0.53\degree\!\pm\!0.09\degree$, 
measured along the Dec. axis.~\cite{cantiello2020} observed a broader (width $\sim\!0.89\!\pm\!0.03\degree$) 
and more extended (length $\sim\!2.6\!\pm\!0.2\degree$) central over-density, with similar main axis inclination and tilt. 

M displays multiple areas of high positive residuals~(Figure~\ref{fig:core_residual_k10}). The most 
extended among these regions of enhanced positive residuals contained within M (labeled with the letter P) 
are discussed in what follows. The main high residuals region is contained in an approximately 
rectangular region (R.A.$\in\![54.75\degree,55.25\degree]$, 
Dec.$\in\![-35.75\degree,-35.25\degree]$), including the positions of NGC,1427, located east of the center of M 
and the major galaxies NGC\,1399, NGC\,1396 and NGC\,1404. The two distinct residual features in this area, 
P1 and P2, are associated with the northern and southern positive residual peaks, respectively. Three other areas
featuring significant GC excess (P3, P4 and P5), all located east of NGC\,1387, are embedded in a larger
region of enhanced residuals that borders on its western and southern sides with the approximately
circular area of mostly neutral residuals centered around NGC\,1387. P3, located S and E of the positions of 
NGC\,1380B and NGC\,1381 respectively, in conjunction with 
P4 and P5, suggests the existence of a larger, coherent residual enhancement extending, along a N-S direction, 
towards NGC\,1389, that can be interpreted as the relic on the Fornax cluster ICGC population of the extended halo 
of the NGC\,1399 and NGC\,1387 GC systems. 

M can be further characterized by distinguishing and discussing separately spatial sub-structures which 
deviate from the simple elliptical model of the structure. In what follows, we will describe the main 
spatial features shown in Figure~\ref{fig:core_residual_k10} in order of decreasing size:

 \begin{itemize}
    \item The largest distinct morphological feature of the core structure (M1) is located in the western 
    section of M, and extends along the SE-NW direction with 
    Declination in the $[53.15\degree,53.75\degree]$ interval. Its jagged, irregular morphology suggests 
    that this spatial structure is the merging of multiple, distinct, roughly circular overdensities 
    connected by thin, filamentary bridges. M1 contains $~\sim\!42$ excess GCs and does not overlap with 
    the positions of bright galaxy, although 
    NGC\,1374 and NGC\,1365 are respectively located $\sim\!0.3\degree$ NE and $\sim\!0.35\degree$ 
    SW of its borders. 
    \item M2 extends radially from M towards the S direction, and spans $\sim\!0.65\degree$ 
    ($[-35.75\degree,-36.4\degree]$) 
    along the Declination axis. M2 is connected to M by a relatively narrow bridge located
    S of the positions of galaxies NGC\,1427A and NGC\,1404, in a region devoid of bright galaxies. 
    Its shape and orientation suggest that M2 is the southern extension of the enhanced positive 
    residual area comprising P1 and P2. M2, whose area is $\sim\!0.11(^{\circ})^{2}$,  
    contains 56 GC candidates, with an excess of 33.0 GCs over the expected number. 
    \item A third spatial structure (M3) is connected to the main body of M 
    in two different points: it extends from the northern side of the main over-density region 
    (containing P1 and P2) towards N, and branches off towards W until it reaches P1 through 
    a thin filamentary substructure. M3 spans $\approx\!0.75\degree$
    and $\approx\!0.5\degree$ along the E-W and S-N directions, respectively, and is comparable
    with M2 in terms of size (its area is $\sim\!0.11(^{\circ})^{2}$) and number of FDS GC 
    candidate (58), but it hosts a slightly larger number of excess GCs than M2 ($\sim\!37.5\!\pm\!4.4$).
\end{itemize}

Outside of the main structure M, four statistically significant, isolated areas of positive residuals
(labeled C1, C2, C3 and C4 in Figure~\ref{fig:core_residual_k10}) are detected within the 
cluster virial radius. The two relatively large structures C1 and C2, whose areas are both 
$\sim0.05(^{\circ})^{2}$, are located SE and NE of M and both extend radially towards the center of the cluster. On the other hand, the shapes of the two structures C3 and C4, located N and W of M, 
are roughly circular: while structure C3 is relatively isolated at $\sim0.45\degree$ 
($\sim\!150$ kpc) NW of NGC\,1380A, C4 sits $\sim0.1\degree$ from M and 
includes NGC\,1428, which likely contributes to the observed GCs excess of 6.7$\!\pm\!$1.9 thanks 
to an unusually high specific frequency for its luminosity. 

The residual structures detected in the eastern side of core of the Fornax cluster match spatial features 
observed in the density map of FDS GC candidates by~\cite{cantiello2020}~(Figure~11). In particular, 
structures M2 and C1 are observed in the same area occupied by feature G in Figure~11, while M3 partially
overlaps structure F, although in~\cite{cantiello2020}, features F and G extend 
further along the N-E and the S-W directions than their counterparts described in this paper. 
The presence of several smaller isolated areas of positive residuals along the two directions and 
visible in Figure~\ref{fig:residual_k10} suggests the existence of an underlying residual enhancement 
approximately matching the shapes and sizes of F and G. Another, weaker association 
can be made between the alignment of residual peaks P3, P5, isolated structure C3 and smaller 
areas of positive residuals surrounding NGC\,1380A with the over-density structure C in Figure~11 
from~\citet{cantiello2020}. 

The only significant area of positive residuals outside of the box delimiting the 
core region in Figure~\ref{fig:residual_k10} is a roughly circular over-density (O1) with 
R.A.$\in[56.25\degree,56/5\degree]$ and Dec.$\in[-37\degree,-36.75\degree]$. Only two galaxies 
belonging to the SGS (FCC299, FCC303) are located within the boundaries of O1. 

\begin{deluxetable}{cccccccccc}   
 	\tablecaption{Properties of the features of the $K\!=\!10$ residual map of
        the spatial distribution of GCs.}
 	\label{tab:residual_maps_features}
 	\tablehead{
            \colhead{Str.\tablenotemark{a}} &
            \colhead{Substr.\tablenotemark{b}} &
 	    \colhead{Area\tablenotemark{c}} &
 	    \colhead{N$^{\mathrm{tot}}_{\mathrm{GCs}}$\tablenotemark{d}} &     
 	    \colhead{N$^{\mathrm{exc}}_{\mathrm{GCs}}$\tablenotemark{e}} &
  	    \colhead{$\overline{\mathrm{m}}_{g}$\tablenotemark{f}} &     
  	    \colhead{$\overline{g\!-\!i}$\tablenotemark{g}} &    
  	    \colhead{N$^{\mathrm{blue}}_{\mathrm{GCs}}$/N$^{\mathrm{red}}_{\mathrm{GCs}}$\tablenotemark{h}} &       
 	    \colhead{Galaxies\tablenotemark{i}}
 	    }
        \startdata
             M       & -         & 0.90  & 990  & 627.1$\pm$74.3& 22.5$\pm$0.6 & 0.92$\pm$0.13& 669/321& {\tiny{\bf NGC\,1399},~{\bf NGC\,1387},~{\bf NGC\,1404},~{\bf NGC\,1427A},~{\bf NGC\,1379},~({\bf NGC\,1427})}                     \\
             M       & M1        & 0.16  & 74   & 41.9$\pm$5.6  & 22.5$\pm$0.6 & 0.88$\pm$0.12& 56/18  & {\tiny FCC\,115,~FCC\,130,~FCC\,136~({\bf NGC\,1374})}                      \\ 
             M       & M2        & 0.11  & 56   & 33.0$\pm$4.6  & 22.4$\pm$0.7 & 0.93$\pm$0.13& 39/17  & {\tiny FCC\,212,~FCC\,233,~FCC\,236,~FCC\,223~({\bf NGC\,1427A},~{\bf NGC\,1404})}                      \\
             M       & M3        & 0.11  & 58   & 37.5$\pm$4.4  & 22.3$\pm$0.6 & 0.92$\pm$0.12& 37/21  & {\tiny FCC\,226,~FCC\,207,~({\bf NGC\,1380B}) }                       \\   
             M       & P1        & 0.03  & 51   & 43.6$\pm$2.7  & 22.4$\pm$0.5 & 0.87$\pm$0.12& 41/10  & {\tiny FCC\,228~(NGC\,1427A,~{\bf NGC\,1404})}                      \\ 
             M       & P2        & 0.01  & 23   & 20.4$\pm$1.6  & 22.7$\pm$0.7 & 0.87$\pm$0.08& 22/1   & {\tiny FCC\,227~({\bf NGC\,1427A},~{\bf NGC\,1404},~{\bf NGC\,1399})}                      \\ 
             M       & P3        & 0.01  & 14   & 12.9$\pm$1.1  & 22.6$\pm$0.7 & 0.91$\pm$0.12& 10/4   & {\tiny FCC\,182,~FCC\,191~({\bf NGC\,1380B},~{\bf NGC\,1381},~{\bf NGC\,1387})}                      \\ 
             M       & P4        & 0.003 & 15   & 12.9$\pm$1.4  & 22.4$\pm$0.7 & 0.86$\pm$0.12& 14/1   & {\tiny (FCC\,188,~{\bf NGC\,1387},~{\bf NGC\,1396},~{\bf NGC\,1399})}                      \\ 
             M       & P5        & 0.003 & 18   & 16.7$\pm$1.1  & 22.6$\pm$0.5 & 0.85$\pm$0.14& 15/3   & {\tiny (FCC\,191,~{\bf NGC\,1387},~{\bf NGC\,1396},~{\bf NGC\,1399})}                      \\ 
             C1      & -         & 0.05  & 20   & 10.5$\pm$2.9  & 22.1$\pm$0.8 & 0.95$\pm$0.15& 13/7   & {\tiny (FCC\,233,~FCC\,246,~FCC\,285)}                      \\
             C2      & -         & 0.05  & 21   & 12.61$\pm$2.9 & 22.1$\pm$0.6 & 0.96$\pm$0.15& 11/10  & {\tiny FCC\,273 }                     \\
             C3      & -         & 0.04  & 22   & 15.0$\pm$2.4  & 22.4$\pm$0.7 & 0.85$\pm$0.13& 18/4   & {\tiny FCC\,153~(FCC\,135)}           \\ 
             C4      & -         & 0.02  & 11   & 6.7$\pm$1.9   & 22.4$\pm$0.6 & 0.90$\pm$0.12& 9/2    & {\tiny NGC\,1428,~(FCC\,266,~{\bf NGC\,1427})}       \\ 
             O1      & -         & 0.03  & 18   & 11.5$\pm$2.6  & 22.3$\pm$0.8 & 0.93$\pm$0.15& 12/6   & {\tiny FCC\,299,~FCC\,303}  \\
        \enddata
     \tablecomments{(a): Label of the residual structure;
                    (b): Label of the residual sub-structure; 
                    (c): Area of the residual structure or substructure [$(^{\circ})^{2}$];
                    (d): Total number of GCs within the residual structure or substructure;
                    (e): Mean and standard deviation of the excess number of GCs within the residual structure or substructure;
                    (f): Mean and standard deviation of the $g$-band magnitude $\mathrm{m}_{g}$ of GCs within the residual structure or substructure;
                    (g): Mean and standard deviation of the $g\!-\!i$ color of GCs within the residual structure or substructure;
                    (h): Number of red and blue GCs within the residual structure or substructure;
                    (i): Galaxies within and in the immediate surroundings of the residual structure or substructure~(MGS galaxies in boldface)
                    } 
\end{deluxetable} 

The areas of negative and neutral residual observed in the $K\!=\!10$ residual map 
(Figure~\ref{fig:core_residual_k10}) around the positions of bright galaxies (most notably 
NGC\,1387, NGC\,1399, NGC\,1381, NGC\,1380B and NGC\,1380), are the result of the 
combined effect of the negligible number of FDS candidate GCs detected at small galacto-centric 
radii in these areas (because of the decreased detection efficiency of GCs with increasing 
surface brightness from the galaxy in the background) and the flattening of the radial 
density profiles in the core of galaxies as per 
the best-fit modified Hubble models~\citep{caso2019,debortoli2022,caso2024} used to simulate the 
radial distribution of the GCSs of these hosts. As discussed in 
Section~\ref{subsubsec:galcomponent}, we quantified the degradation of the detection efficiency of candidate
GCs in FDS data by determining elliptical avoidance area whose major axis (aligned with the axis of the 
elliptical model of the host galaxy) is equal to the galacto-centric distance of the innermost GCs. The 
simulated GCs positioned in these areas are neither used for the determination of the residual maps nor 
tallied for the general statistics of the simulated sample of GCs. Even after the removal of the simulated
GCs in the avoidance regions, a deficit of observed candidate GCs can be seen in the 
vicinity of the bright galaxies. 

This deficit is typically represented by the approximately 
circular areas of neutral residuals surrounding very small regions (few cells) occupied by 
negative residuals in the center of the galaxies. The size of this effect, mostly evident around galaxies
NGC\,1387 and NGC\,1399, is influenced by the value of $K$ used to determine the density maps that generate 
the residual maps. For values of $K$ smaller than $K\!=\!10$
(see Figure~\ref{fig:residuals_k_values} in Appendix~\ref{sec:appendix4}), 
the area of the neutral residuals regions surrounding the center of the bright MGS galaxies decreases, 
as additional, usually filamentary spatial structures are observed at small galacto-centric distances to 
account for smaller-scale fluctuations in the density of observed GC candidates. For $K\!>\!10$ values, 
the circum-galactic regions of neutral/negative residuals shrink and disappear as the density values 
used to calculate the residual maps are smoothed out over larger and larger areas. Our 
attempts to eliminate this pattern by arbitrarily tweaking the behavior 
of the modified Hubble radial profiles in the core of the galaxies and
changing the values of the b$^{(\mathrm{mH})}$ indexes were unsuccessful. No values were able to 
completely remove or significantly attenuate the regions of neutral residuals, suggesting that 
the inefficiency of the candidate GC detections is non-negligible at larger galacto-centric distances
than the avoidance radii. Since these artifacts do not affect the residual maps and the 
results discussed in this paper on the degree-wide, larger spatial scale characteristics of the 
ICGCs population in the Fornax cluster, we decided not to further pursue the issue. 
In Appendix~\ref{sec:appendix5}, we describe the reconstruction of the residual map used to derive
the results described in this manuscript, and show (Figure~\ref{fig:mosaic_residuals_sims}) the positions 
of the observed FDS candidate GCs and 
the simulated GCs from one of the simulations, superimposed to the $K\!=\!10$ residual map the whole 
f.o.v. and two zoomed-in areas in the center of the cluster.

\subsection{Colors of Residual Structures}
\label{subsec:colorsresidualstructures}

We do not attempt to obtain color-specific residual maps of the distribution of ICGCs in Fornax by 
separately simulating the spatial distributions of red and blue GCs because of the lack of homogeneous
color-specific models of the brightness and radial distributions for all MGS GCSs, and 
color-specific correlations between the brightness of the host galaxies and the properties of 
their GCSs for the SGS. 

\begin{figure*}[ht]
    \centering
    \includegraphics[width=0.575\linewidth]{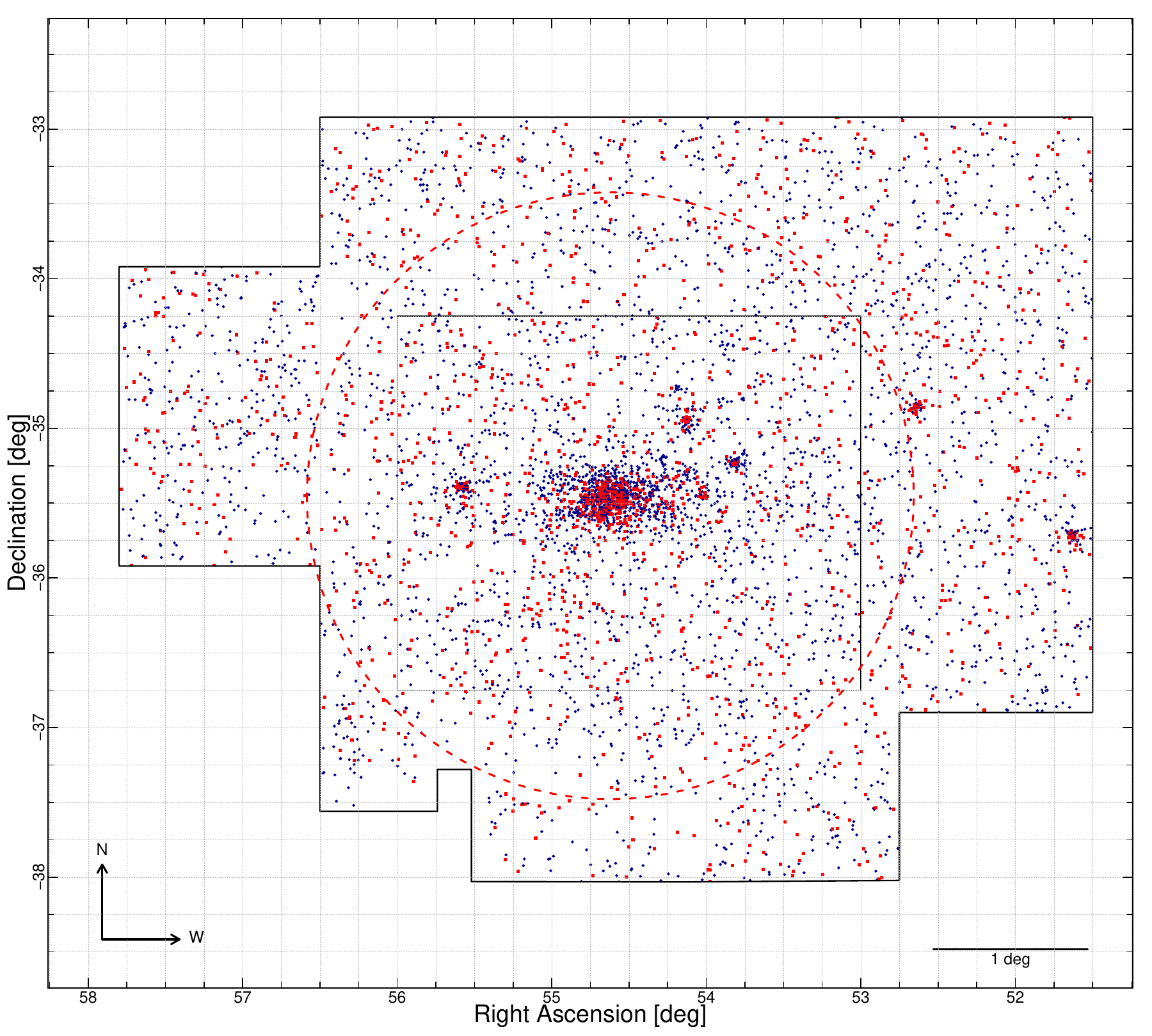}
    \includegraphics[width=0.375\linewidth]{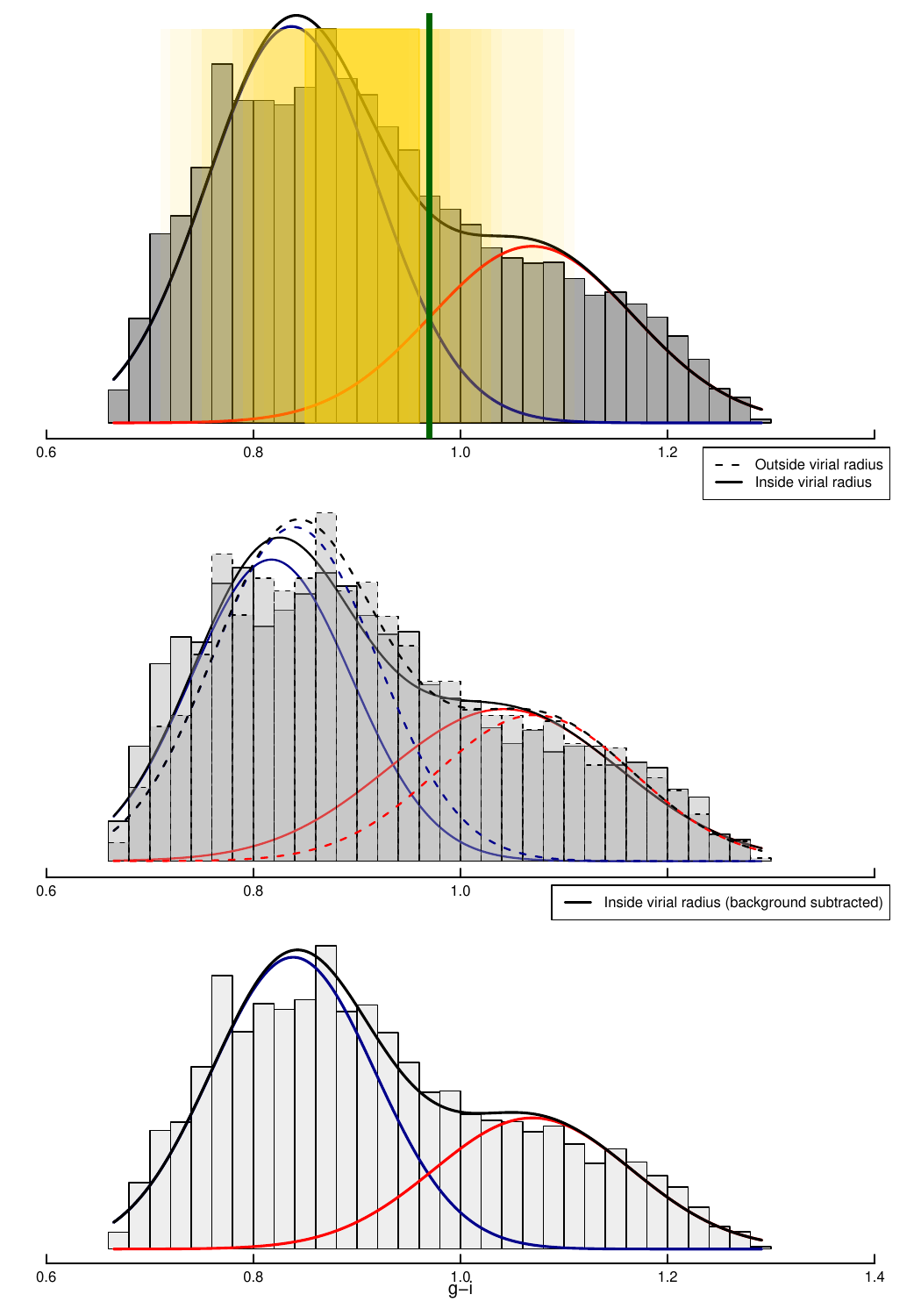}\\
    \includegraphics[width=0.75\linewidth]{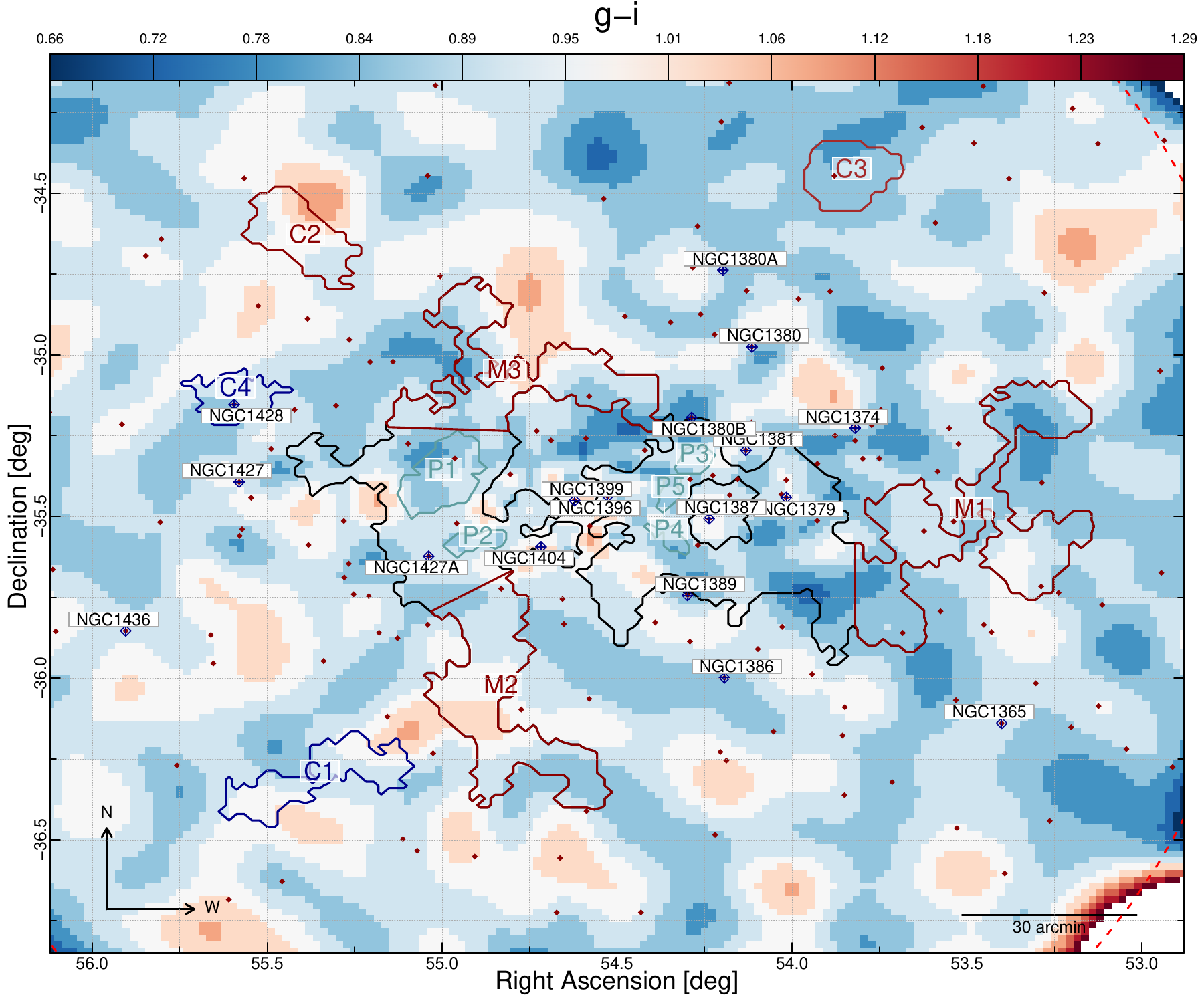}
    \caption{Upper left: positions of FDS candidate GCs split in red and 
    blue sub-classes using the threshold $g\!-\!i\!=\!0.97$. Upper right: histogram of the 
    $g\!-\!i$ color distribution of all FDS GCs (upper panel), with two best-fit 
    Gaussian models corresponding to the red and blue sub-classes and the color value 
    $g\!-\!i\!=\!0.97$ used as threshold (green vertical line). 
    The interval of average $g\!-\!i$ color spanned by the GCs included in the residual 
    structures~(Table~\ref{tab:residual_maps_features}) are displayed by the 
    yellow rectangle. The mid panel shows the color distribution for GC candidates 
    within and outside the virial radius of the cluster (solid and dashed histograms, respectively), 
    with corresponding best-fit blue and red gaussian subclasses, while the lower panel
    displays the color distribution within the virial radius after the color distribution of GC 
    candidates located outside the virial radius has been statistically subtracted (see 
    Section~\ref{subsec:colorsresidualstructures} for details).
    Lower: smoothed cell-wise, background-subtracted average color map of GCs candidates 
    in the core of the Fornax cluster, obtained on the same grid employed for the residual 
    maps of the spatial distribution of observed GC candidates. The contours of the residuals 
    structures derived from the spatial distribution of GC candidates and 
    discussed in Section~\ref{sec:results} (solid lines) are overplotted.}
    \label{fig:colors}
\end{figure*}

We investigate the color properties of the residual structures described
in the previous Section by comparing the colors of the GCs whose projected positions are located within 
the boundaries of the residual structures, to the global
color distribution of the whole sample of FDS candidate GCs, and their locations relative to 
the spatial features of the density maps of red and blue GCs. Using a Gaussian Mixture Modeling 
method~\citep{muratov2010}, we determined that the $g\!-\!i$ distribution of the general sample 
of FDS GCs is bimodal with blue and red components whose best-fit parameters are 
$\mu^{\mathrm{blue}}_{\mathrm{(g-i)}}\!=\!0.84,\sigma^{\mathrm{blue}}_{\mathrm{(g-i)}}\!=\!0.08$ and 
$\mu^{\mathrm{red}}_{\mathrm{(g-i)}}\!=\!1.07,\sigma^{\mathrm{red}}_{\mathrm{(g-i)}}\!=\!0.10$ 
respectively~(upper right plot in Figure~\ref{fig:colors}). We also 
investigated the color distribution of GC candidates outside of the Fornax cluster virial radius and 
far from all FCC galaxies
(the same sample used to determine the density of the homogeneous component), and of GC candidates located 
within the virial cluster. In both cases, 
the Gaussian Mixture Modeling method favors a bimodal distribution. The best-fit 
parameters of the normally distributed GCLFs for the sample of GC candidates outside 
and inside the virial radius are ($\mu^{\mathrm{blue,out}}_{\mathrm{(g-i)}}\!=\!0.82\!\pm\!0.08$,
$\mu^{\mathrm{red,out}}_{\mathrm{(g-i)}}\!=\!1.05\!\pm\!0.11$), 
and ($\mu^{\mathrm{blue,core}}_{\mathrm{(g-i)}}\!=\!0.84\!\pm\!0.08$, 
$\mu^{\mathrm{red,core}}_{\mathrm{(g-i)}}\!=\!1.07\!\pm\!0.11$), respectively. The associated Gaussians are shown in the 
mid panel of the upper right plot in Figure~\ref{fig:colors}. 
Since the peak of the red and blue components for these two subsets are compatible within the uncertainties, 
we decided to adopt $(g\!-\!i)_{\mathrm{thresh}}\!=\!0.97$, the value where the probabilities of belonging 
to the red and blue components, derived from the whole
sample of FDS GC candidates, are equal as the fiducial threshold separating the two color sub-classes. 
Using this value, the total sample of 
FDS GCs is split in 3804 blue GCs ($\sim\!67.3\%$) and 1849 red GCs ($\sim\!32.7\%)$; the positions of 
red and blue GCs in the whole FDS field and the core region of the cluster 
are shown in the upper left plot of Figure~\ref{fig:colors}. While the color 
distributions of individual GCs located within the
spatial structures discussed in the previous Section cover the whole interval of $g\!-\!i$ color, 
their average colors~(column $g$ in Table~\ref{tab:residual_maps_features}) span a range 
of values mostly consistent with the blue component of the global GCs color distribution (shaded yellow 
rectangles in upper right plot of
Figure~\ref{fig:colors}) and approximately delimited by the peaks of the blue component 
$\mu^{\mathrm{blue}}_{\mathrm{(g-i)}}$ and the threshold $(g\!-\!i)_{\mathrm{thresh}}$. 
The numbers (although with large variance) of red and blue GCs for each structure reported in 
Table~\ref{tab:residual_maps_features} also confirm that the GCs whose projected positions match 
the residual structures discussed in this manuscript, are in majority blue GCs.

To verify that the color distribution of the excess of GC candidates 
in the core of the Fornax cluster is not
affected by background and foreground contaminants, we statistically 
subtracted the colors of the GC candidates located outside of the virial radius of the Fornax cluster 
from the colors distribution of the excess of GC candidates in the core of the cluster. 
We calculated the expected number of GCs belonging to the homogeneous component 
within the virial radius 
$N_{\mathrm{GCs}}^{(\mathrm{hom})}(r_{\mathrm{cluster}}\!\leq\!r_{\mathrm{vir}})\!=\!2291.7\!\pm\!263.8$
using the density and its uncertainty derived from the candidate GCs located outside of 
the virial radius of the cluster and excluding sources nearby FCC galaxies (Section~\ref{subsubsec:homocomponent}). 
$\hat{N}_{\mathrm{GCs}}^{(\mathrm{hom})}(r_{\mathrm{cluster}}\!\leq\!r_{\mathrm{vir}})$ color values (where
$\hat{N}$ is randomly drawn from a Gaussian with standard deviation set to the uncertainty on the number of 
homogeneous component GCs within the virial radius of the cluster) are obtained 
by sampling the $g\!-\!i$ color distribution of GCs outside of the virial radius~(Figure~\ref{fig:colors}, 
right plot, mid-panel), perturbed by adding a random uncertainty term drawn 
from the $[-0.15, 0.15]$ interval, where 0.15 is the typical uncertainty on the $g\!-\!i$ color of the FDS GC candidates. 
Then, the GC candidate within the virial radius with the closest $g\!-\!i$ value to the 
simulated color is removed from the full list 
of $N_{\mathrm{GCs}}^{\mathrm{(all)}}(r_{\mathrm{cluster}}\!\leq\!r_{\mathrm{vir}})\!=\!3587$ GC candidates
within the Fornax cluster $r_{\mathrm{vir}}$, unless the difference between the simulated color and the 
closest observed color exceeds $\pm0.15$, in which case no observed GC is removed from the list. 
The histogram of the background-subtracted color distribution of GC
candidates inside the virial radius, was obtained by repeating this process 
100 times and averaging the resulting distributions, is shown in the lower 
panel on the right of 
Figure~\ref{fig:colors}, with superimposed the best-fit Gaussians for the red and blue components.
The two color components for the background-subtracted population of excess GC candidates in the core of the cluster
are compatible within the uncertainties with the models for the whole population of GC candidates in the core 
of the cluster ($\mu^{\mathrm{(blue,core,back-sub)}}_{\mathrm{(g-i)}}\!=\!0.84\!\pm\!0.08$,
$\mu^{\mathrm{(red,core,back-sub)}}_{\mathrm{(g-i)}}\!=\!1.1\!\pm\!0.1$).

The lower plot of Figure~\ref{fig:colors} displays the average smoothed map of the per-cell GC candidates color for 
the 100 background-subtracted GC candidates samples obtained by using the two-dimensional LOcally Estimated Scatterplot 
Smoothing (LOESS) method~(with degree of the fitting polynomial $d_{\mathrm{LOESS}}\!=\!2$ and 
span that regulates the locality of the fit set to $s_{\mathrm{LOESS}}\!=\!8$ cells) on the same grid used to 
derive the residual map of the spatial distribution of FDS GC candidates. 
The core of the Fornax cluster is populated by GC candidates 
bluer that the average of the FDS sample, even when the contribution
of the homogeneous component is taken into account. Most of
the residual structures detected in the spatial 
distribution of FDS GC candidates are in areas where the average GC color is bluer than the 
$(g\!-\!i)_{\mathrm{thresh}}\!=\!0.97$ (blue hues in the lower plot of Figure~\ref{fig:colors}). 
Very blue average colors are detected around the P1 over-density peak and 
within the central section of the structure M, in particular E of NGC\,1387, around NGC\,1380 and
S of NGC\,1379. The spatial features M1, M2 and M3 are also dominated by regions of average blue GC candidates
but partially overlap or border areas of neutral to red average $g\!-\!i$ colors.
C2 is the only spatial structure characterized by a significant red
GC candidates density enhancements. Two additional regions of average red GC color are
found between M2 and C1, and north of M3. The mostly red GCs located in the inner regions of 
the GCSs of bright ETGs are not visible in the color map because they are smoothed out 
by the larger spatial components mostly populated by blue GC candidates. NGC\,1399 and NGC\,1404, where 
patches of red GCs density excess are detected near the centers (likely due to contribution of 
of the GCSs of NGC\,1404 and NGC\,1396, that are typically dominated by red GCs, to the extended, bluer
ICGCs population) are the only exception. LOESS smoothed color maps 
of the background-subtracted population of GC candidates obtained
with larger values of the span parameter highlight the same large-scale 
picture, while smaller spans ($s_{\mathrm{LOESS}}\!=\!\{3,4,5\}$ cells)
display larger fluctuations on smaller spatial scales and more 
extended regions with red GCs density exceeding the density of blue GCs in the same areas highlighted 
by the LOESS map obtained with $s_{\mathrm{LOESS}}\!=\!8$. The smaller areas with red average color
surround the positions of some bright galaxies in the core of the Fornax cluster, likely corresponding 
to the inner regions of these hosts' GCSs which are redder than GCs at larger galacto-centric distances and ICGCs. 

These results are consistent with the large spatial coverage of FDS data that allows the exploration of intra-cluster
areas far away from bright hosts where GCs are known to be blue~\citep[cp.][]{schuberth2008,durrell2014,cantiello2020}. 
The abundance of blue GCs in the ICGCs population can be explained as outer GCs, that are more metal-poor than inner 
GCs in most GCS~\citep{zepf1993,brodie2006}, are more easily stripped than inner 
GCs during galaxy interactions. This mechanism, confirmed by simulations~\citep{ramos2015,ramosalmendares2018},
contributes to the over-representation of blue GCs in the ICGCs. Another potential process
affecting the average color of the ICGCs population observed in the high-density regions of clusters of galaxies
is the contribution of the typically blue GCs associated with dwarf galaxies~\citep{lotz2004} that combine with 
the ICGCs after the destruction of their hosts. Since tidal friction causes more massive galaxies to sink 
towards the cluster center more quickly than dwarf galaxies, they deposit the GCs stripped from their
blue and red GCSs there, while the blue GCs stripped from dwarf galaxies are deposited further our in the 
intra-cluster space.

The selection effects on the GC candidates used in this paper should be taken into account when interpreting 
the results presented in this Section. While completeness functions for different color classes are not
available for the FDS GCs, we can qualitatively infer how the color-related bias affects the sample. 
Since at fixed $gri$ magnitudes, red GCs close to the detection limit in the $u$-band are under-detected 
compared to blue GCs, we expect the red component of GCs population to suffer from a larger incompleteness 
than the blue component. However, since red GCs are also more centrally concentrated around the host galaxies 
than blue GCs, the differential incompleteness that favors the detection of blue GCs is less relevant at large
galacto-centric distances where most of the residual structures discussed in this paper are located. 

\section{Discussion}
\label{sec:discussion}

\subsection{Comparison with Intra-Cluster Light and Dwarf Galaxies}
\label{subsec:icl}

We compare the spatial properties of the ICGCs population observed in the core 
of the Fornax cluster with the Intra-Cluster Light (ICL) therein detected. The ICL 
is the relic of the stellar material stripped from the halos of their original hosts 
via gravitational interactions as they infall towards the center of the cluster 
potential well~\citep{delucia2007,contini2021,montes2022}, and it retains features that are shaped 
by earlier interactions between galaxies~\citep{kluge2024}. Since both stars and GCs are subject to 
the same stripping mechanism from their parent galaxies, ICGCs are usually good 
tracers of the ICL and the spatial features of both components should bear 
similarities. 

\begin{figure*}[ht]
    \centering
    \includegraphics[width=0.95\linewidth]{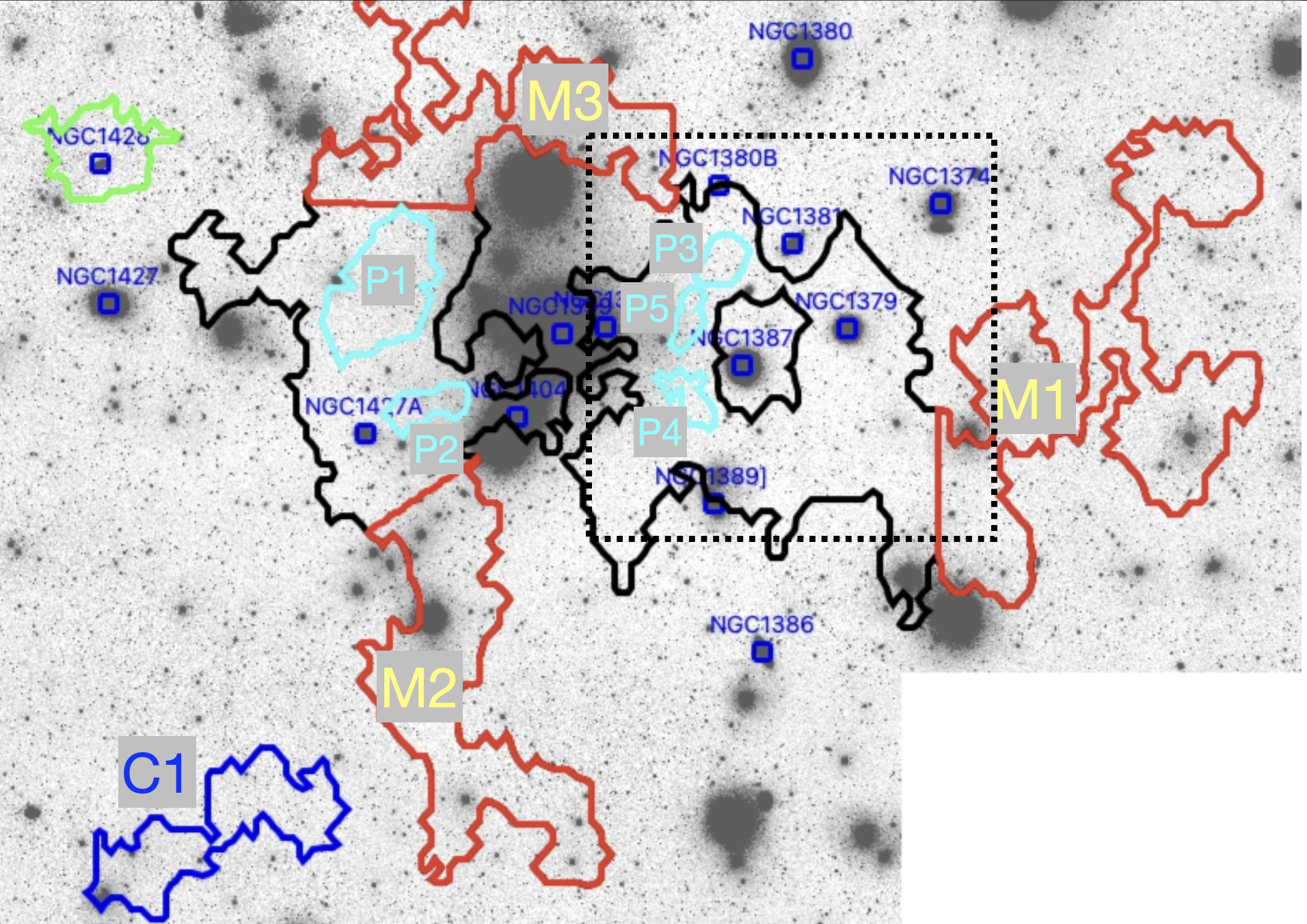}\\
    \includegraphics[width=0.95\linewidth]{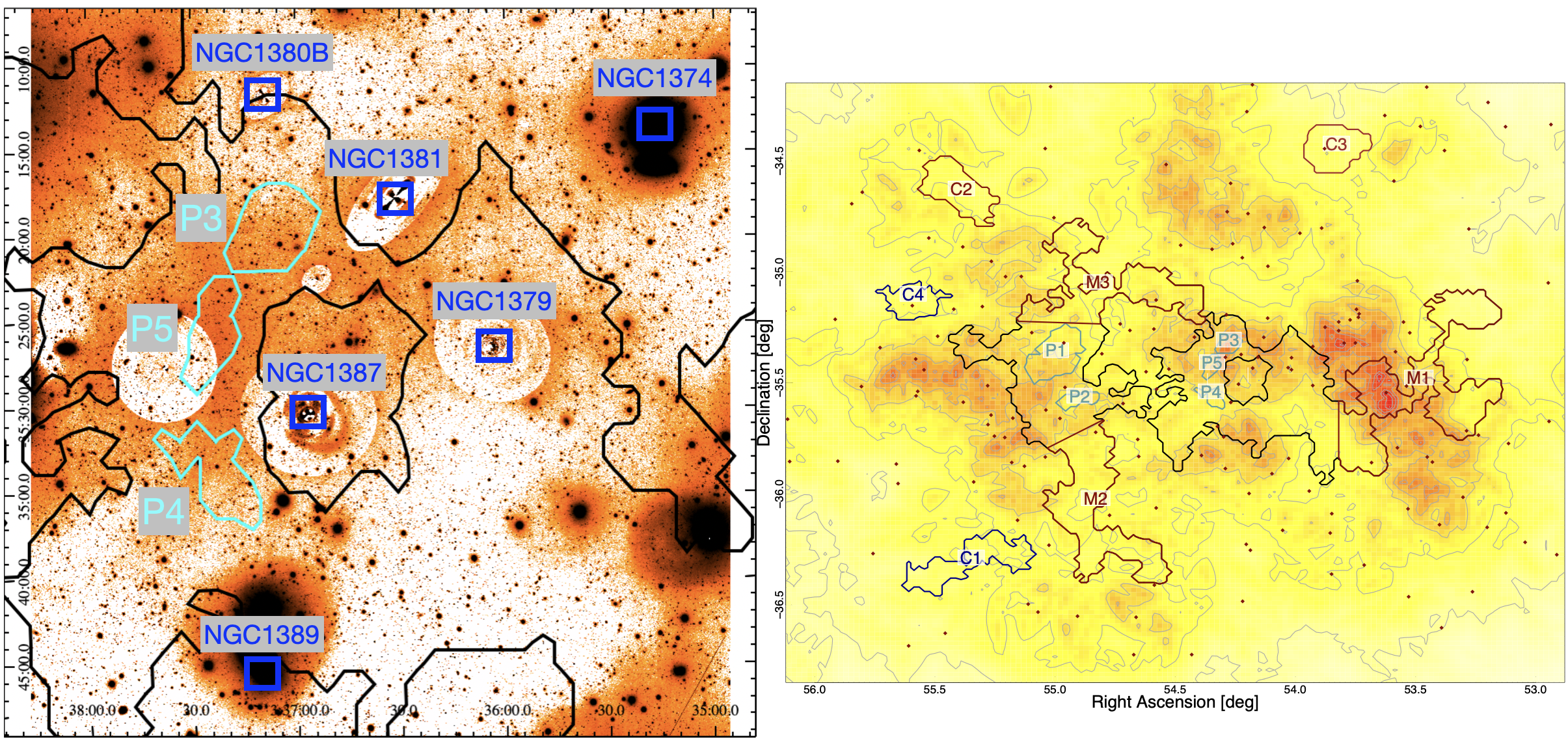}
    \caption{Upper: FDS $r$-band image of the core of the Fornax 
    cluster~\citep{peletier2020}, with the contours of the spatial structures
    and areas of large positive residuals from the $K\!=\!10$ GCs residual maps overplotted 
    (contours of structures and high residual areas are color-coded as in 
    Figure~\ref{fig:core_residual_k10}) and FCC 
    galaxies (blue squares). The black dotted rectangle highlights the region where 
    the ICL enhancements discovered by~\cite{iodice2019b} are located. 
    Lower left: 41$\arcmin$ x 43$\arcmin$ $g$-band VST mosaic of a region of the 
    Fornax cluster core where significant 
    patches of ICL are detected~\citep{iodice2017} with, superimposed, the $K\!=\!10$ 
    contours of the residual structures and peaks discovered in the GC candidates distribution. 
    The brightest galaxies in the region were modeled and 
    subtracted from the parent image. The brightest cluster galaxy, NGC\,1399, is located (not
    visible) on the left of the field.
    Lower right: $K\!=\!10$ KNN density map (galaxies/$(\arcmin)^2$) of the distribution of dwarf 
    galaxies~\citep{su2021,venhola2022} in the core 
    of the Fornax cluster with the main GC candidates spatial features labeled
    (color-coded as in Figure~\ref{fig:core_residual_k10}). Brown diamonds show
    the positions the dwarf galaxies.}
    \label{fig:rimage_dwarf_residuals}
\end{figure*}

FDS imaging data were obtained with a step-dither observational strategy designed to 
optimize the evaluation of the sky background around bright and extended sources. Taking advantage
of this approach, several studies~\citep{iodice2016,iodice2017,iodice2019b,spavone2020} have used 
FDS data to explore the faint features ($\mu_{g}\geq28$ mag arcsec$^{-2}$) of the ICL detected in 
the core of the Fornax cluster. The upper plot in Figure~\ref{fig:rimage_dwarf_residuals} 
shows the $K\!=\!10$ isoresidual contours obtained from the GC spatial distribution in 
the core of the cluster superimposed to the FDS $r$-band mosaic. The black dotted rectangle highlights
the region of the cluster, located to the west of NGC\,1399 and approximately centered on NGC\,1387,
where significant ICL was detected by~\cite{iodice2017}.

As shown in the lower left of Figure~\ref{fig:rimage_dwarf_residuals}~(based on Figure~1 
from~\citealt{iodice2017}), the large and statistically significant area of positive residuals
corresponding to the section of M found west of NGC\,1399 where 
residual peaks P3, P4 and P5 are located, overlaps with the faint 
($\mu_{g}\sim28-29$ mag arcsec$^{-2}$) stellar bridge connecting NGC\,1399 
to NGC\,1387~\citep{bassino2006,iodice2016}, confirming the results obtained 
by~\cite{dabrusco2016} and~\cite{cantiello2018} by inspecting the density distribution 
of FDS GCs. This finding further 
reinforces the hypothesis that these two galaxies are currently interacting, with 
NGC\,1399 stripping field stars and GCs from the eastern side halo 
of NGC\,1387. Another area of enhanced
ICL emission located north and west of NGC\,1379 partially matches a smaller region of positive 
residuals found NW of the galaxy and is entirely enclosed within the boundaries of M. 
No ICL excess is observed S and SW of NGC\,1387, and S of NGC\,1379, 
where, conversely, another area of enhanced GCs 
positive residuals, belonging to the main spatial structure M, is detected 
between the positions of NGC\,1387 and NGC\,1389.

The comparison between the spatial distribution of the ICL patches discovered by~\cite{iodice2017} 
in the central area of the Fornax cluster core and the GCs residual structures shows that the GCs 
excess matches only in part the regions of ICL discovered in the same region, 
and does not provide evidence of a global correlation. 
The differences may stem from limitations in the method used to derive the residual map 
of the GC candidates, that could over-estimate GC candidates excess in certain areas, 
and/or from the presence of ICL features that are too faint ($\mu_{g}>28$ mag arcsec$^{-2}$) 
to be detected even in the optimized FDS imaging data. 
The link between the ICL and GCs can be indirectly examined through
Planetary Nebulae (PNe), discrete probes of the same stellar populations that, 
stripped from the halos of their host galaxies, contribute to the ICL observed in high-density cluster 
environments~\citep[see][and references therein]{longobardi2015}. Using a sample of 1635 PNe 
extending up to 200 kpc ($\sim\!2000\arcsec$) from NGC\,1399,~\cite{spiniello2018} 
confirmed the existence of two kinematically defined PNe streams in the core of the 
Fornax cluster, one located N of NGC\,1399 (and already reported by~\citealt{mcneil2010}), 
and the other associated with the NGC\,1399-NGC\,1387 ICL bridge detected by~\cite{iodice2017} 
and overlapping with the residual structures P3, P4 and P5. Such PNe sub-structures 
match kinematic features discovered by~\cite{chaturvedi2022} in the phase-space distribution of
spectroscopic GCs located in the same area.~\cite{spiniello2018} also observed 
that the line-of-sight (l.o.s.) velocity dispersion radial profile of PNe located in a 
rectangular region centered on NGC\,1399 and extending west in the direction of NGC\,1387, 
is consistent with the v$_{\mathrm{los}}$ dispersion profile of {\it bona-fide}, 
spectroscopically-observed GCs~\citep{schuberth2010,chaturvedi2022}. In particular, at 
virial-centric distances $[400\arcsec,1000\arcsec]$, the radial profile of the PNe 
velocity dispersion correlates better with the profile of blue GCs, thought 
to belong to the ICGC component, than the profile of red GCs, dynamically 
linked to the central galaxy. Above $\sim\!1000\arcsec$, the PNe 
$\sigma_{\mathrm{los}}^{(\mathrm{PNe})}$ radial profile flattens to $\sim\!300$ 
km$^{-1}$s, $\sim\!80$ km$^{-1}$s smaller than the systemic velocities dispersion 
of galaxies in the Fornax cluster~\citep{drinkwater2001}, 
in accordance with the scenario where the intra-cluster 
PNe are subject to the same cluster-wide potential as the virialized, bright galaxies 
in the Fornax cluster, but follow a different density profile~\citep{spiniello2018}.

While a quantitative assessment of the degree of correlation among the kinematical 
properties of PNe and GCs, and the spatial distributions of the detected 
ICL and the GCs over-density structures
is difficult on the entire core of the Fornax cluster because of the 
varying spatial coverage of the three populations, the aforementioned results 
confirm that the PNe, which
correlate with the ICL features detected in the vicinity of NGC\,1399,
trace both the potential of the Fornax cluster at 
large virial-centric distances, and match the kinematical properties of the 
ICGCs population at distances $\geq\!400\arcsec$.

In addition to tidal stripping of GCs from the outskirts of bright host galaxies, 
the disruption of dwarf galaxies has been proposed
as one of the mechanisms responsible for 
the production of ICL in clusters of galaxies~\citep{purcell2007,martel2012,annunziatella2016}.
This would suggest a correlation between the spatial properties of dwarf galaxies currently
detected in Fornax cluster and the distribution of the diffuse population of GCs.
The lower right image in Figure~\ref{fig:rimage_dwarf_residuals}
shows the $K\!=\!10$ KNN density map in the core of the Fornax cluster of 
a complete sample of 829 dwarf galaxies detected in FDS data and obtained by merging the 
catalogs from~\cite{su2021} and~\cite{venhola2022}, with the 
$K\!=\!10$ isoresidual contours of the distribution of FDS GC candidates and their
main spatial structures overplotted. 
As already reported by~\cite{venhola2019} using FDS data and~\cite{ordenesbriceno2018} using data from the 
Next Generation Fornax Survey~\citep{munoz2015}, the spatial distribution of dEs in the
core of the cluster features a complex morphology with several local enhancements. The two main regions
of dwarfs over-density are located in the eastern and western sections of the core of the 
Fornax cluster. The western over-density, characterized by an elongated shape, extends 
along a NE to SW direction from the position of NGC\,1374 towards 
NGC\,1365, in a region largely devoid of bright galaxies, and significantly overlaps 
with the FDS GC candidates residual structure M1. The eastern area of dwarfs 
over-density is mostly located outside of the main 
GC excess in the core of the Fornax cluster although it partially overlaps with the eastern section 
of M and borders with the residual positive enhancements P1 and P2.
A smaller region of enhanced dwarfs density connected to the western, major
over-density and roughly located N of C\,1387, partially overlaps the central section of the 
main residual structure M and the residual peaks P3 and P5. Additionally, another less 
statistically significant but larger region of high
density for dwarf galaxies can be found north of the residual structure M, extending
east of the position of NGC\,1380A, in a region with no significant GC residual structures. 

The disruption of low-mass dwarf galaxies in the center of cluster of 
galaxies~\citep{peng2008,harris2013} has been identified as one of the most efficient 
channels of ICGCs formation~\citep[see][for the Coma cluster]{peng2011} because of the 
typically high specific frequency of dEs and their tendency to be
easily tidally destroyed while interacting with massive galaxies and the cluster 
potential. The direct comparison of the spatial density of dwarfs detected in the core of the Fornax 
cluster and the residual map of the spatial distribution of GC 
candidates~(lower right panel in Figure~\ref{fig:rimage_dwarf_residuals}) does not suggest 
a general, statistically robust, direct spatial correlation between the two populations. 
Differently from the western half of the Fornax core, where dwarfs galaxies over-density areas
and residual structures/peaks in the GC candidates distribution significantly 
overlap (around peaks P3, P4 and P5 and across the spatial structure M1), 
in the eastern side of the core the principal regions of enhanced density of dwarf galaxies 
are adjacent to the main residual 
structure M and residual peaks P1 and P3, but do not substantially overlap with them. 
This scenario suggests the existence of spatially 
complex interplay between the two populations, that can be summarized as follows:

\begin{itemize}
    \item The absence of statistically significant regions of dwarfs over-density 
    in the central $\sim\!0.55\degree$ of the Fornax cluster and east of NGC\,1399 
    towards the nearby residual structures M2, M3 and residual features P1 and P2 could be
    explained with the partial depletion of an original, more abundant population of 
    dwarf galaxies that might have 
    occurred in the past, as dEs were tidally disrupted by the gravitational potential 
    of NGC\,1399 and the other massive galaxies in the core of the cluster. 
    The blue average color of GC candidates within the boundaries of the 
    spatial structures in the eastern side of the GC candidates residual 
    map~(Section~\ref{subsec:colorsresidualstructures})
    supports this scenario and suggests that a large fraction of the excess 
    ICGC population observed in this area could have initially belonged to the GCSs 
    of destroyed dwarfs, known to contain significantly smaller fraction of red GCs compared 
    to more luminous ETGs~\citep{peng2006}.
    \item Tidal disruption of dEs has likely not been the driving 
    mechanism behind the growth of the ICGCs in the western half of the 
    core of the Fornax cluster and around NGC\,1387. In this region,
    the most significant area of dwarfs over-density overlap the residual structure M.
    No evidences of correlation between under-dense areas in the dwarfs spatial 
    distribution and the GCs density enhancements,
    that would be the indication of a depletion of a pre-existing dEs population (in particular
    around NGC\,1387, where the three significant GC candidates residual peaks P3, P4 and P5 
    are superimposed to a secondary region of relatively high dEs density) can be observed.  
    Therefore, the GCs excess in the positive residual spatial structures
    and the average blue GCs colors in this area cannot be uniquely 
    attributed to  
    the contribution from the GCSs of dwarf galaxies, but would require other 
    mechanisms to replenish the abundant ICGC population detected, such as 
    stripping of GCs from the halos of bright galaxies (NGC\,1387, NGC\,1381, 
    NGC\,1380B) by the gravitational potential of NGC\,1399, as suggested 
    by~\cite{bassino2006} for NGC\,1387, and/or tidal stripping by the Fornax 
    cluster potential of preferentially blue GCs at large galacto-centric 
    distances~\citep{ramos2015,ramosalmendares2018}.
\end{itemize}

Additional constraints on the origin of the ICGCs and their relationship with other
observables can be imposed by making 
reasonable conjectures about the nature of their progenitors. If the whole population
of ICGCs originated from a single class of progenitors, 
their current S$_{N}^{(\mathrm{ICGCs})}$ would be equal to that of their progenitors. 
By assuming that all ICGCs derived from the GCSs of tidally disrupted low-mass dwarf 
galaxies located at small virial-centric distances, where environmental effects favor 
dEs with larger specific frequencies than at larger distances~\citep{peng2008,lim2018,marleau2024}, 
one can adopt a typical value for dEs S$_{N}^{(\mathrm{dEs})}\!=\!8$, as observed 
for high S$_{N}$ dwarfs in the Virgo and Fornax clusters~\citep[c.p.][]{millerlotz2007,peng2008}.
In order to produce the ICGCs population in M, after completeness correction,
the destruction of {\bf$\sim\!3,600$ to $\sim\!8,400$} dEs with absolute magnitudes 
$M_{V}^{(\mathrm{dEs})}\!=\!-13.3$ ($L_{V}^{(\mathrm{dEs})}\!\simeq\!2\!\cdot\!10^{7}L_{\odot}$) 
and $M_{V}^{(\mathrm{dEs})}\!=\!-12.5$ ($L_{V}^{(\mathrm{dEs})}\!\simeq\!8.5\!\cdot\!10^{6}L_{\odot}$) 
respectively, would be needed. On the other hand, by assuming a lower specific frequency 
S$_{N}^{(\mathrm{ICGCs})}\!=\!1.5$, 
typical of L$^{\star}$ intermediate-luminosity, early-type galaxies or GCs-poor dwarfs in the 
outskirts of cluster of galaxies~\citep{georgiev2010}, the total luminosity of the ICGCs 
progenitors  needed to produce the entire ICGCs population in the central overdensity M
would be $\sim\!3.8\!\cdot\!10^{11}L_{\odot}$ (with uncertainties $\pm30\%$). This value 
is $\sim$50 times larger than the luminosity 
$L_{V}^{(\mathrm{ICL})}\!\simeq\!7\times10^{9}$ L$_{\sun}$ derived by~\cite{iodice2017} from the 
total integrated absolute magnitude of the all ICL discovered 
in the core of the Fornax cluster near NGC\,1387, NGC\,1379, NGC\,1381 and NGC\,1380B (in a 
roughly rectangular region of area $\sim\!432(\arcmin)^2$ around FCC\,182), and is comparable to 
the luminosity of several ICL regions detected around bright ETGs in the Virgo cluster~\citep{mihos2017}.  

While dEs disruption has been proved to be an inefficient formation channel for 
ICL compared to the dominant process of stellar stripping from intermediate/massive 
galaxies~\citep{montes2014,contini2021}, the simple estimations presented above favor ICGCs 
progenitors which display significantly larger values of S$_{N}$ over large interval of 
luminosity and regardless of the density and dynamical state of the cluster environment where 
they reside~\citep{harris2013}. An upper limit on the S$_{N}$ of the progenitors can be determined 
by assuming that all the integrated ICL luminosity and the observed excess number of GCs 
located in the same region occupied by the ICL patch derive from the same progenitors. 
The observed excess of candidate GCs $N_{\mathrm{GCs,ICL}}^{\mathrm{(obs,exc)}}\!=\!105.9\!\pm\!12.3$, 
corresponding to a total number of ICGCs $N_{\mathrm{ICGCs,ICL}}\!=\!1138.2\!\pm\!145.7$, after 
incompleteness correction, can be used to infer a 
high specific frequency $\sim\!13.7$ for the systems whose GCSs contributed to the growth of 
the local ICGCs population. If the number of blue ICGCs, corrected for completeness, in the 
ICL region $N^{(\mathrm{blue})}_{\mathrm{ICGCs,ICL}}\!=\!796.7\!\pm\!102$ is used instead (since, as~\citealt{iodice2017}
reported, their morphology more closely resemble the shape of the ICL emission than red ICGCs), a 
specific frequency of the progenitors of S$_{N}\!=\!9.6$ would be required to produce the observed ICL total
luminosity. These specific frequency values are not uncommon for high-luminosity dEs and ultra-diffuse 
galaxies (UDGs) in high-density environments at small cluster-centric distances~\citep{peng2008,lim2018}, but
since the disruption of dwarfs alone is unlikely to assemble the amount of ICL discovered in 
the Fornax cluster core, other mechanisms (stripping of GCs from the GCSs of bright galaxies and, 
as secondary channels, mergers and pre-processing) that produce ICL more efficiently, are likely 
co-contributors to the ICGCs detected in the same region as the ICL excess. 

\subsection{Line-of-sight Velocities of the Residual Spatial Structures}
\label{subsec:spectrogcs}

\begin{figure*}[ht]
    \centering
    \includegraphics[width=\linewidth]{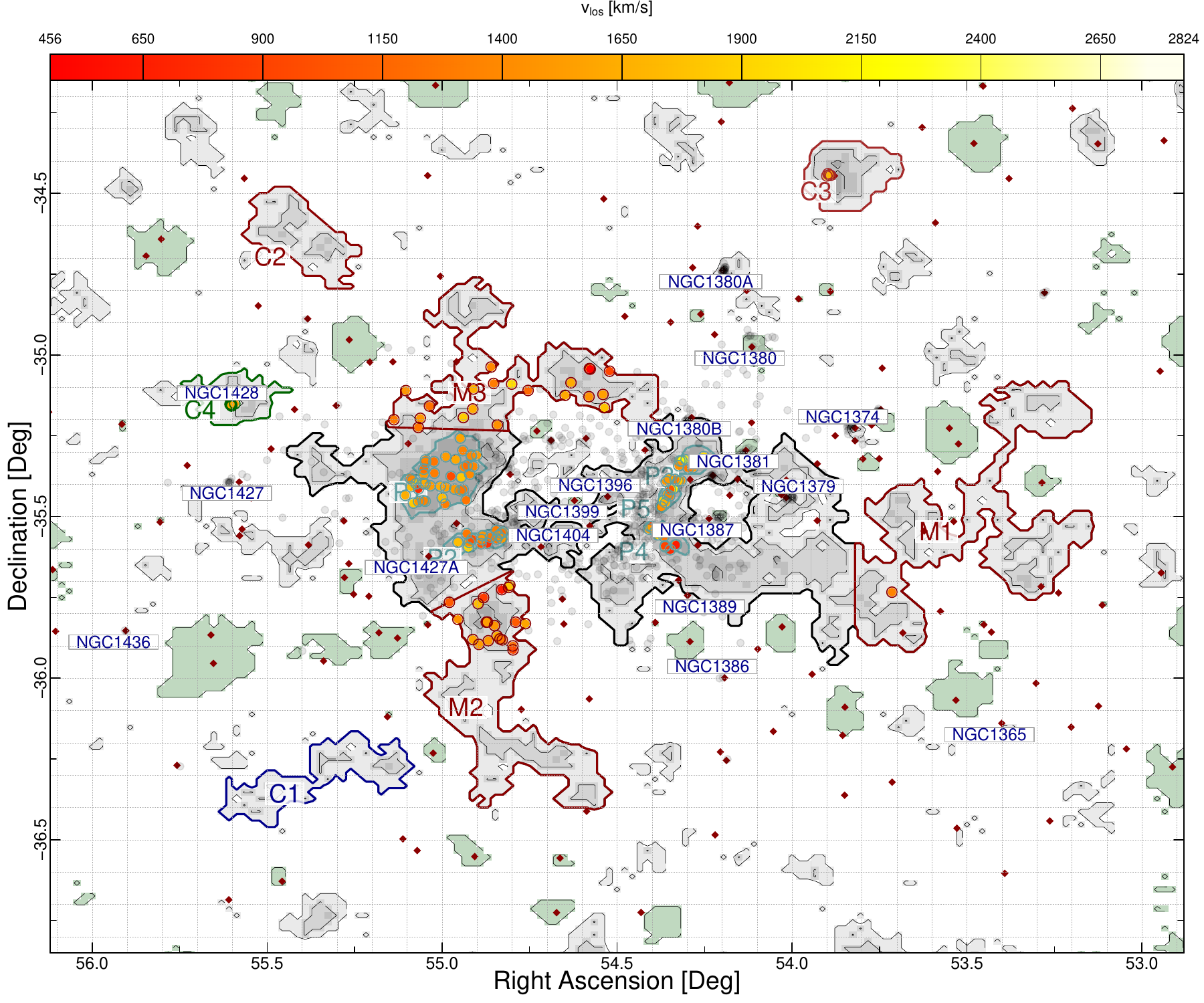}
    \caption{Positions of the ICGCs with velocity measurements~\citep{chaturvedi2022} (gray points) 
    in the core of the Fornax cluster. The ICGCs located within the spatial feature of the residual
    map of the FDS GCs are color-coded according
    to their v$_{\mathrm{los}}$. The $K\!=\!10$ residual map of FDS candidate GCs is in the background 
    (positive residual are expressed in gray tones).}
    \label{fig:core_spectro}
\end{figure*}

The kinematics of the GCSs of galaxies in the Fornax cluster has been thoroughly 
investigated in the literature using optical spectra obtained through multiple 
observational campaigns for a large number of 
GCs~\citep{mieske2004,dirsch2004,bergond2007,firth2007,schuberth2010,pota2018,fahrion2020,chaturvedi2022}. 
By taking advantage of these efforts, we can probe the nature of the spatial features 
identified in the residual maps of FDS ICGCs in the line-of-sight (l.o.s.) velocity v$_{\mathrm{los}}$ 
space of a subset of confirmed GCs. 
We use the sample of spectroscopically-selected GCs with v$_{\mathrm{los}}$ measurements 
collected by~\cite{chaturvedi2022}. This dataset, based on VLT/VIMOS observations 
of the central square degree of the Fornax cluster augmented by additional data from existing 
literature, comprises 2288 sources selected as {\it bona fide} GCs based on the 
absence of emission lines and $v_{\mathrm{los}}$ in the $[450,2500]$ 
km s$^{-1}$ interval~\citep{schuberth2010}. Following~\cite{chaturvedi2022}, we select ICGCs 
among GCs in the spectroscopic sample by discarding all sources located within
two effective radii from the location of major galaxies~\citep{iodice2019a} and whose 
$v_{\mathrm{los}}$ is within $\pm2\sigma_{v_{\mathrm{los}}}$ the l.o.s. velocity of 
the closest galaxy, 
where $\sigma_{v_{\mathrm{los}}}$ is the standard deviation of the GCSs v$_{\mathrm{los}}$ 
of the galaxy. These criteria return 890 GCs with measured $v_{\mathrm{los}}$. The positions
of these spectroscopically selected ICGCs relative 
to the residual maps in the core of the Fornax cluster are shown in 
Figure~\ref{fig:core_spectro}. The number of spectroscopic GCs and ICGCs 
in each of the spatial structures of the FDS GC residual maps, the average 
l.o.s velocities and their standard deviations are listed in 
Table~\ref{tab:residual_maps_spectro}.

\begin{deluxetable}{cccccccccc}
	\tablecaption{Properties of the spectroscopically selected GCs
        from~\cite{chaturvedi2022} located within the spatial features of the FDS GCs residual map.}
 	\label{tab:residual_maps_spectro}
 	\tablehead{
            \colhead{Str.\tablenotemark{a}} &
            \colhead{Substr.\tablenotemark{b}} &
 	    \colhead{N$^{\mathrm{v_{los}}}_{\mathrm{GCs}}$\tablenotemark{c}} &
            \colhead{$\bar{v}_{\mathrm{los}}^{(GCs)}$\tablenotemark{d}} &
            \colhead{N$^{\mathrm{v_{los}}}_{\mathrm{ICGCs}}$\tablenotemark{e}} &
            \colhead{$\bar{v}_{\mathrm{los}}^{(\mathrm{ICGCs})}$\tablenotemark{f}} &
            \colhead{N$^{\mathrm{v_{los}}}_{\mathrm{ICGCs,FDS}}$\tablenotemark{g}} &
            \colhead{$\overline{\mathrm{m}}_{g}^{(\mathrm{ICGCs,FDS})}$\tablenotemark{h}} &
            \colhead{$\overline{g\!-\!i}^{(\mathrm{ICGCs,FDS})}$\tablenotemark{i}} &
            }
        \startdata
             M       & M2    & 20    & 1371.4$\pm$310.8      & 20    & 1371.4$\pm$310.8    & 10  & 22.5$\pm$0.3 $[22.4\!\pm\!0.7]$& 0.85$\pm$0.04 $[0.93\!\pm\!0.13]$ \\
             M       & M3    & 21    & 1452.1$\pm$336.6      & 21    & 1452.1$\pm$336.6    & 11  & 22.1$\pm$0.3 $[22.3\!\pm\!0.6]$& 0.89$\pm$0.07 $[0.92\!\pm\!0.12]$                \\
             M       & P1    & 40    & 1434.0$\pm$228.7      & 40    & 1434.0$\pm$228.7    & 27  & 22.3$\pm$0.5 $[22.4\!\pm\!0.5]$& 0.86$\pm$0.11 $[0.87\!\pm\!0.12]$                \\
             M       & P2    & 16    & 1352.2$\pm$282.8      & 16    & 1352.2$\pm$282.8    & 13  & 22.5$\pm$0.7 $[22.7\!\pm\!0.7]$& 0.88$\pm$0.08 $[0.87\!\pm\!0.08]$                \\
             M       & P3    & 9     & 1717.9$\pm$339.3      & 9     & 1717.9$\pm$339.3    & 6   & 22.4$\pm$0.7 $[22.6\!\pm\!0.7]$& 0.90$\pm$0.12 $[0.91\!\pm\!0.12]$                \\ 
             M       & P4    & 15    & 1287.0$\pm$416.7      & 15    & 1287.0$\pm$416.7    & 9   & 22.2$\pm$0.6 $[22.4\!\pm\!0.7]$& 0.88$\pm$0.09 $[0.86\!\pm\!0.12]$                \\
             M       & P5    & 12    & 1591.5$\pm$309.0      & 12    & 1591.5$\pm$309.0    & 9   & 22.3$\pm$0.4 $[22.6\!\pm\!0.5]$& 0.85$\pm$0.08 $[0.85\!\pm\!0.14]$                \\ 
             C3      & -     & 14    & 1625.3$\pm$76.0       & 4     & 1602.4$\pm$67.6     & 1   & 22.63        $[22.4\!\pm\!0.7]$& 0.96          $[0.85\!\pm\!0.13]$                 \\ 
             C4      & -     & 8     & 1599.1$\pm$28.8       & 7     & 1597.8$\pm$30.9     & -   & -                & -                                  \\
        \enddata
     \tablecomments{(a): Label of the residual structure;
                    (b): Label of the residual sub-structure; 
                    (c): Total number of GCs with measured v$_{\mathrm{los}}$ within the residual structure or substructure;
                    (d): Mean and standard deviation of the v$_{\mathrm{los}}$ of GCs within the residual structure or substructure;
                    (e): Total number of ICGCs with measured v$_{\mathrm{los}}$ within the residual structure or substructure; 
                    (f): Mean and standard deviation of the v$_{\mathrm{los}}$ of ICGCs within the residual structure or substructure;
                    (g): Total number of ICGCs with measured v$_{\mathrm{los}}$ matched to FDS GCs within the residual structure or substructure; 
                    (h): Mean and standard deviation of the FDS $g$-band magnitude m$_{g}$ of spectroscopic
                    ICGCs matched to FDS GCs (in parenthesis, the same values for the full 
                    sample of FDS GCs located within the structure); 
                    (i): Mean and standard deviation of the FDS $g\!-\!i$ color of spectroscopic
                    ICGCs matched to FDS GCs (in parenthesis, the same values for the full 
                    sample of FDS GCs located within the structure); 
                    } 
\end{deluxetable} 

\begin{figure*}[ht]
    \centering
    \includegraphics[width=\linewidth]{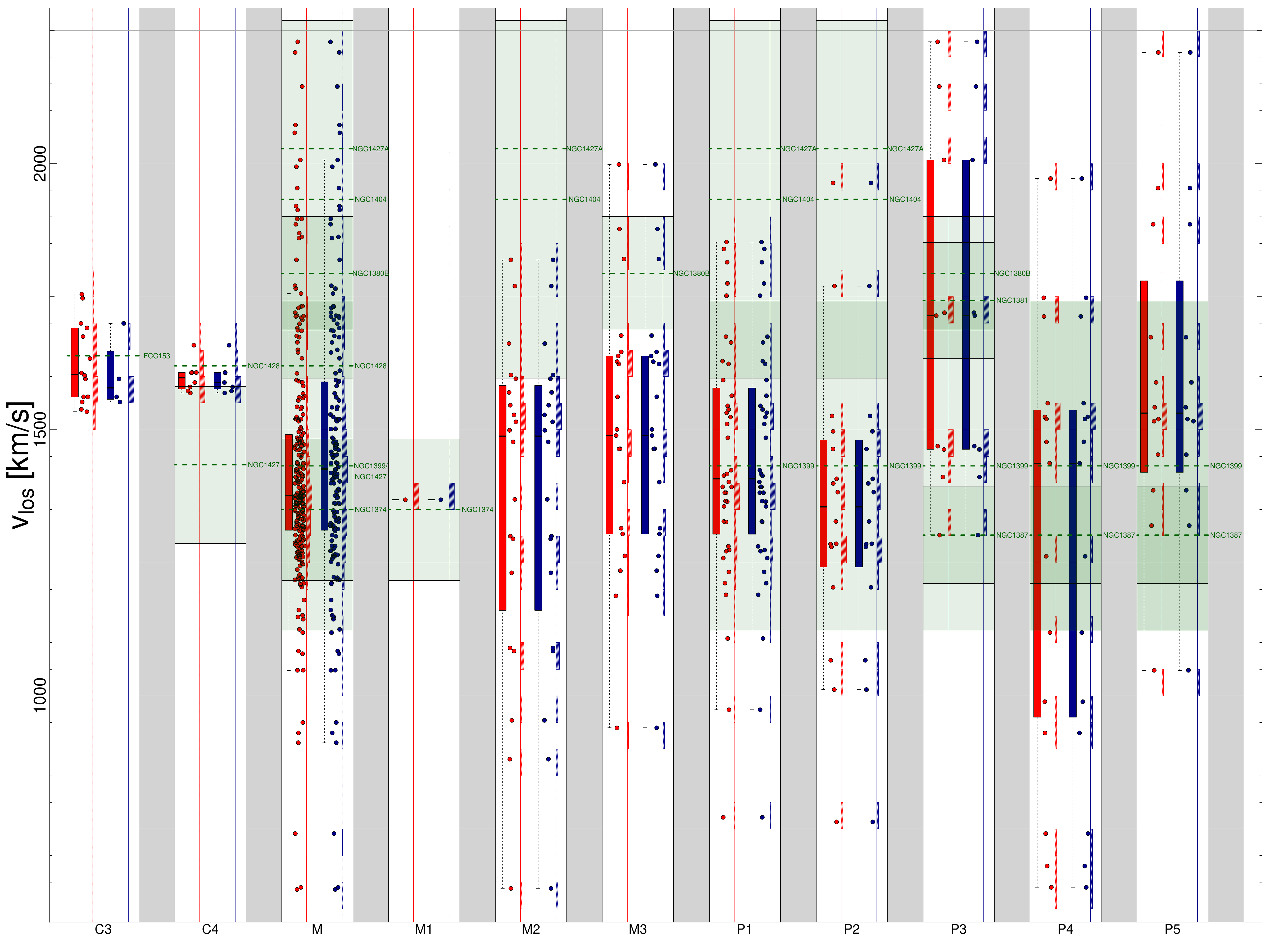}
    \caption{Line-of-sight velocity $v_{\mathrm{los}}$ distribution of confirmed GCs 
    from~\cite{chaturvedi2022} (see Section~\ref{subsec:spectrogcs} for details) 
    located within the spatial structures and areas of positive residual peaks 
    detected in the residual map of FDS GCs  and described in Section~\ref{sec:results}. 
    The distribution of $v_{\mathrm{los}}$ of all spectroscopically observed GCs located 
    within the main over-density M in the core of the Fornax cluster is also displayed 
    for reference. Boxplots, histograms and distinct v$_{los}$ values for each GCs are shown for 
    all spectroscopically selected GCs (red) and ICGCs only (blue). 
    The horizontal dashed lines show the systemic velocities of the main nearby galaxies,
    while the green shaded areas represent the $\pm1\sigma$ v$_{\mathrm{los}}$ 
    dispersion of the GCSs of the corresponding galaxies, based on the~\cite{chaturvedi2022} 
    sample and calculated within two effective radii.} 
    \label{fig:core_spectro_vhel}
\end{figure*}

The v$_{\mathrm{los}}$ distributions for both the whole~\cite{chaturvedi2022} sample of
spectroscopically-selected GCs (red) and ICGCs (blue) located within the residual structures of 
the FDS GCs spatial distribution are shown in Figure~\ref{fig:core_spectro_vhel}. The
dispersion of the $v_{\mathrm{los}}$ distribution of the GCSs of nearby galaxies, calculated
using all~\cite{chaturvedi2022} GCs within 2r$_{e}$ of the galaxy, is also displayed. 
The number of spectroscopic GCs in the spatial structures M2 and M3 is small 
(20 and 21, respectively) because they are located far from bright galaxies and the 
area surrounding NGC\,1399 where most of the targets of the spectroscopic 
campaigns can be found. M1 was not included because it only contains
one GCs with measured l.o.s. velocity. The numbers of spectroscopically observed GCs
in the residual enhancement areas P1, P2, P3, P4 and P5 are 40, 16, 9, 15 and 12, respectively.
For all the residual structures mentioned so far, all GCs with $v_{\mathrm{los}}$ are also
selected as ICGCs, unlike for spatial structures C3 (14 spectroscopic GCs and 4 ICGCs) 
and C4 (8 total GCs and 6 ICGCs).

By comparing the v$_{\mathrm{los}}$ of GCs in the features of the residual 
maps described in this paper with the systemic velocities 
of the galaxies~\citep{iodice2019a} within or nearby each residual spatial structure, 
and the dispersion of the v$_{\mathrm{los}}$ of GCSs associated with these galaxies, we can 
comment on the nature of the residual structures. 

\begin{itemize}
    \item The bulk of the v$_{\mathrm{los}}$ distributions of spectroscopic GCs 
    located within structures M2 and M3 does 
    not match the systemic velocities of any nearby galaxies (NGC\,1404, NGC\,1427A
    for M2 and NGC\,1380B for M3), although partial overlaps with the 
    v$_{\mathrm{los}}$ span of GCs in  
    NGC\,1404, at $\sim\!1900$ km s$^{-1}$ and the v$_{\mathrm{los}}$ interval 
    covered by the NGC\,1380B GCS (centered around $\sim\!1800$ km s$^{-1}$). While the number of 
    GCs and ICGCs with velocity measurements is relatively small for both
    structures and the GCs in the~\cite{chaturvedi2022}
    sample only cover a small, distinct northern area of M2, the majority
    of the v$_{\mathrm{los}}$ in both residual structures are compatible with the systemic
    velocity of NGC\,1399 and the dispersion of its GCS. 
    \item The residual enhancement P1 contains the largest number ($\sim$40) of 
    spectroscopic GCs among
    all the spatial residual features investigated in this paper. The v$_{\mathrm{los}}$ 
    distribution extends from 
    $\sim\!600$ km s$^{-1}$ to $\sim\!1850$ km s$^{-1}$ and features two sub-peaks 
    approximately located at $\sim\!$1350 km s$^{-1}$ and $\sim\!$1550 km s$^{-1}$ respectively, both well
    within the range of l.o.s. velocities spanned by the GCS of NGC\,1399. While characterized by 
    a large v$_{\mathrm{los}}$ variance, the GC systems of
    NGC\,1427A and NGC\,1404, the two galaxies located closes to the residual peak, do not seem to 
    significantly contribute to the observed distribution of v$_{\mathrm{los}}$ in P1, except for 
    v$_{\mathrm{los}}\geq\!1750$ km s$^{-1}$. These results suggest that this over-density is 
    mostly the result of the accumulation of GCs stripped from the eastern side of the 
    very large halo of NGC\,1399. Similar conclusions can be drawn for P2, although 
    the number of spectroscopically observed GCs is smaller than for P1.
    \item The distributions of v$_{\mathrm{los}}$ for GCs within residual peaks P3 and P4, 
    although affected by small statistics, are consistent with the superpositions of 
    the velocity distributions of GCs around
    NGC\,1399, NGC\,1387, NGC\,1380B, NGC\,1381 for P3 and NGC\,1399 and NGC\,1387 for P4, with
    the exception of a few GCs with v$_{\mathrm{los}}\!<\!1100$ km s$^{-1}$ in P3.
    \item The residual peak P5 features a small sample of spectroscopic GCs with 
    v$_{\mathrm{los}}$ whose values are compatible with the GC velocity ranges of the GCSs 
    of the nearby galaxy NGC\,1399.
    \item The distributions of v$_{\mathrm{los}}$ of the spectroscopically observed GCs within 
    spatial structures C3 and C4 are compatible with the systemic velocities of the galaxies 
    lying within their boundaries, i.e. FCC\,153 and NGC\,1428 respectively, indicating that the 
    observed excess of GCs in these two regions can be entirely associated with the extended GCSs
    of the two galaxies.
\end{itemize}

In summary, the analysis of the distribution of the v$_{\mathrm{los}}$ of
GCs located within the boundaries of the residual structures derived from FDS GCs  
strongly signals that most spatial structures can be kinematically correlated with the GCSs of 
nearby, bright galaxies. In particular, we observe large fractions of GCs 
whose line-of-sight velocities are compatible with the range of values typical of the 
NGC\,1399 GCS in all spatial structures (except for C3 and C4) and areas of positive 
residual peaks P1 and P2, which are all $\geq\!0.3\degree$ ($\geq\!100$ kpc) away
from the center of NGC\,1399. The comparison of the v$_{\mathrm{los}}$ of spectroscopic
GCs inside the spatial structures and the residual peaks with the systemic 
velocities of the main galaxies in their immediate surroundings seems to suggest that 
the GC over-densities inside the core of the Fornax cluster may be the result 
of the superposition of kinematically distinct subgroups of GCs, i.e. A) GCs 
still belonging to the GCSs of such
galaxies and located at galacto-centric distances $\geq\!2\mathrm{r}_{e}$, 
B) GCs stripped from their original host by the gravitational interaction with NGC\,1399, 
and A) members of the NGC\,1399 GC systems initially located at very large 
galacto-centric distances. Caveats to these interpretations
stem from the differences in the photometric selection criteria used to extract 
the samples of FDS candidate GCs and the targets of the spectroscopic observations, 
although the average $g$ magnitudes and $g\!-\!i$ colors
of the subsets of~\cite{chaturvedi2022} GCs spatially matched with FDS GCs
are similar to the same properties for the whole sample of FDS GCs in each residual 
structure~(cp. columns {\it h} and {\it i} from Table~\ref{tab:residual_maps_spectro}). 
Another potential limitation of this analysis is the uneven number of spectroscopic sources 
within the residual structures, caused by the geometrical incompleteness of the spectroscopic
surveys: the fraction of the number of matched 
spectroscopically-selected GCs to total number of FDS GCs ranges in the $\sim\!25\%$ to 
$\sim\!60\%$ interval for residual peak 
regions P1, P2, P3, P4 and P5 and hovers around the $\sim\!15\%$ level for spatial structures 
M2 and M3, where spectroscopic GCs with measured v$_{\mathrm{los}}$ are also concentrated
along the boundaries with the main body of the spatial structure M~(Figure~\ref{fig:core_spectro}).

\subsection{The Assembly History of the Fornax cluster}
\label{subsec:assembly}

The Fornax cluster has been long believed to be relaxed and in an advanced evolutionary stage, 
mostly based on the ratio of ETGs to late-type galaxies (LTGs) being larger than in other nearby, 
unrelaxed clusters of galaxies~\citep{cappellari2011}. Over time, distinct observational
approaches have painted a more detailed picture of the Fornax cluster assembly history, which
hints at multiple infalls of galaxy sub-structures towards the center
of the cluster gravitational potential, occurring at different times. 

The morphology of the large-scale X-ray emission of the core of the Fornax cluster as
observed by ROSAT has been modeled using three different, asymmetric components whose 
centroids do not match the optical center of the cluster~\citep{paolillo2002}, likely as 
the result of the bright 
galaxies in the core sloshing within the dark matter halo of the cluster.~\cite{paolillo2002} 
also found surface brightness inhomogeneities of filamentary shape while, more recently, 
in the central region of the cluster,~\cite{su2017} discovered sloshing cold fronts  
in the diffuse emission observed by {\it Chandra} and XMM-{\it Newton} and likely 
associated with the infall of NGC\,1404. The displacement between the center of the 
cluster-wide component of the extended ROSAT X-ray emission and the center of the 
cluster of galaxies (NGC\,1399) suggests that the Fornax 
cluster might not be relaxed and that a merger event may be ongoing~\citep{paolillo2002}. 
On smaller scales,~\cite{sheardown2018} used hydro-dynamical and N-bodies simulations to 
reproduce the sloshing patterns observed in the X-ray surface brightness around NGC\,1404; 
they found that these spatial features are produced by a second or third infall 
of NGC\,1404 towards the center of the cluster. The small area of  
density enhancement visible in the GC candidates residual maps in the position of 
NGC\,1404, if not an artifact produced by the method used to derive the residual maps 
(see discussion in Section~\ref{sec:results}), would not suggest a significantly 
depleted NGC\,1404 GCS due to multiple interactions with the deeper potential of 
NGC\,1399, although this scenario seems to be favored by independent observations 
highlighting that NGC\,1404 has very low S$_N$ for its luminosity~\citep{richtler1992} and compared 
to other ETGs in the Fornax cluster~\citep{kisslerpatig1997}, and by simulations of the interaction
with NGC\,1399~\citep{bekki2003}. On the other hand, the larger, morphological complex 
region of moderately positive residuals observed between NGC\,1404 and NGC\,1399 could 
result from localized GCs excess linked to the stripping of the NGC\,1404 GCS during its 
long-duration interaction with NGC\,1399. 

\begin{figure*}[ht]
    \centering
    \includegraphics[width=\linewidth]{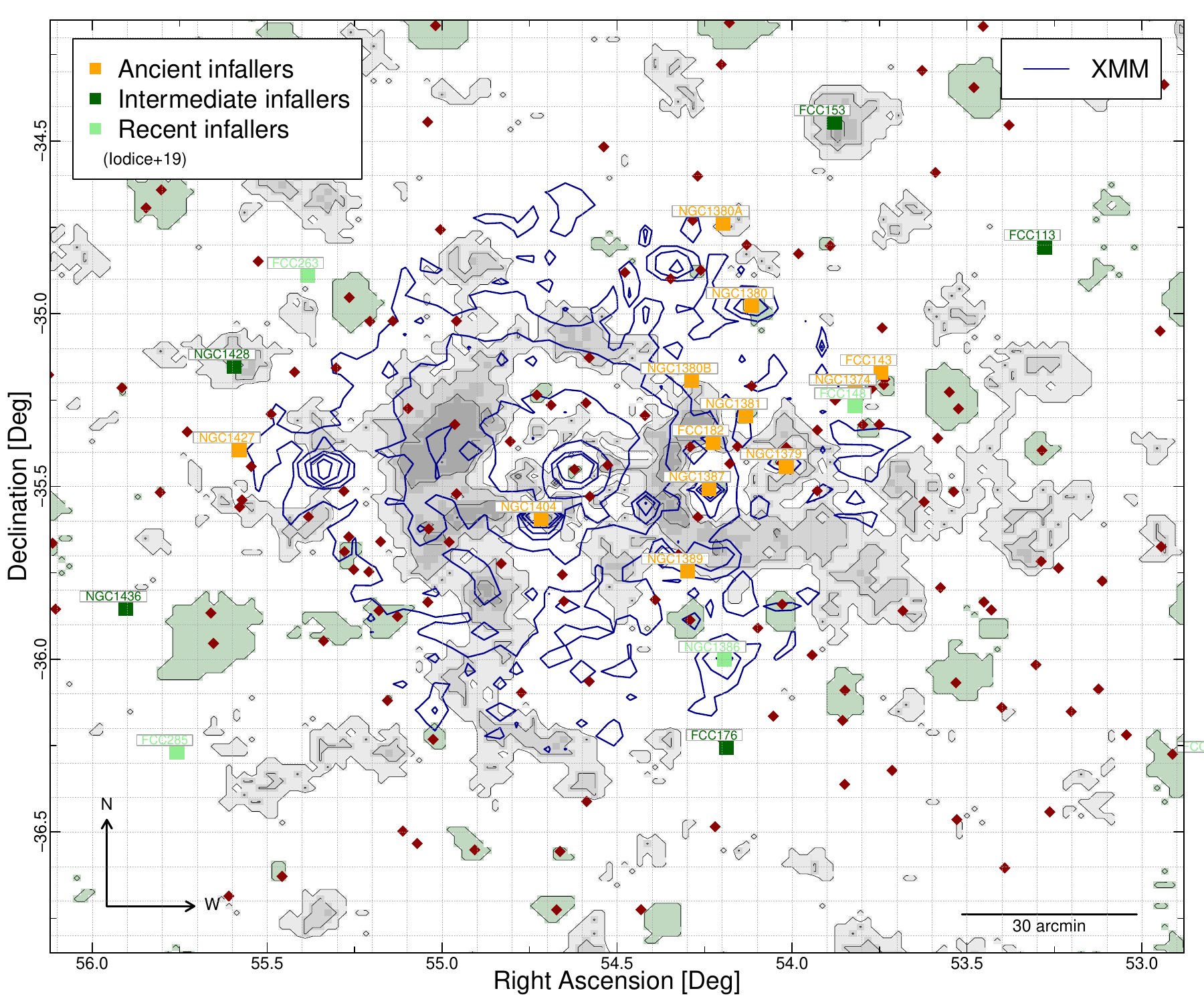}
    \caption{Residual map obtained with $K\!=\!10$ from the spatial distribution
    of GCs in the core of the Fornax cluster (positive residuals are color-coded in 
    grayscale). The blue lines represent the X-ray contours from XMM-Newton images 
    in the 0.4-1.3 KeV 
    energy band~\citep{frank2013}. The orange, dark green and light green squares represent
    the ancient, intermediate and recent infallers respectively~\citep{iodice2019a}.}
    \label{fig:core_xrays}
\end{figure*}

\cite{iodice2019a} provided additional details of the assembly history of the cluster 
by jointly investigating the Projected Phase-Space~(PPS) distribution~\citep{rhee2017} 
of all bright Fornax galaxies within one virial radius of the cluster
and their structural features (morphology, kinematics, colors and stellar 
population).~\cite{iodice2019a} found three classes of galaxies with
distinct spatial distributions and global properties: the core galaxies, the North-South
clump and the infalling galaxies. The positions of the latter two groups in the PPS
diagram suggest that the N-S clump galaxies infalled more than $\sim$8 Gyr ago, while
the infalling galaxies include both intermediate and recent infallers, which entered
the cluster $\leq$ 4 Gyr ago. Figure~\ref{fig:core_xrays} shows the positions of the 
three classes of galaxies identified by~\cite{iodice2019a} in the core of the Fornax cluster, 
with the contours of the diffuse X-ray emission observed by 
XMM-{\it Newton}~\citep{frank2013} (in the background, the $K\!=\!10$ residual map of the 
FDS candidate GCs spatial distribution). 

The X-ray emission in the eastern side of the cluster core 
approximately matches the shape of the main spatial over-density M and overlaps with 
the positions of the residual peaks P1 and P2. In the 
western half of the core, the X-ray brightness remains high in the 
region of large GC candidates positive residuals P3, P4 and P5 detected 
between NGC\,1399 and NGC\,1387, before sharply falling to levels compatible 
with the background emission immediately 
west of NGC\,1387. The analysis of recent SGR/eROSITA data of the Fornax
cluster confirms the asymmetry in the X-ray diffuse emission observed in the core of the 
cluster and a sharp fall of the X-ray brightness W of NGC\,1387 (although lower
surface brightness emission whose morphology resembles the spatial distribution
of ICGCs, is detected beyond the virial radius~\citep{reiprich2025}.
The similarities between the diffuse X-ray emission and the spatial distribution 
of ICGCs in the eastern and central regions of the Fornax cluster core contrast with the 
lack of measurable X-ray emission in the western section, where the large spatial feature 
M1 is located. These differences reinforce the scenario according to which different dominant
physical mechanisms have driven the formation of ICGCs detected in 
opposite sides of the Fornax core. The western GCs over-density has been likely 
produced by galaxy-to-galaxy interactions that have occurred in comparatively shallow 
gravitational potential, which is less favorable to the accumulation and heating of 
the hot, dense Intra-Cluster Medium (ICM) responsible for the X-ray emission
then the core of a rich cluster. According to this scenario, the ICGCs was deposited in the western
region of the Fornax cluster core, where they are currently observed, 
at a later stage. This non-local assembly of ICGCs would mimic what
is also observed for the ICL in clusters, for which there is strong observational 
evidence that diffuse light formed in groups can contribute
up to $\sim\!30\%$ of the total ICL luminosity observed in 
clusters~\cite[i.e. {\it pre-processing}, see][]{contini2014,iodice2017,montes2022}.
The western edge of the spatially extended X-ray emission matches the area occupied by
the ancient infallers located within the N-S clump,  
and the extended region of enhanced GCs residuals surrounding 
NGC\,1387 and between NGC\,1399 and NGC\,1387, where the GCs residual peaks 
P3, P4 and P5 are found. As suggested by~\cite{nasonova2011}, these galaxies 
are distributed along the direction of the 
cosmic web filament of the Fornax-Eridanus large-scale complex. This alignment indicates
that the assembly of the Fornax cluster may be still ongoing and that the N-S clump might have 
formed as a distinct sub-group currently merging with the Fornax core. Assuming that
the regions of ICL~\citep{iodice2017} and the residual over-densities associated with 
excess of ICGCs matching the position of the N-S clump both originated in the merging 
sub-group, it would possible to constrain the formation mechanism and time of the currently 
observed ICGCs excess.

\section{Conclusions}
\label{sec:conclusions}

In this paper, we investigated the spatial distribution of the
ICGCs in the core of the Fornax cluster. The properties of the ICGCs population have been estimated 
by simulating and subtracting from the observed distribution of FDS candidate GCs the sum of two
components: a homogeneous, spatially isotropic component of fore- and background contaminants, and 
the contributions of GCSs of all bright galaxies (the contribution from dwarfs was not modeled)
in the observed field. The GCSs associated with galaxies in the MGS, for 
which detailed GCLFs and radial profiles are available in the literature, and fainter galaxies (SGS) for 
which correlations between the host magnitude and the properties of their GCSs can be used, were modeled 
separately. 

A summary of our main results includes the following:

\begin{itemize}
    \item We found that the core of the Fornax cluster hosts a substantial excess of 
    GCs associated with the population of ICGCs observed in previous 
    works~\citep{bassino2003,dabrusco2016,cantiello2020}. The main GC residual 
    enhancement M detected in this area is roughly centered on NGC\,1399, 
    has an area of $\sim0.9(^{\circ})^{2}$ and spans {$\sim\!(2.35\!\pm\!0.08)^{\degree}$}
    along the W-E direction. Its complex geometrical structure is characterized by
    several spatial features that deviate from a simple elliptical model and by a
    two-dimensional profile with several, high statistically-significant residual peaks. 
    This intricate morphology suggests a long and active history of interactions among galaxies 
    which have shaped the observed spatial distributions of ICGCs on the spatial scale 
    of the whole Fornax cluster core.
    \item The central over-density M contains an observed excess number of GCs 
    N$_{\mathrm{GCs}}^{(\mathrm{obs,exc})}\!=\!627.1\!\pm\!74.3$
    (N$_{\mathrm{GCs}}^{(\mathrm{obs,exc})}\!=\!989.5\!\pm\!138.3$ if the four 
    control fields are used for the estimation of the density of the homogeneous 
    component). This estimate corresponds to a total 
    ICGCs population N$_{\mathrm{ICGCs}}\!=\!6750\!\pm\!552$ 
    (N$_{\mathrm{ICGCs}}\!=\!10650\!\pm\!1090$ for the lower homogeneous component density value) 
    when the excess number is corrected for flux incompleteness, using a Gaussian model for the GCLF whose parameters 
    are set to the mean values of the turn-over magnitudes and dispersions 
    of the ACSFCS GCLFs for MGS galaxies~\citep{villegas2010}, and geometric incompleteness due
    to the avoidance regions in the centers of the bright host galaxies~(Section~\ref{subsubsec:galcomponent}).
    \item We found that the $g\!-\!i$ color distribution of the full sample of FDS candidate GCs is
    bimodal. After statistically subtracting the contribution of 
    candidate GCs located outside the virial radius of the cluster from the colors of GC candidates inside
    the virial radius of the cluster, we confirmed that the color distribution
    of the ICGCs excess is still bimodal, with red and blue components compatible with the models from
    all GCs observed in the same area. The smoothed spatial color map, averaged over 100 realizations of the
    background-subtracted distributions of GCs within the Fornax r$_{\mathrm{vir}}$, shows that the residual 
    structures 
    in the core of the cluster tend to be mostly populated by an excess of blue GCs, with the exception
    of structure M3m M3 and C2, where significant areas of red GC candidates are detected. This scenario seems to 
    agree with the 
    general observational knowledge of the color properties of ICGCs and the fact that most of the spatial features 
    and residual peaks are relatively distant from bright galaxies. The average blue color of candidate 
    GCs in the ICGCs spatial structures also seems to hint at a scenario where ICGCs originates 
    from the combined contributions of stripped GCs in the outskirts of the radially extended GCSs of bright galaxies, 
    and the tidal disruption of low-mass dwarf galaxies, whose GCS tends to be on average bluer 
    than those of more luminous hosts~\citep{peng2006}.
    \item The comparison of the positions of the residual structures of the FDS GCs 
    spatial distribution with the ICL emission discovered in the core 
    of the cluster by~\cite{iodice2017} reveals a complex picture. While the majority of ICL 
    light detected between NGC\,1399, NGC\,1387, NGC\,1381 and NGC\,1380B overlaps with the 
    main residual structure M and, in particular, with the positions of the residual 
    enhancement region P3, P5 and P4, no indications of ICL with surface 
    brightness $\mu_{g}\!\geq\!28$ mag arcsec$^{-2}$ are found west of NGC\,1389 and S of NGC\,1379.
    Even if the direct comparison of the two components does not support 
    the existence of a global correlation between the two
    on the whole region occupied by the ICGC over-density, other considerations seem to 
    indirectly confirm that ICL and ICGCs both trace the same underlying gravitational potential.
    By investigating the kinematics of a large sample of PNe in the same 
    area,~\cite{spiniello2018} found that intra-cluster PNe, which spatially match 
    the ICL emission detected by~\cite{iodice2017}, have velocity dispersion radial 
    profile similar to spectroscopically observed blue GCs, likely members of the ICGC population, 
    for virial-centric distances larger than 400$\arcsec$; beyond this radius, the kinematics of 
    PNe indicates that they follow 
    the same cluster-wide potential as bright Fornax galaxies, albeit with a different density
    profile.
    \item The density map of dwarf galaxies detected in the core of 
    the Fornax cluster features several enhancements and does not generally 
    resembles the geometry of ICGCs as described by the residual maps of the
    GC candidates spatial distribution. The 
    more extended and statistically significant over-densities are located  
    on the west side, where they overlap with the spatial feature M1 and the 
    central section of the main structure M, in the region surrounding NGC\,1387 
    where peaks P3, P5 and P4 are located. Conversely, the absence of dEs over-density
    overlapping with the eastern section of M and within $\sim\!0.55\degree$ from NGC\,1399 points to the
    effects of the
    tidal disruption of a pre-existing, more abundant dwarfs population (whose remnants are 
    still visible in the eastern boundaries of M) that may have contributed to the local ICGCs 
    (in agreement with the blue average color of the excess GCs located 
    within the residual structures across the whole Fornax cluster core, as discussed above). 
    Assuming a specific
    frequency S$_{N}^{(\mathrm{ICGCs})}\!=\!8$, typical of GC-rich dEs located in the centers
    of cluster, the total ICGC population within M, corrected for completeness, could be produced
    by the disruption of $\sim\!3,600$ $M_{V}^{(\mathrm{dEs})}\!=\!-13.3$ dwarf galaxies,
    while a lower specific frequency value S$_{N}\!=\!1.5$ would require the disruption of progenitors
    with total luminosity $\sim\!3.8\cdot10^{11}L_{\sun}$. If the assembly of the ICL detected in the 
    core of the Fornax cluster west of NGC\,1399 and the accumulation of the local ICGCs excess are
    driven by the same processes, as 
    suggested by their morphological similarities noticed by~\citep{iodice2017}, the total ICL luminosity, 
    the absence of additional significant ICL in other regions of the cluster and the inefficiency
    of dEs disruption as production channel of ICL in cluster of galaxies suggest
    a scenario according to which a mix of formation mechanisms, including the stripping of GCs 
    from the outskirts of intermediate-luminosity galaxies by the NGC\,1399, other galaxy-to-galaxy 
    interactions and the contribution of pre-processed material, that can generate more efficiently 
    ICL in the intra-cluster space, is responsible for the growth of the ICGCs population in the core of 
    the Fornax cluster.
    \item We used the most comprehensive sample of spectroscopically-selected GCs observed in the 
    core of the Fornax cluster~\citep{chaturvedi2022} to compare the v$_{\mathrm{los}}$ distribution
    of GCs within the residual features with the systemic velocities of the GCSs associated 
    with bright galaxies in the same region. 
    This analysis shows that most spatial structures are kinematically correlated with the GCSs of 
    nearby, bright galaxies, with sizable fractions of GCs whose l.o.s. velocities are compatible with 
    the range of values occupied by members of the NGC\,1399 GCS (and consistent with the systemic 
    velocity of the Fornax cluster) in almost all spatial structures (except for C3 and 
    C4), and positive residual peaks P1 and P2. The relatively small size of the spectroscopic 
    sample and its 
    uneven spatial coverage can bias the interpretation 
    of the properties of their l.o.s. velocity distributions. A further complicating factor in the 
    interpretation of the kinematics of the GCs included in residual structures can be
    the presence of kinematically
    cold streams that are not related to any existing galaxy, as they were left over from the total 
    disruption of former host 
    galaxies~\citep[see][for examples in the Fornax cluster]{napolitano2022}.
    \item The analysis of the potential mechanisms that have determined the 
    observed ICGCs spatial distribution in the context
    of the current understanding of the assembly history of Fornax cluster reinforces
    the notion that different mechanisms have contributed to the growth of the ICGCs population
    in different sections of the Fornax cluster core. The fact that the morphologies of the 
    spatial distribution 
    of ICGCs in the eastern and central region of 
    the Fornax cluster and the large-scale X-ray emission detected by 
    ROSAT~\citep{paolillo2002} and XMM-Newton~\citep{frank2013} are similar
    suggests that the geometry of 
    ICGCs population in this area may not be affected by the assembly 
    events that have shaped the recent growth of the cluster. 
    On the other hand, the spatial match between
    the large, statistically significant residual enhancement region straddling the 
    NGC\,1399-NGC\,1387 region and encompassing peaks P3, P4 and P5 with the N-S clump of galaxies 
    identified as ancient ($\geq8$ Gyr) infallers by~\cite{iodice2019a}, and the steep fall-off 
    in the brightness of the diffuse X-ray emission west of NGC\,1387~\citep{reiprich2025} 
    associated with the ICGC spatial structure M1 seem to indicate that, 
    at least on sub-cluster scales, the geometry of ICGCs traces relatively 
    old merging events, and that pre-processing may have contributed to the
    observed excess of ICGCs. The qualitative agreement between the morphology and alignment
    of the elongated spatial features in the large-scale distribution of ICGCs within 
    the Fornax cluster virial radius (structures C1, M2 and M1 described in 
    this paper, and G and F from~\citealt{cantiello2020}) and the finger-like features detected 
    in the diffuse X-ray emission observed by SRG/eROSITA~\citep{reiprich2025} supports the hypothesis 
    that the ICGCs may have been partially pre-processed in infalling systems and deposited along 
    the accretion directions that are also highlighted by other probes of the 
    cluster assembly history (e.g. X-ray bright gas and ICL).
\end{itemize}

In the future, new observational facilities will allow to improve the analysis presented 
in this paper in the transitional region surrounding the bright galaxies and, in turn, 
will produce more accurate maps of the 
spatial distribution of the ICGCs populations at all spatial scales probed, that are
not accessible to ground-based telescopes. For example,
the detailed, 
two-dimensional characterization of the spatial profiles~\citep{dabrusco2022} of the GCSs 
of all the most luminous galaxies in Fornax obtained by using 
ACSFCS data~\citep{jordan2009,jordan2015} will be leveraged 
to replace the assumption of elliptical geometry from the diffuse stellar light
of the host galaxies, adopted in this paper, for the azimuthal distribution
of the simulated GCSs with realistic models~\citep{jordan2007}.
These improvements will allow to refine the simulations of the galaxy contribution to the global
observed GCs population and describe more accurately the properties of the ICGCs in the 
central area of the cluster and in the vicinity of bright galaxies. In particular, 
the initial results based on {\it Euclid} ERO-F data~\citep[][Saifollahi et al. in prep.]{saifollahi2024}, 
Perseus cluster data~\citep{kluge2024,marleau2024}
and the predictions for {\it Euclid} Wide Survey~\citep{euclid2024voggel} show how well-suited 
{\it Euclid} imaging is for the identification of deep, spatially extended samples of 
GCs in cluster of galaxies, that  will allow, for the first time, 
the modeling of the spatial properties of the GCSs of bright galaxies and of the 
surrounding ICGCs within the same dataset.

\section{Acknowledgments}

\noindent R.D'A. acknowledges support from the {\it Chandra} X-ray Center, which is operated by 
the Smithsonian Institution under NASA contract NAS8-03060.
EI and MC acknowledge the support by the Italian Ministry for Education University and Research 
(MIUR) grant PRIN 2022 2022383WFT “SUNRISE”, CUP C53D23000850006. EI, MS, MC acknowledge funding 
from the INAF through the large grant PRIN 12-2022 "INAF-EDGE" (PI L. Hunt). M.P. acknowledges 
the ﬁnancial contribution from PRIN-MIUR 2022 and from the Timedomes grant within the "INAF 2023 
Finanziamento della Ricerca Fondamentale".\\ 
This paper is based on observations collected at the European Southern Observatory under ESO 
programs 094.B-0496, 094.B-0512 and 092.B-0744, and data obtained from the ESO Science Archive 
Facility with DOI(s) under \href{https://doi.org/10.18727/archive/23}{https://doi.org/10.18727/archive/23}.\\ 
Data analysis for this work was performed by using TOPCAT~\citep{taylor2005} and 
the~\href{https://www.R-project.org/}{R language}.

\appendix

\section{Specific frequency for the Secondary Galaxy Sample}
\label{sec:appendix1}

The values of the specific frequency of the members of SGS used to simulate their GCSs (see Section~\ref{subsubsec:galcomponent} for
details) are randomly extracted from the observed 
range of $S_{N}$ values from Figure~11 of~\citet{harris2013} as a function of the hosts' visible absolute magnitude M$_{V}$. The M$_{V}$
intervals used are shown in Table~\ref{tab:specfreq}. 
Different interval of $S_{N}$ were used for galaxies classified as either Elliptical or dwarf Elliptical (\textit{E*} or~\textit{dE*}), 
lenticular (\textit{S*} or \textit{dS*}), or spiral/irregular according to the FCC~\citep{ferguson1989}. 

\begin{deluxetable}{cccc}
 	\tablecaption{Specific frequency intervals used to draw the simulated values for galaxies in the 
	SGS (Section~\ref{subsubsec:galcomponent}) as a function of the absolute magnitude M$_{V}$ 
        and morphological classification of the host galaxy~\citep[Figure~11]{harris2013}.}
 	\label{tab:specfreq}
 	\tablehead{
            \colhead{M$_{V}$ range\tablenotemark{a}} &
 	    \colhead{S$_N$ intervals (\textit{E},\textit{dE})\tablenotemark{b}} &
 	    \colhead{S$_N$ intervals (\textit{S},\textit{dS})\tablenotemark{c}} &     
	    \colhead{S$_N$ intervals (spirals,irregulars)\tablenotemark{d}}   
	    }
        \startdata
        [-11,-12[   &    -	    	    &	  -   		& - 		\\\relax
        [-12,-13[   &     [0,5]	        &	  -    		& -      	\\\relax        
        [-13,-14[   &     [0,20]	    &	  [0,20]   	& -       	\\\relax
        [-14,-15[   &     [0,15]	    &	  [0,10]  	&  -     	 \\\relax
        [-15,-16[   &     [0,15]	    &	  [0,10]   	&  -      	\\\relax
        [-16,-17[   &     [0,10]	    &	  [0,5]   	&   [0,5]     \\\relax
        [-17,-18[   &     [0,10]	    &	  [0,5]    	&   [0,10]     \\\relax
        [-18,-19[   &     [0,5]	        &	  [0,5]    	&   [0,5]    	 \\\relax     
        [-19,-20[   &     [0.5]	        &	  [0,5]    	&   [0,5]     \\\relax     
        [-20,-21[   &     [0,10]	    &	  [0,5]    	&   [0,5]     \\\relax     
        [-21,-22[   &     [0,15]	    &	  [0,5]    	&   [0,15]     \\\relax     
        [-22,-23[   &     [0,15]	    &	  [0,5]    	&   [0,10]     \\\relax 
        [-23,-24[   &     [0,15]	    &	  -    		&   [0,5]     \\\relax 
        [-24,-25[   &     [0,10]	    &	  -    		&   -     	\\    
        \enddata
        \tablecomments{(a): Bins of absolute magnitude in the visible band of the host galaxy;
                       (b): Range of specific frequency for elliptical and dwarf elliptical galaxies; 
                       (c): Range of specific frequency for lenticular and dwarf lenticular galaxies;
                       (d): Range of specific frequency for spiral and irregular galaxies.
                       }       
\end{deluxetable} 

\section{Radial profiles of GC systems for the Secondary Galaxy Sample}
\label{sec:appendix2}

\begin{figure*}[ht]
     \centering
     \includegraphics[width=\linewidth]{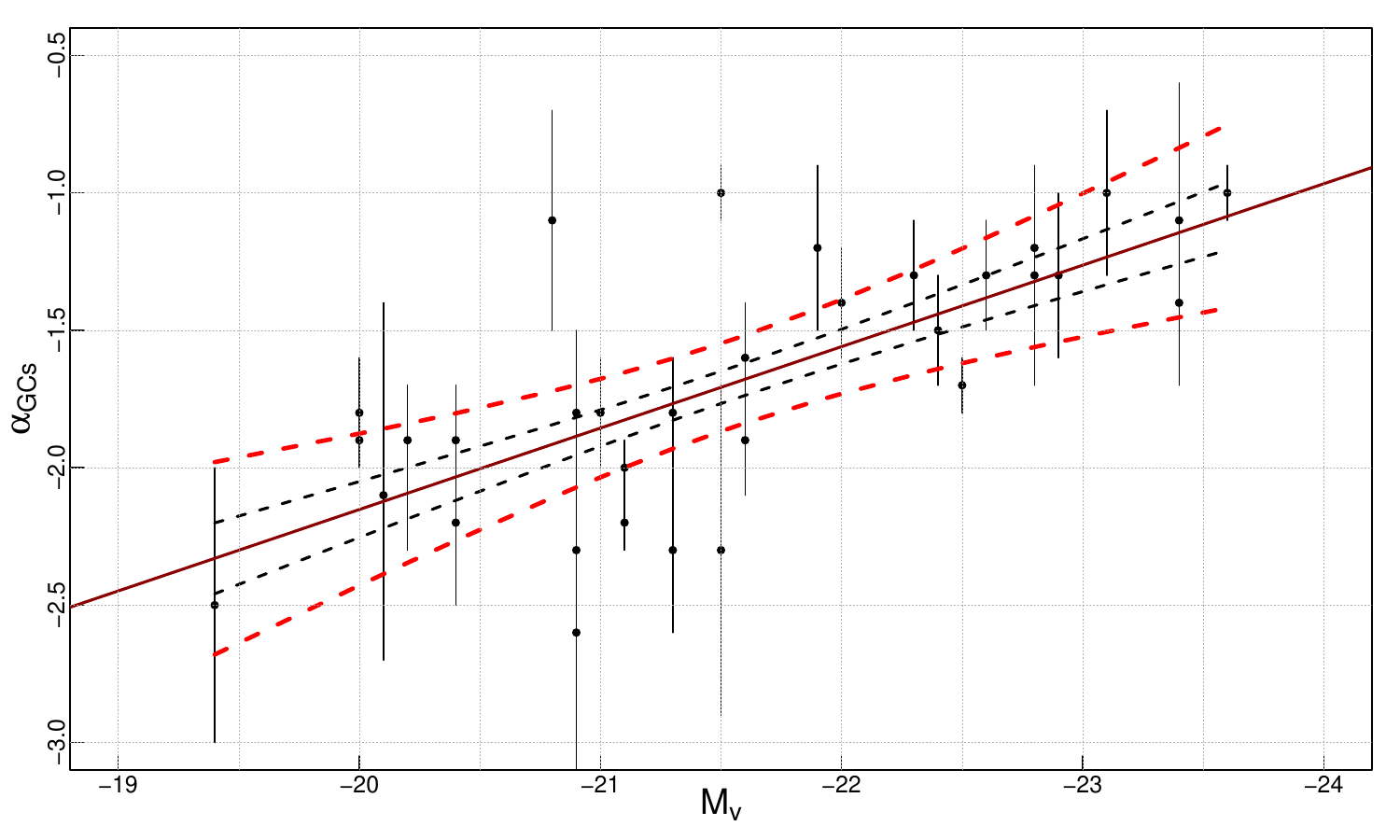}\\	
     \caption{Scatterplot of the host galaxies' absolute magnitudes  M$_{V}$ vs the observed slopes of the radial profiles of
     the GC systems modeled with a power-law from Figure~5.2 of~\citet{ashman2008}. The brown solid line represent the 
     best-fit linear profile ($\alpha_{\mathrm{GCs}}\!=\!-8.079-0.296 M_{V}$). The local uncertainty 
     affecting the linear relation, measured as the local regression curves
     for $\pm$1$\sigma$ (black dashed lines) and $\pm3\sigma$ (red dashed lines), is also shown.} 
     \label{fig:mvsalpha}
\end{figure*}

As described in Section~\ref{subsubsec:galcomponent}, the radial profiles of the simulated GCSs of SGS 
members are assumed to be described by power-law models whose slope depends on the visible absolute magnitude
M$_{V}$ of the host galaxy. For each host galaxy with given M$_{V}$ and each simulation the slope of the 
radial profile is determined by using the best-fit regression between the absolute magnitudes 
and the observed slopes of the radial profiles for a sample of galaxies from Figure~5.2 of~\cite{ashman2008}.
In order to take into account the large uncertainties on the observed values of the slopes of the radial profile 
and the scatter of the measured values, for each galaxy with brightness M$_{V}$ 
the slope used in a simulation is drawn from a Gaussian with mean equal to 
the value of the linear interpolation for M$_{V}$, and dispersion equal to the locally evaluated standard deviation 
of the best-fit linear relation. Figure~\ref{fig:mvsalpha} shows the original data from~\citet{ashman2008}, the 
best-fit linear relation $\alpha_{\mathrm{GCs}}\!=\!-8.079-0.296 M_{V}$ and the 
local regression curves associated with $\pm1\sigma$ and $\pm3\sigma$ from the 
predicted value from the linear relation. 

\section{Effect of different $K$ values and cell sizes on residual maps}
\label{sec:appendix4}

\begin{figure*}[ht]
    \centering
    \includegraphics[width=0.48\linewidth]{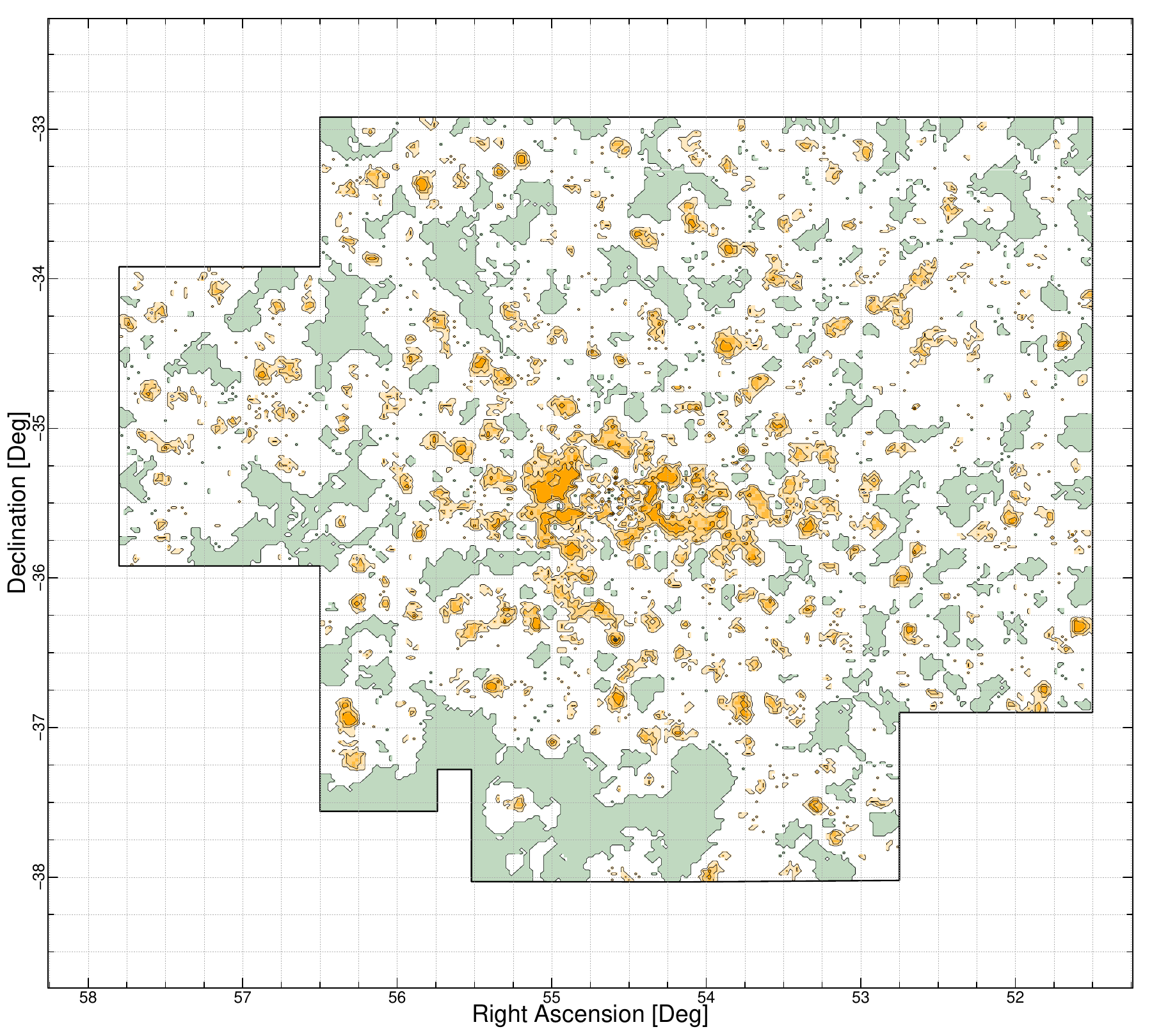}
    \includegraphics[width=0.48\linewidth]{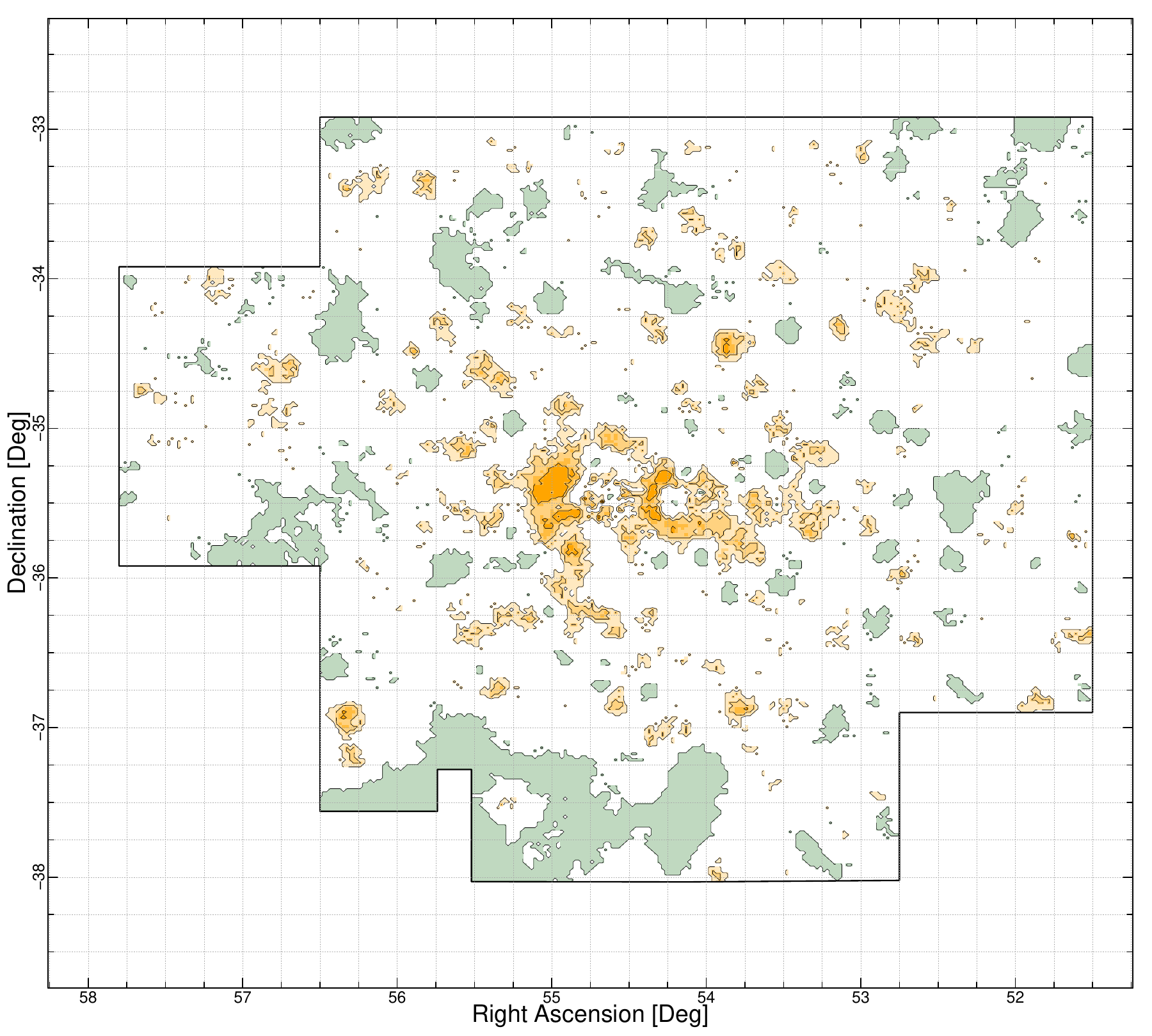}\\  
    \includegraphics[width=0.48\linewidth]{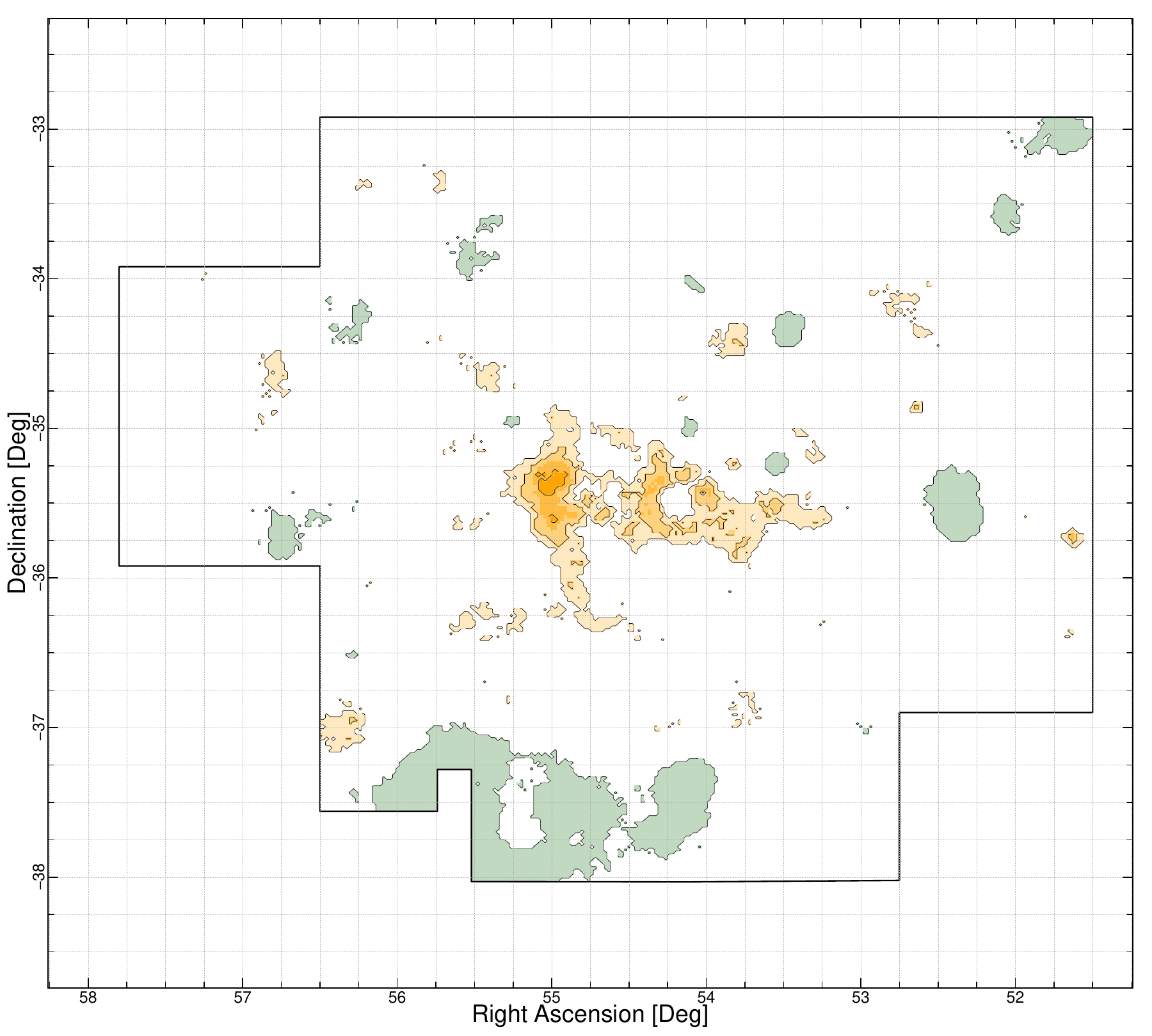}
    \includegraphics[width=0.48\linewidth]{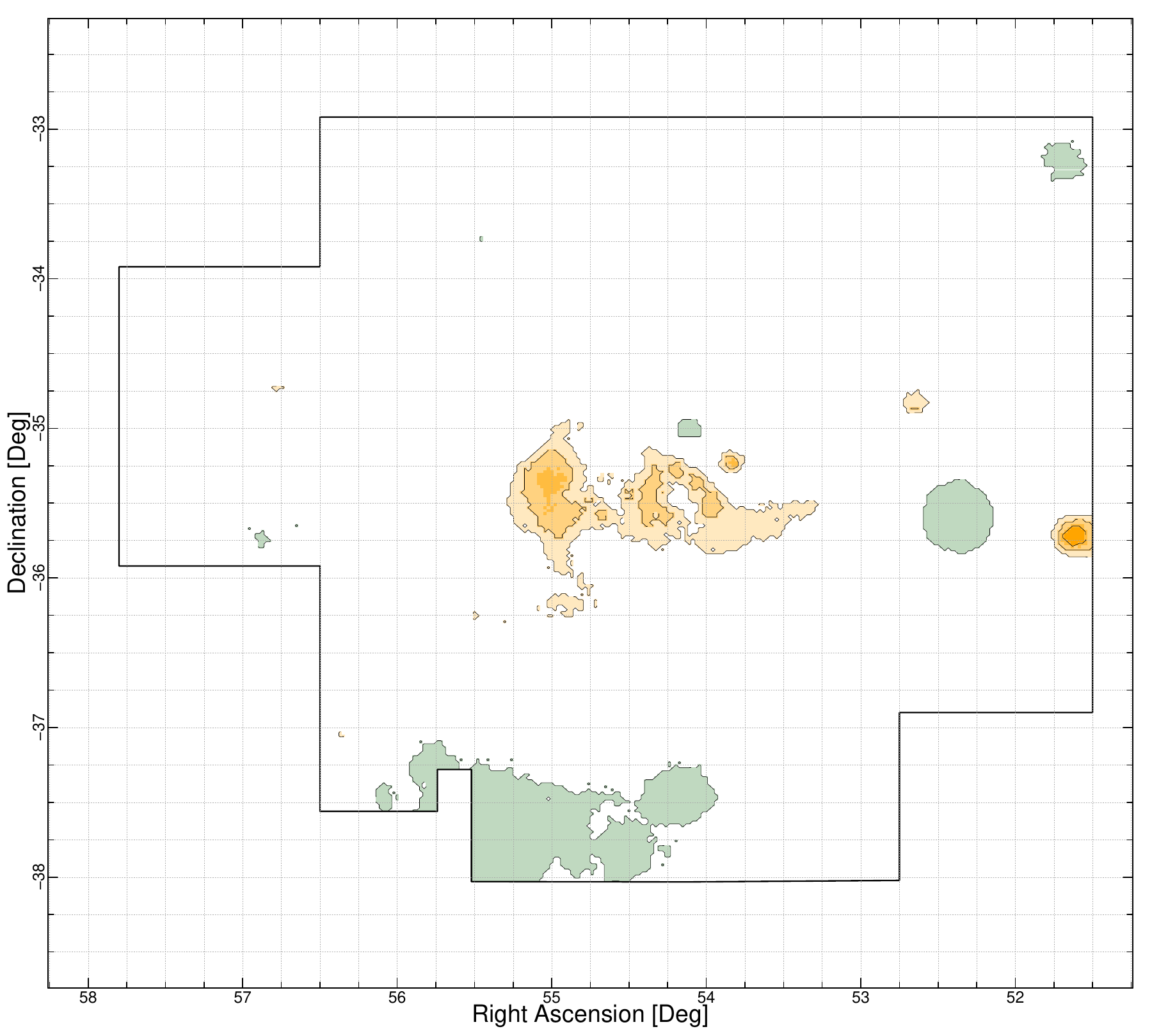}\\ 
    \includegraphics[width=0.48\linewidth]{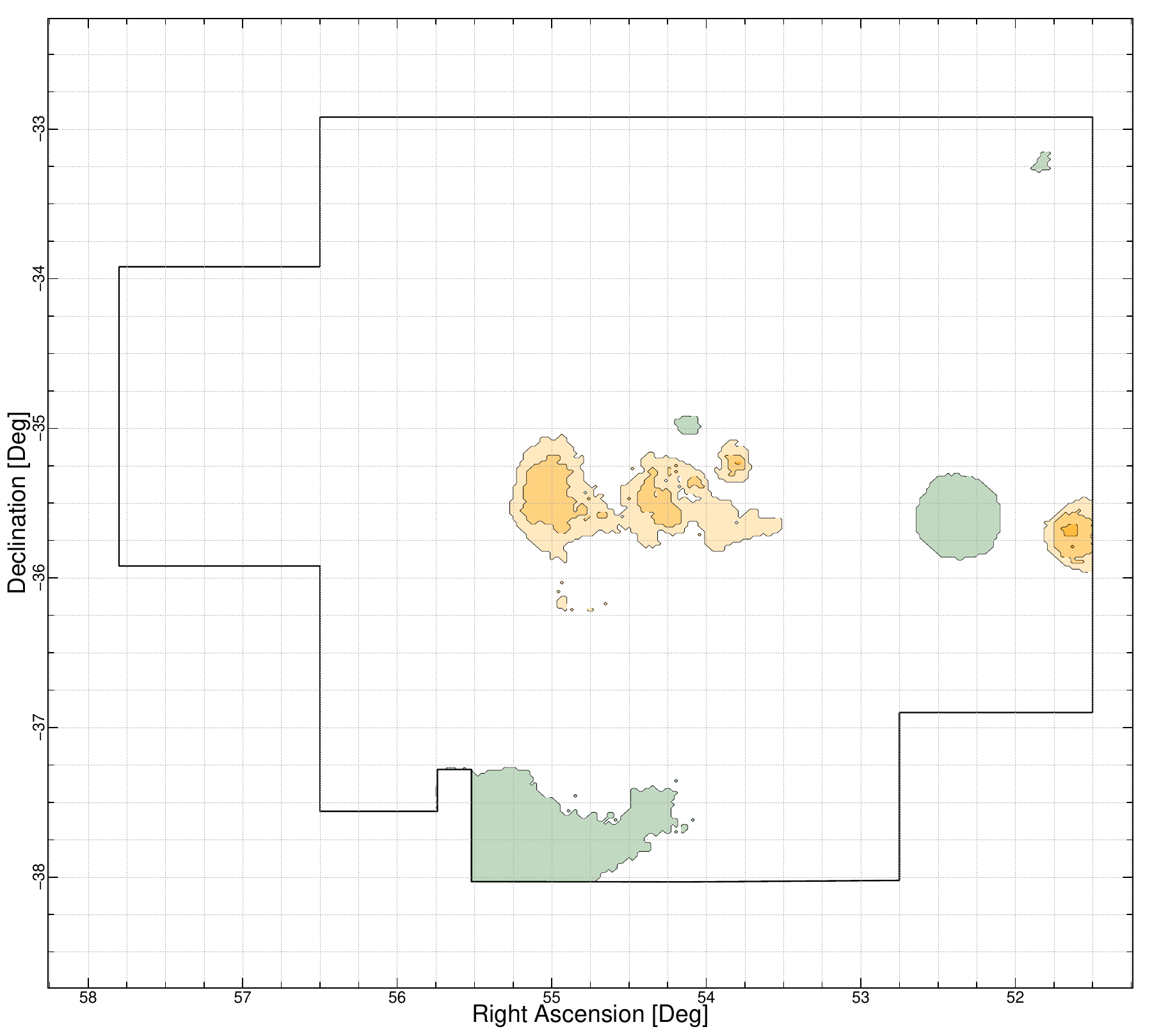}
    \includegraphics[width=0.48\linewidth]{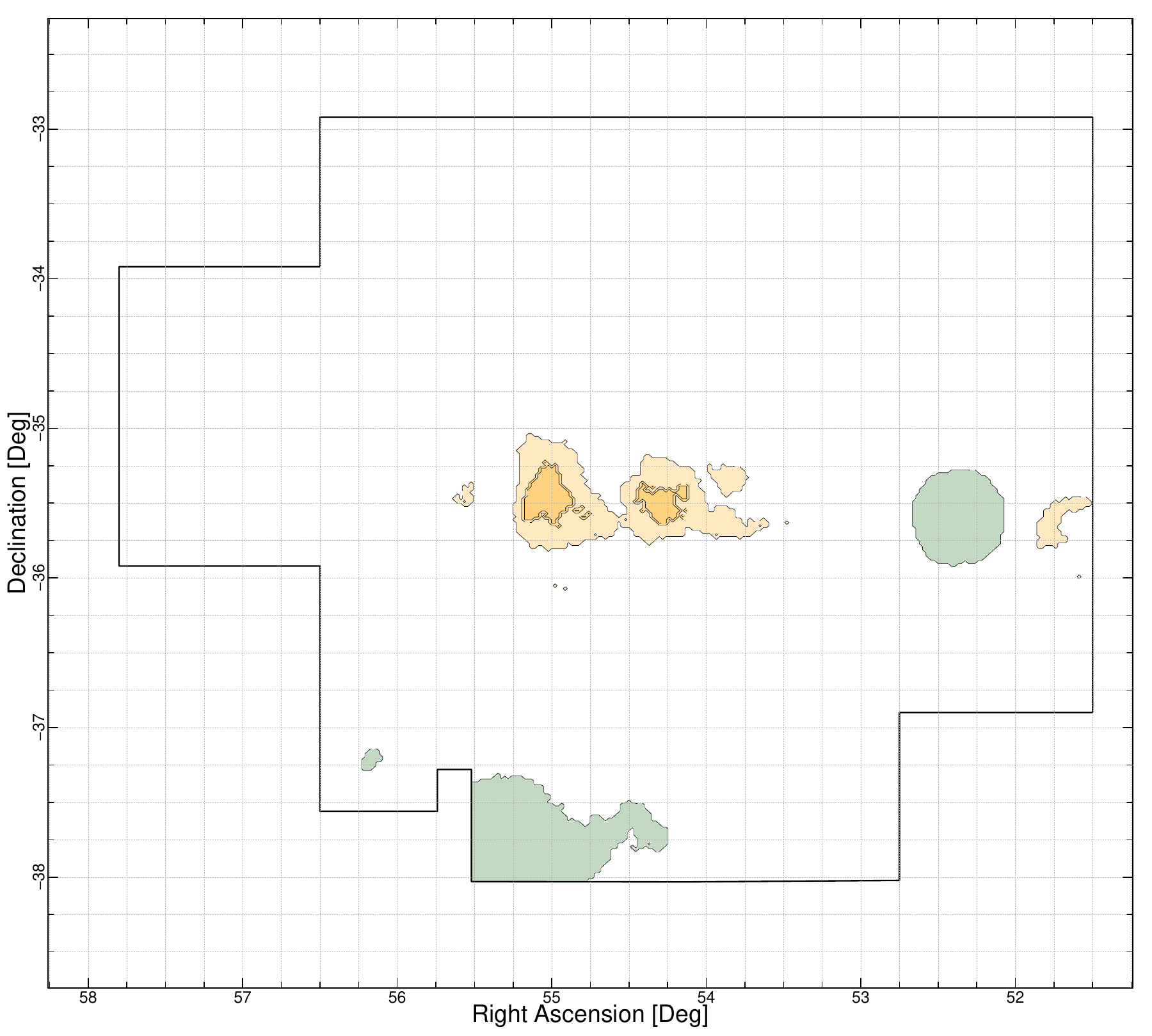}\\     
    \caption{Residual maps as in Figure~\ref{fig:residual_k10} from KNN density maps obtained 
    (from upper left to lower right) with $K\!=\!(5,10,25,50,75,100)$.}
    \label{fig:residuals_k_values}
\end{figure*}

The method for the determination of the residual map of the spatial distribution of GC candidates is
described in Section~\ref{subsec:residuals}. Figure~\ref{fig:residuals_k_values} shows the residual maps
obtained for values of $K$ used in this paper, i.e. $K\!=\!(5,10,25,50,75,100)$, with all other
parameters fixed to the\ standard values of the main experiment described above 
(10,000 simulations, cell size $\approx\!1.25\arcmin$, density of the homogeneous component 
estimated outside of the cluster virial radius, modified Hubble model for the 
radial profile and host galaxy ellipticity for the azimuthal distribution of MGS GCs component). 
Details on the approach employed to choose the $K$ value used to produce the residual 
map whose spatial features have been discussed in this paper, can be found in 
Section~\ref{sec:results}. 

Large $K$ values highlight large spatial structures corresponding to areas of either positive or 
negative residuals, as smaller features are smoothed out. As shown in Figure~\ref{fig:residuals_k_values}, 
the main over-density located in the central region of the Fornax cluster is clearly visible across 
residual maps for all $K$ values, although its morphology gets less complex with increasing $K$. 
In general, the area of the persistent spatial structures increases with increasing $K$ while their pixel-based 
statistical significance decreases: both effects can be explained by noticing that, depending on level of 
clustering of candidate GCs in the field-of-view and the anisotropy of their distribution, the average 
distance of the $K$-th neighboring GCs candidate at some point starts exceeding the typical spatial scale
of the over-density structures. When this happens, the presence of small, high-density regions occupied 
by several GC candidates increases the density also in the surrounding regions, while at the same 
time the KNN density within the area occupied by the GCs excess is diluted by the increased
distance of the $K$-th neighboring source.

\begin{figure*}[ht]
    \centering
    \includegraphics[width=0.195\linewidth]{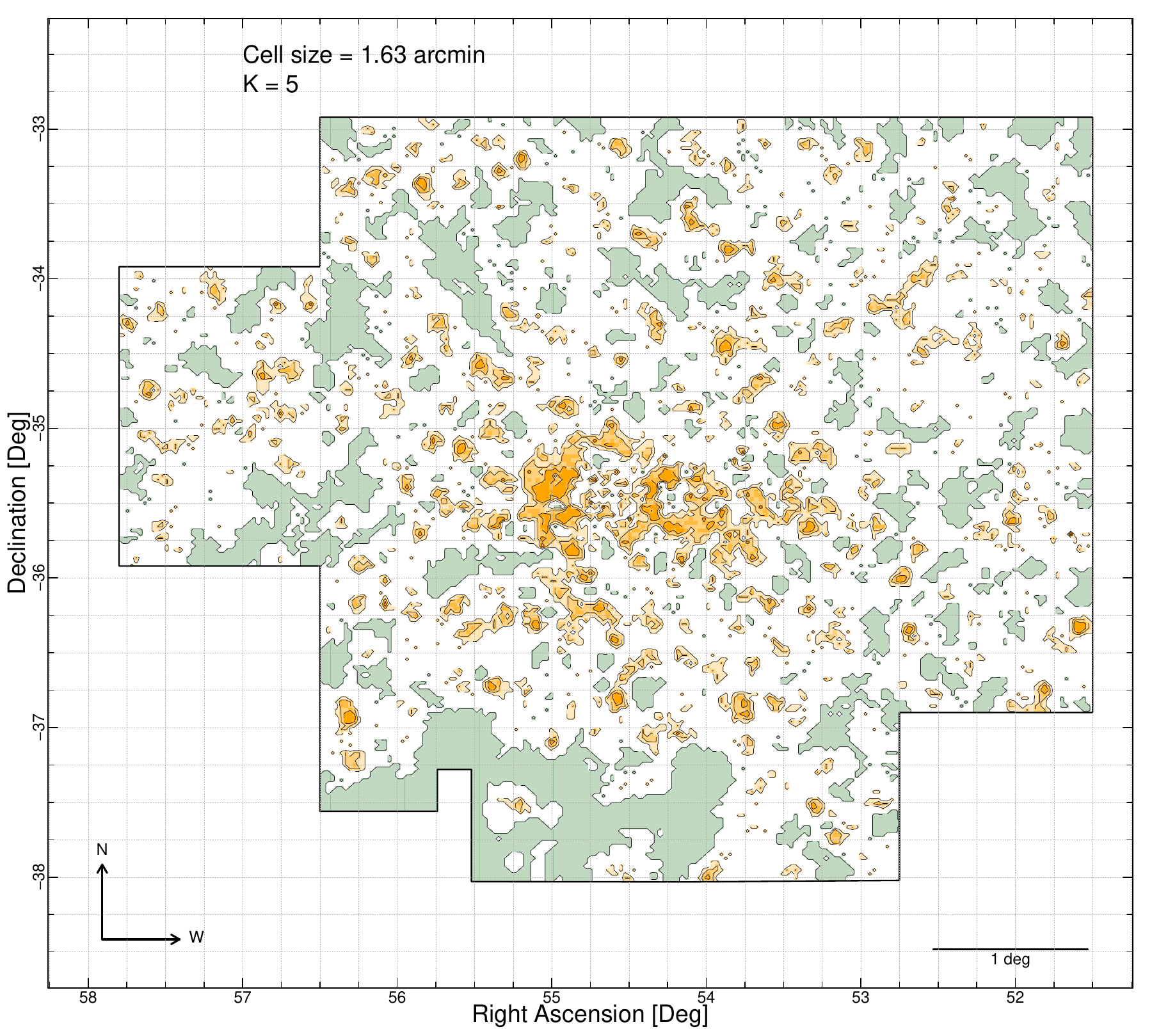}
    \includegraphics[width=0.195\linewidth]{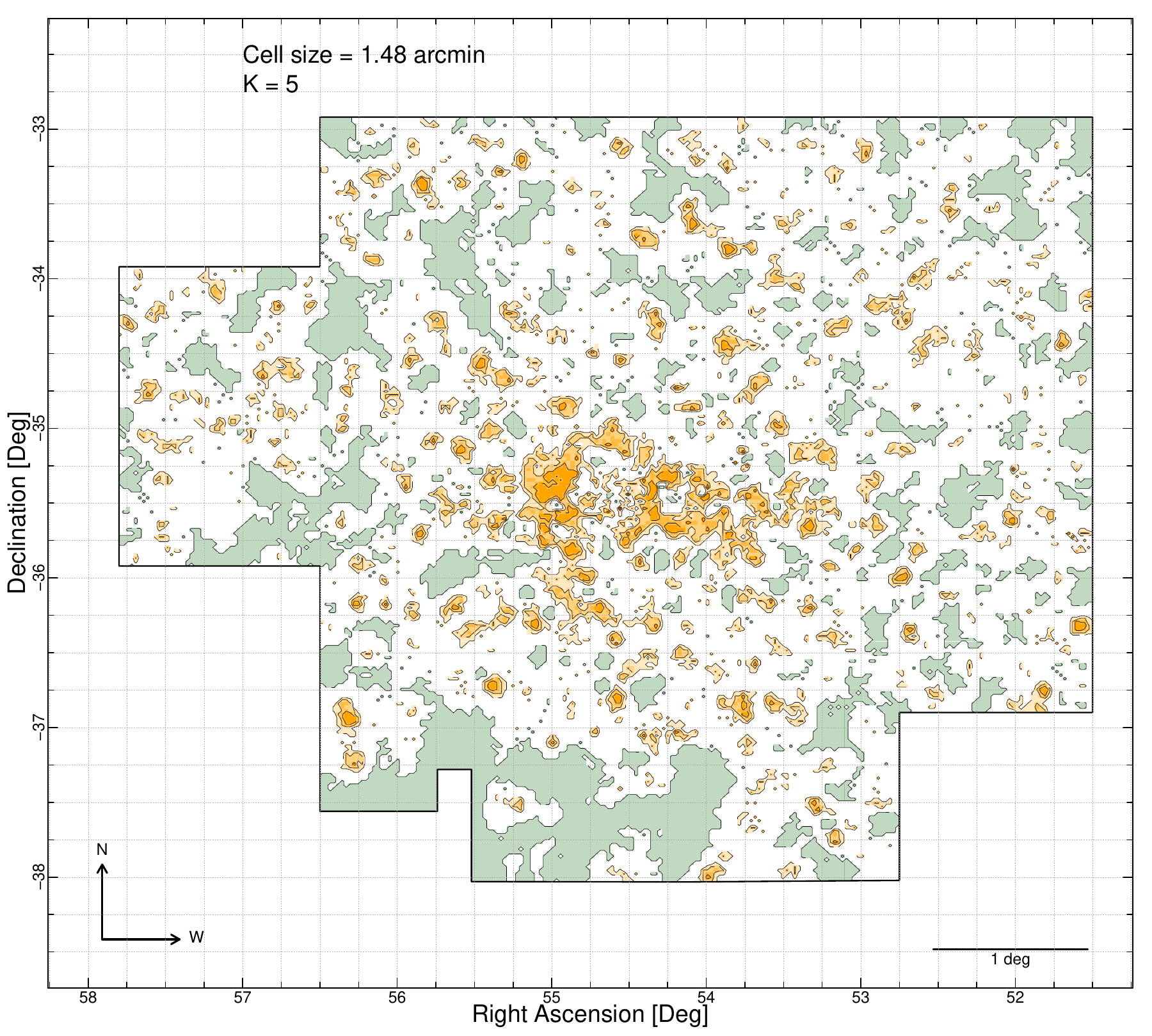}
    \includegraphics[width=0.195\linewidth]{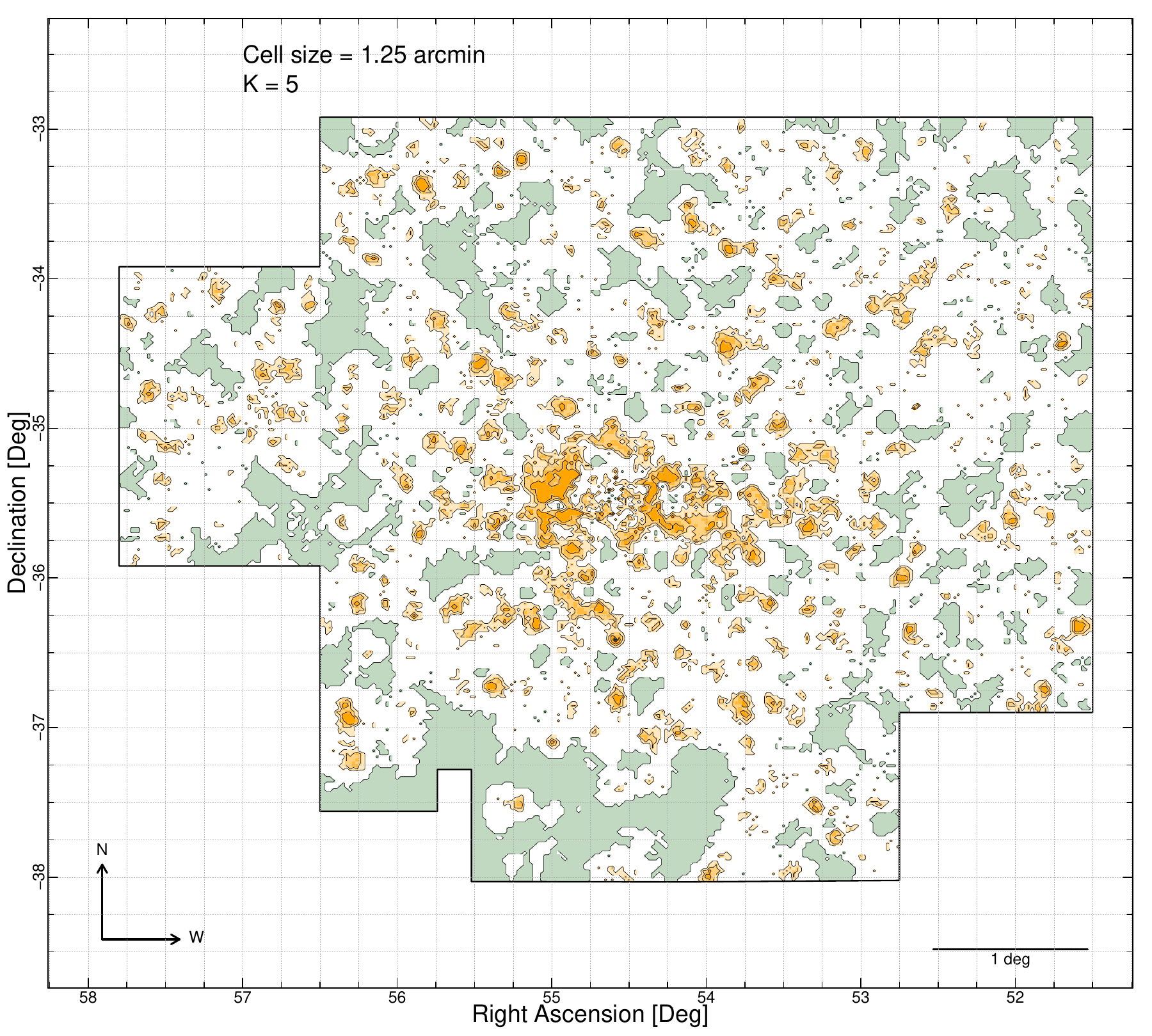}
    \includegraphics[width=0.195\linewidth]{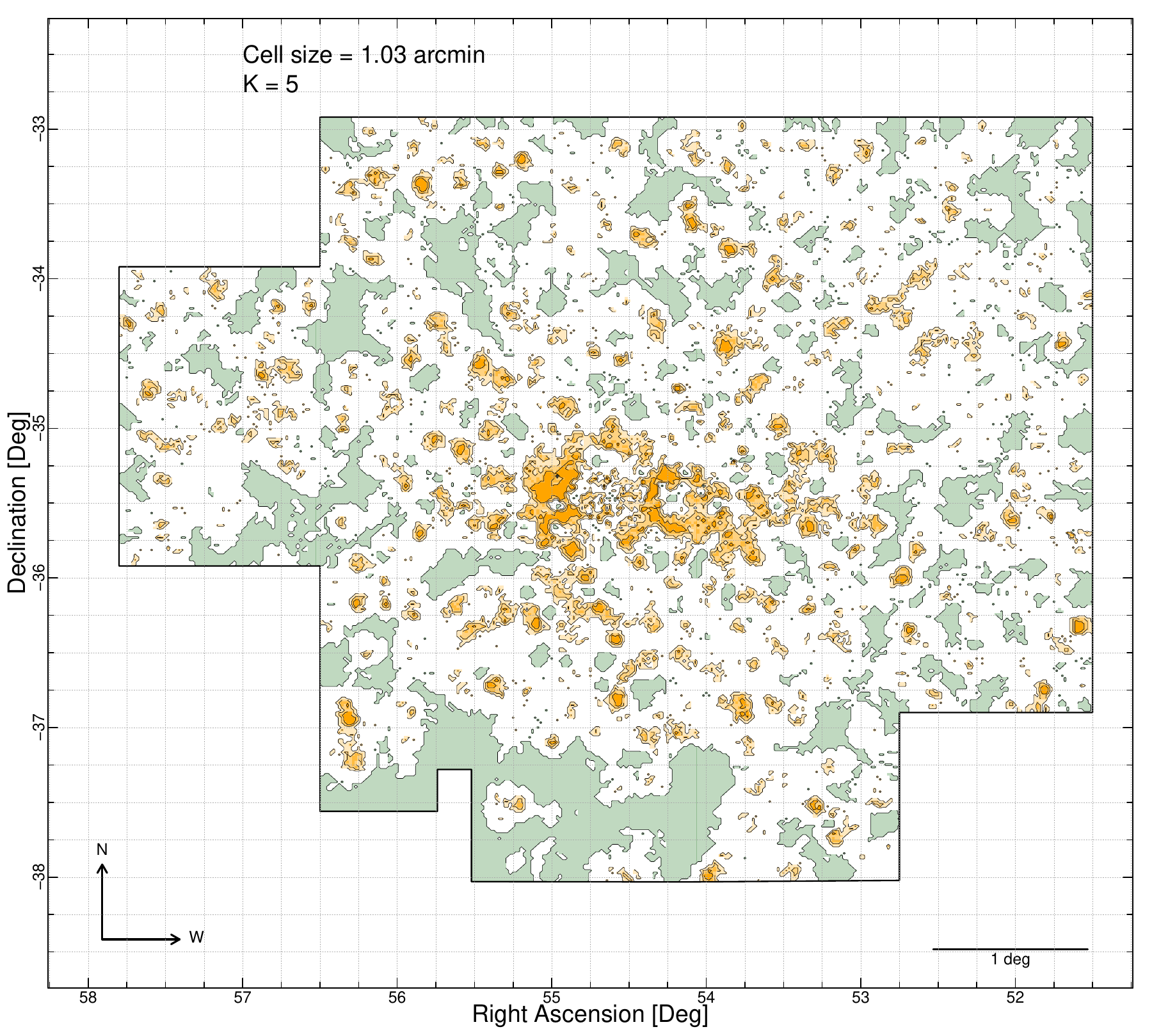}    
    \includegraphics[width=0.195\linewidth]{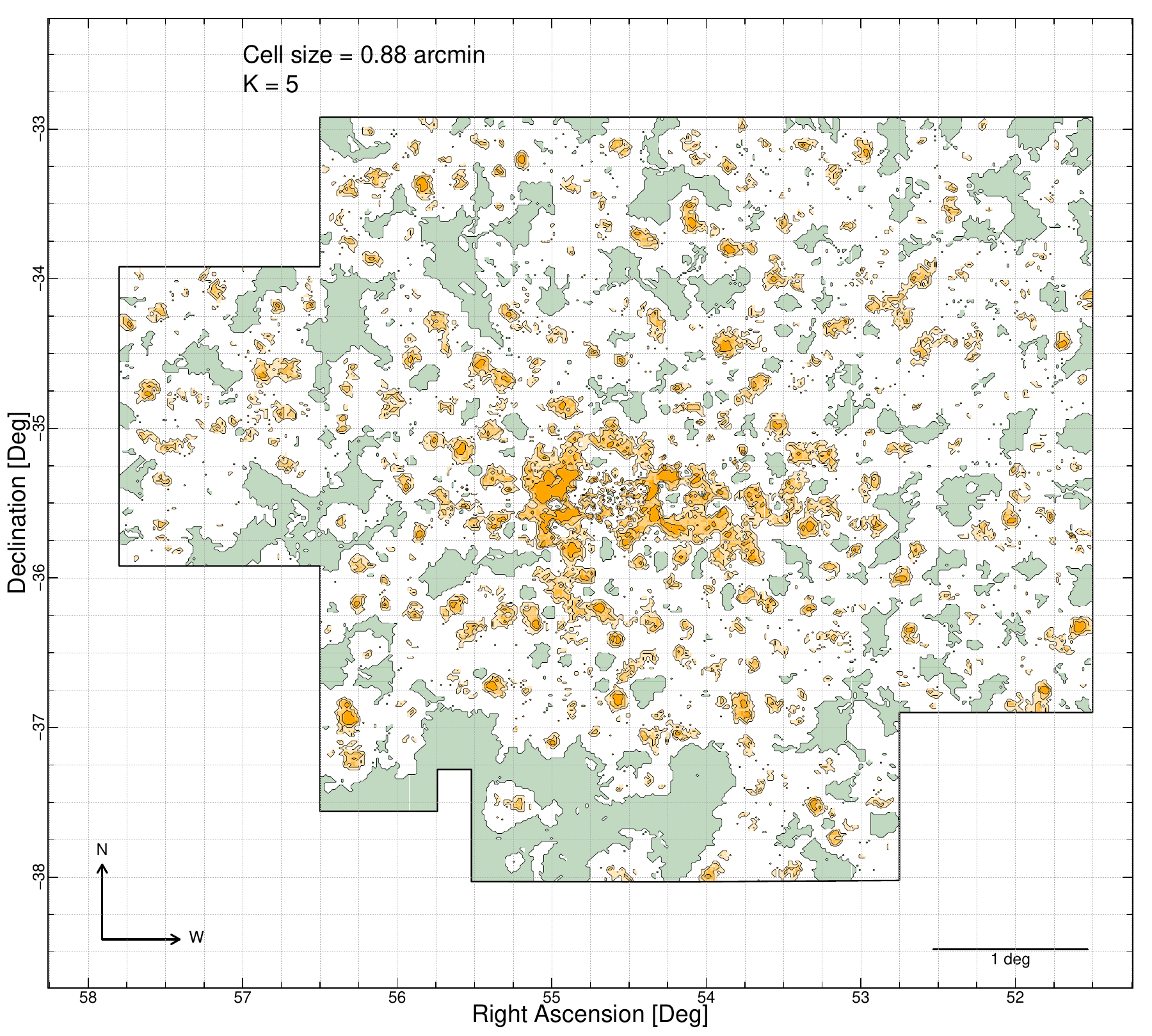}\\  
    \includegraphics[width=0.195\linewidth]{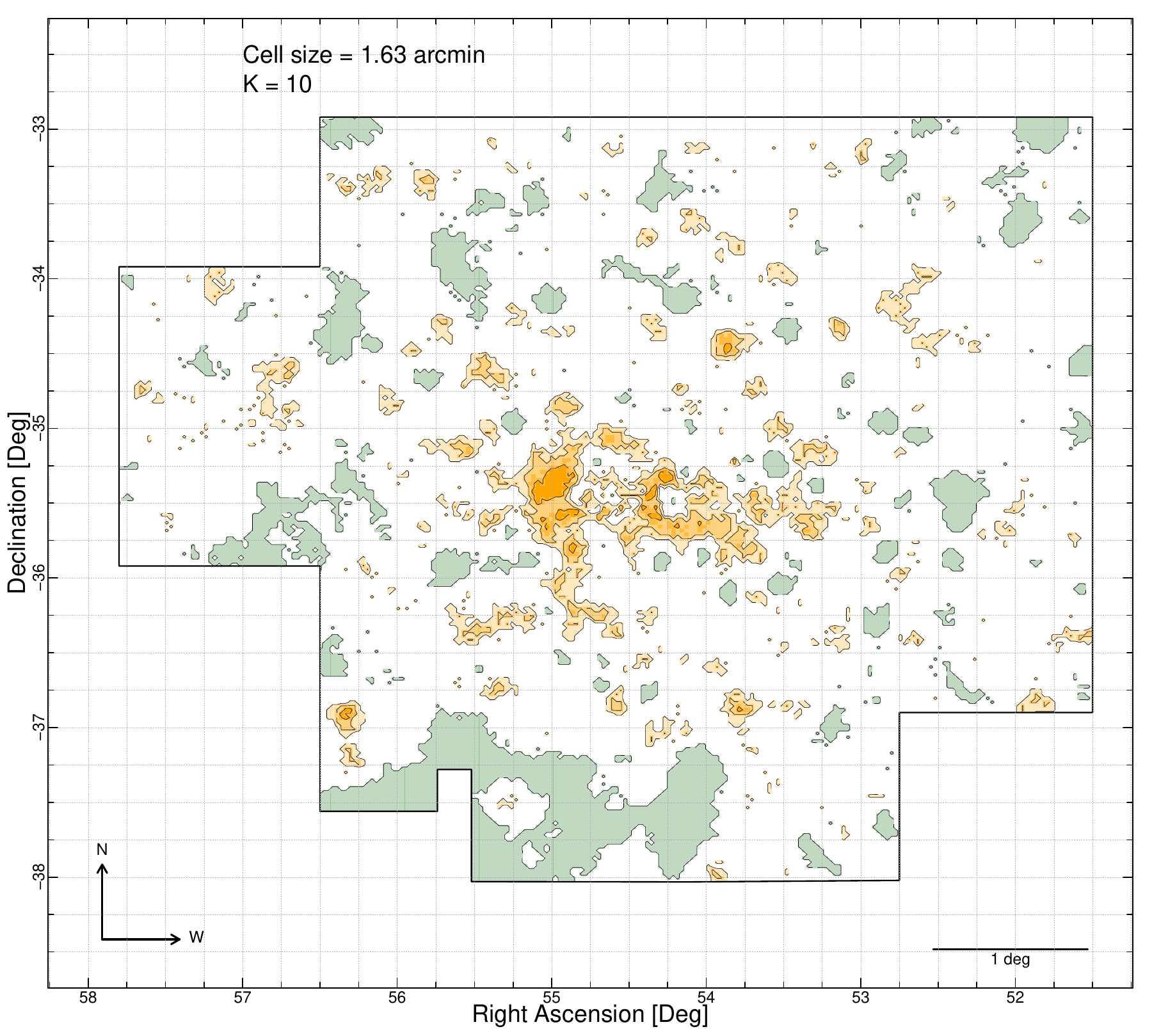}
    \includegraphics[width=0.195\linewidth]{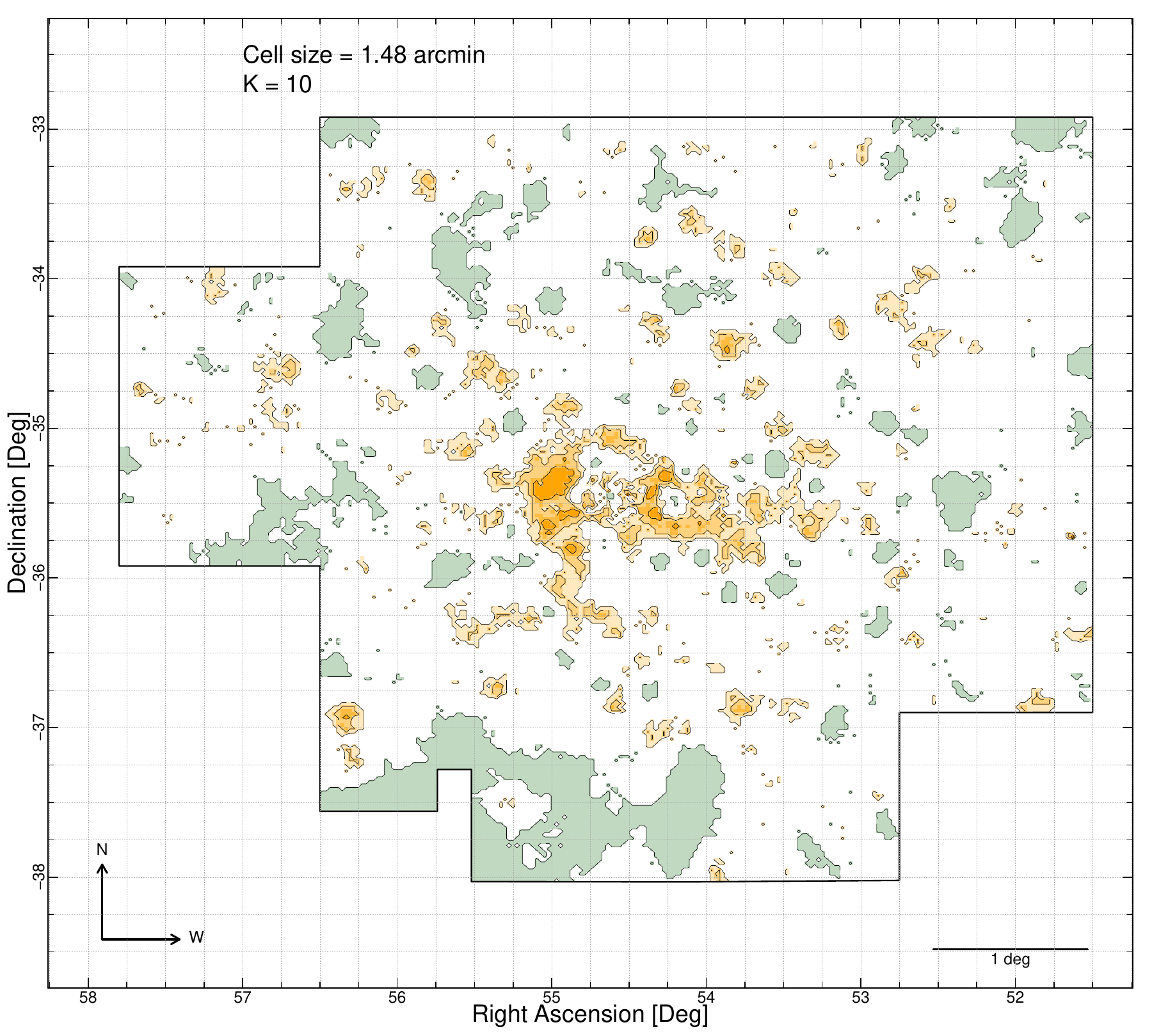}
    \includegraphics[width=0.195\linewidth]{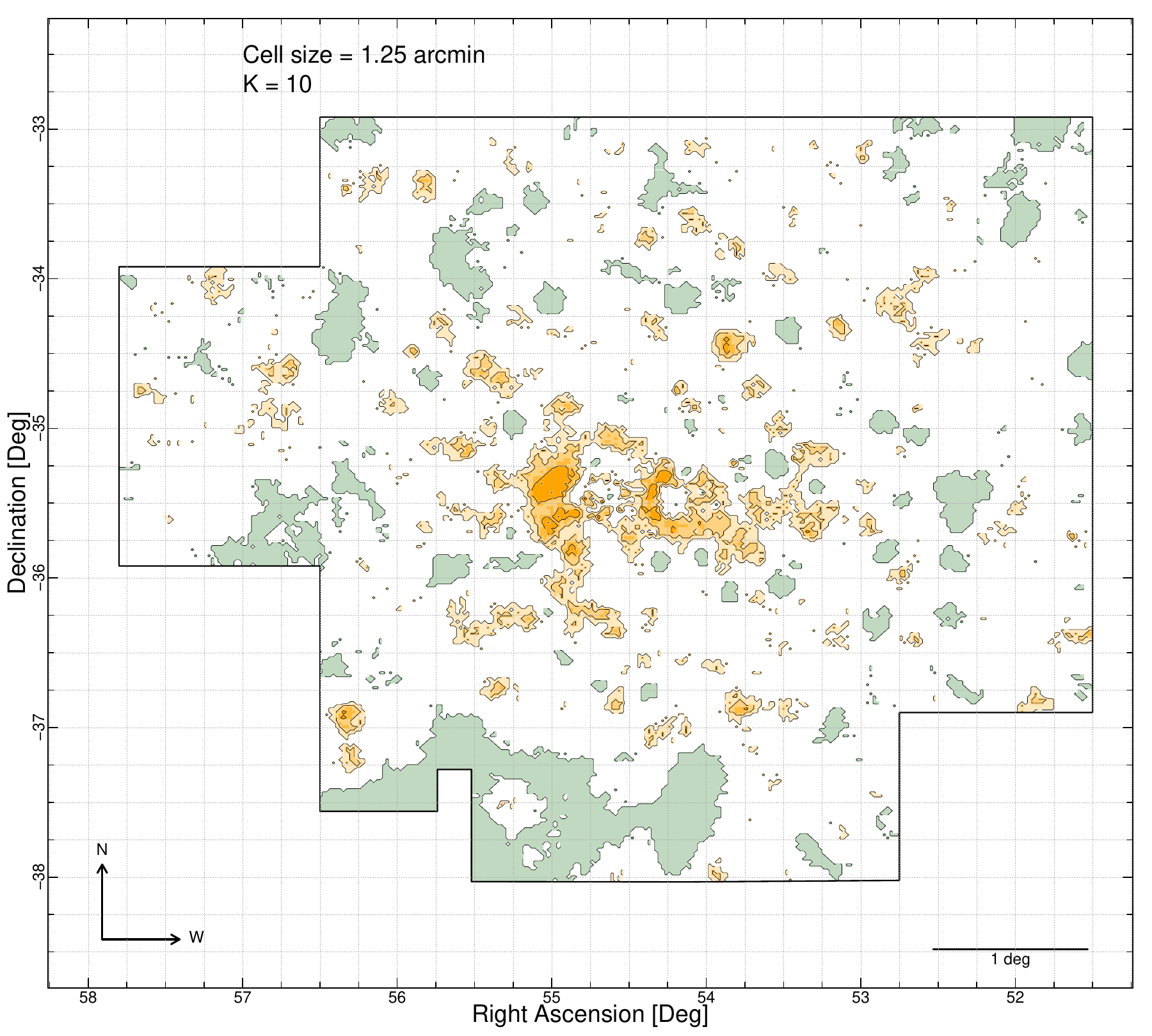}
    \includegraphics[width=0.195\linewidth]{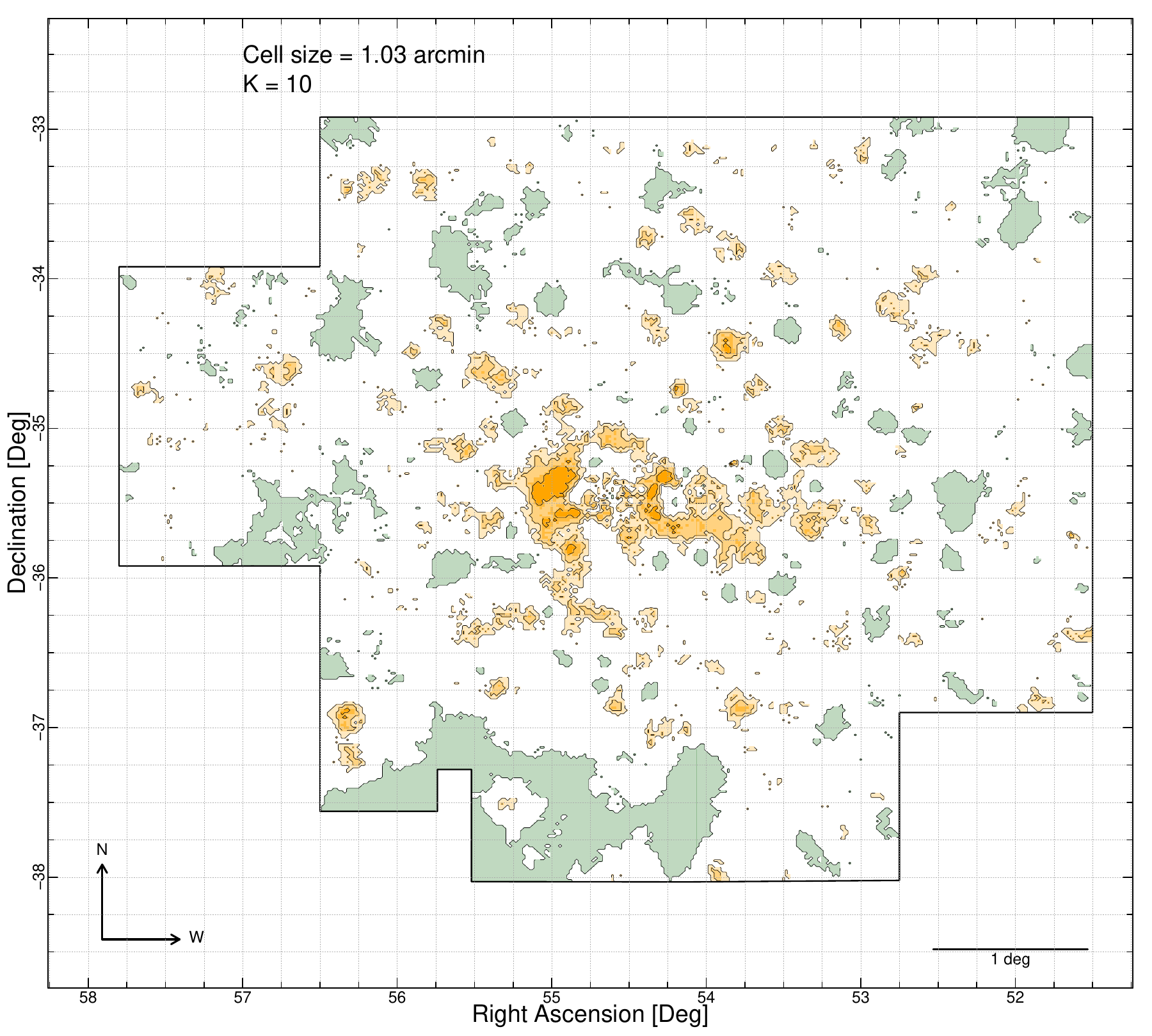}    
    \includegraphics[width=0.195\linewidth]{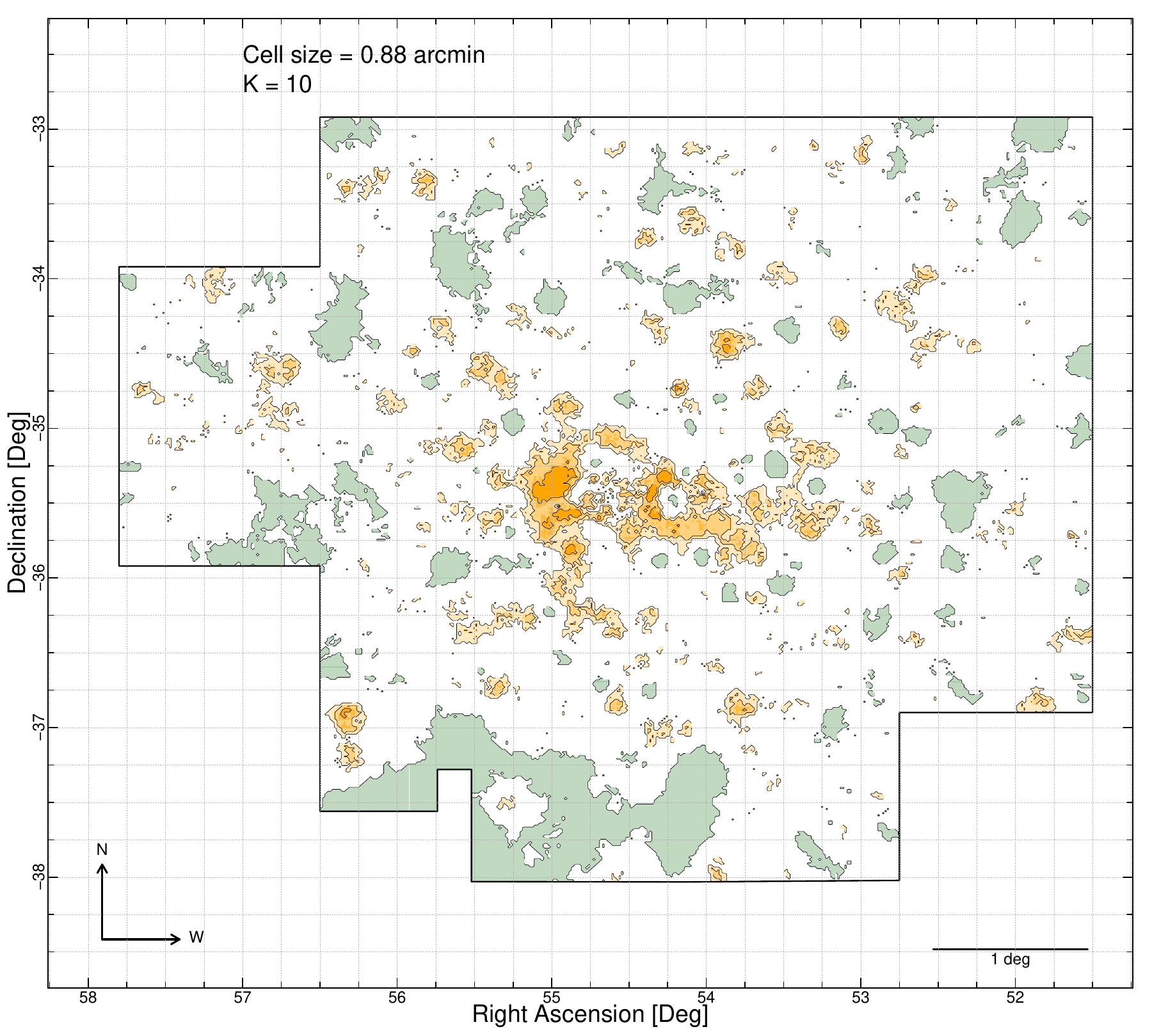}\\  
    \includegraphics[width=0.195\linewidth]{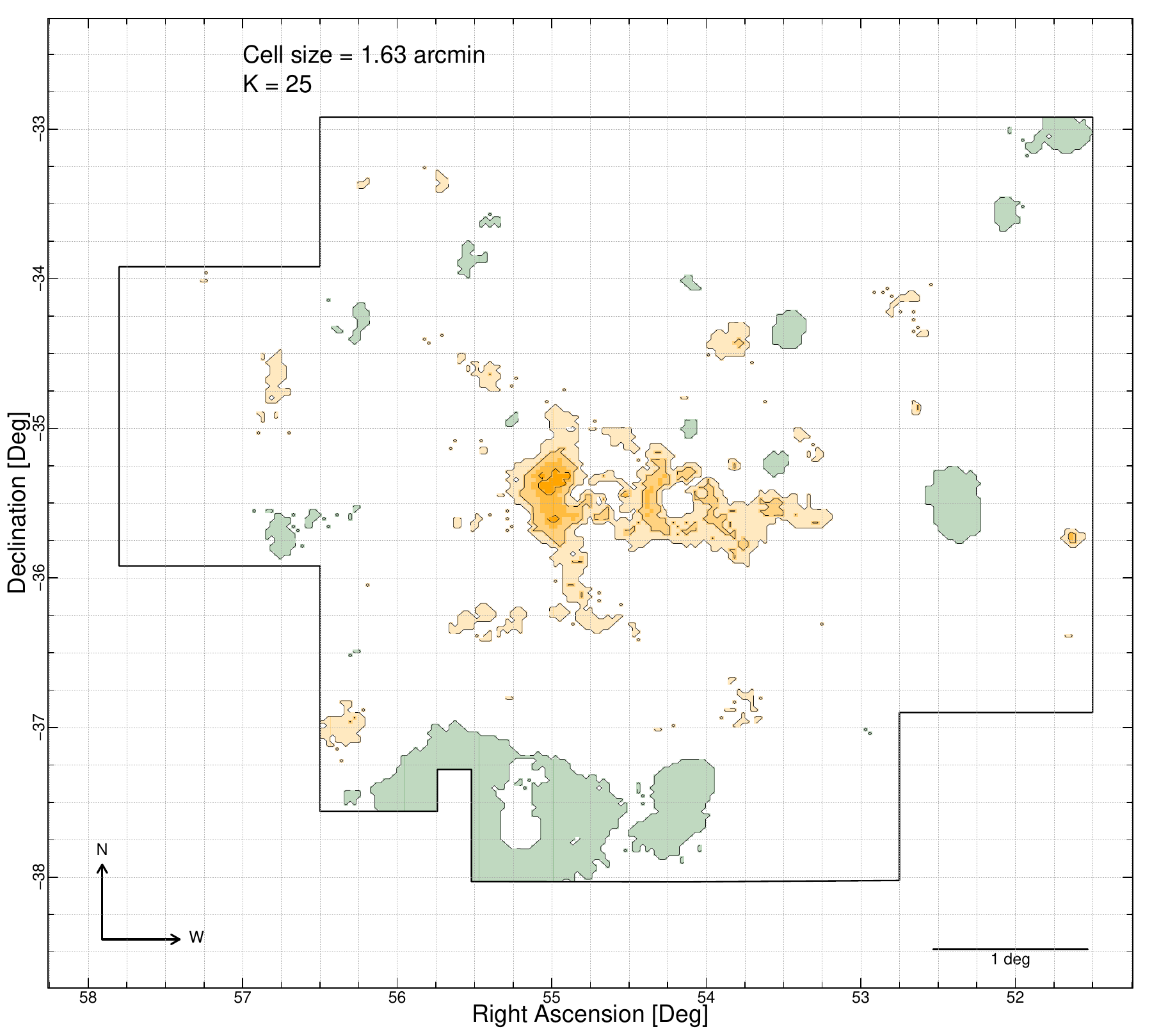}
    \includegraphics[width=0.195\linewidth]{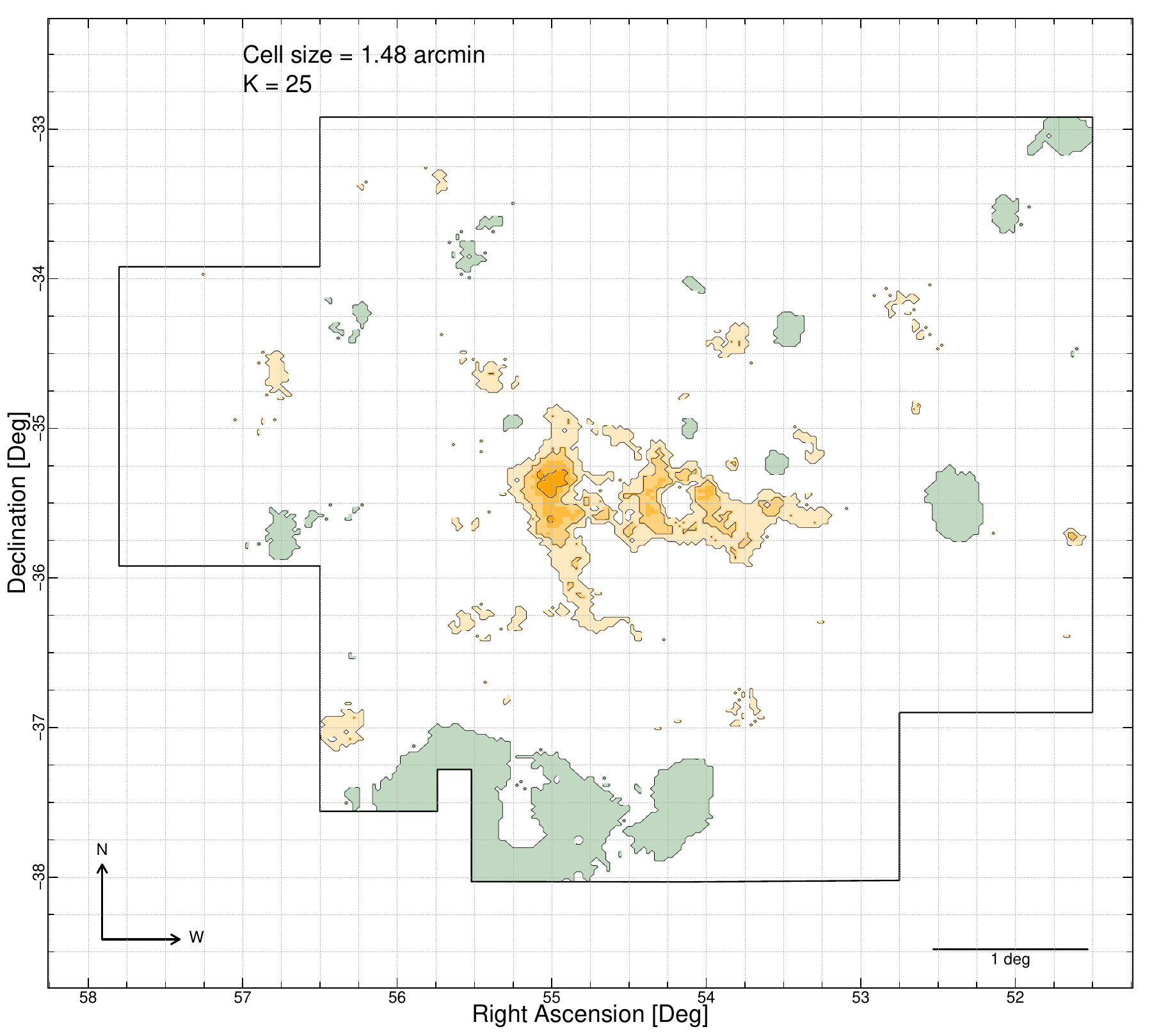}
    \includegraphics[width=0.195\linewidth]{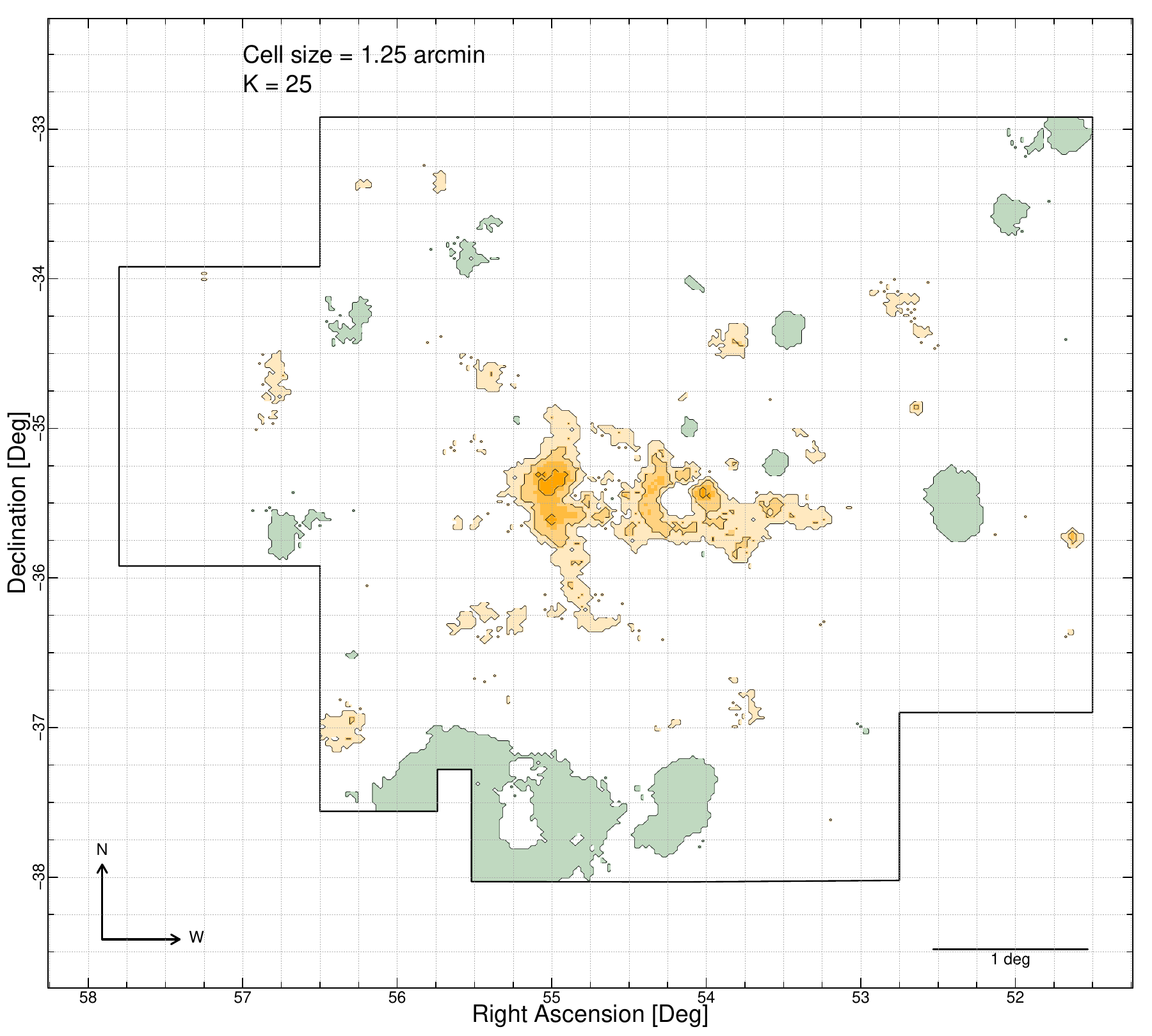}
    \includegraphics[width=0.195\linewidth]{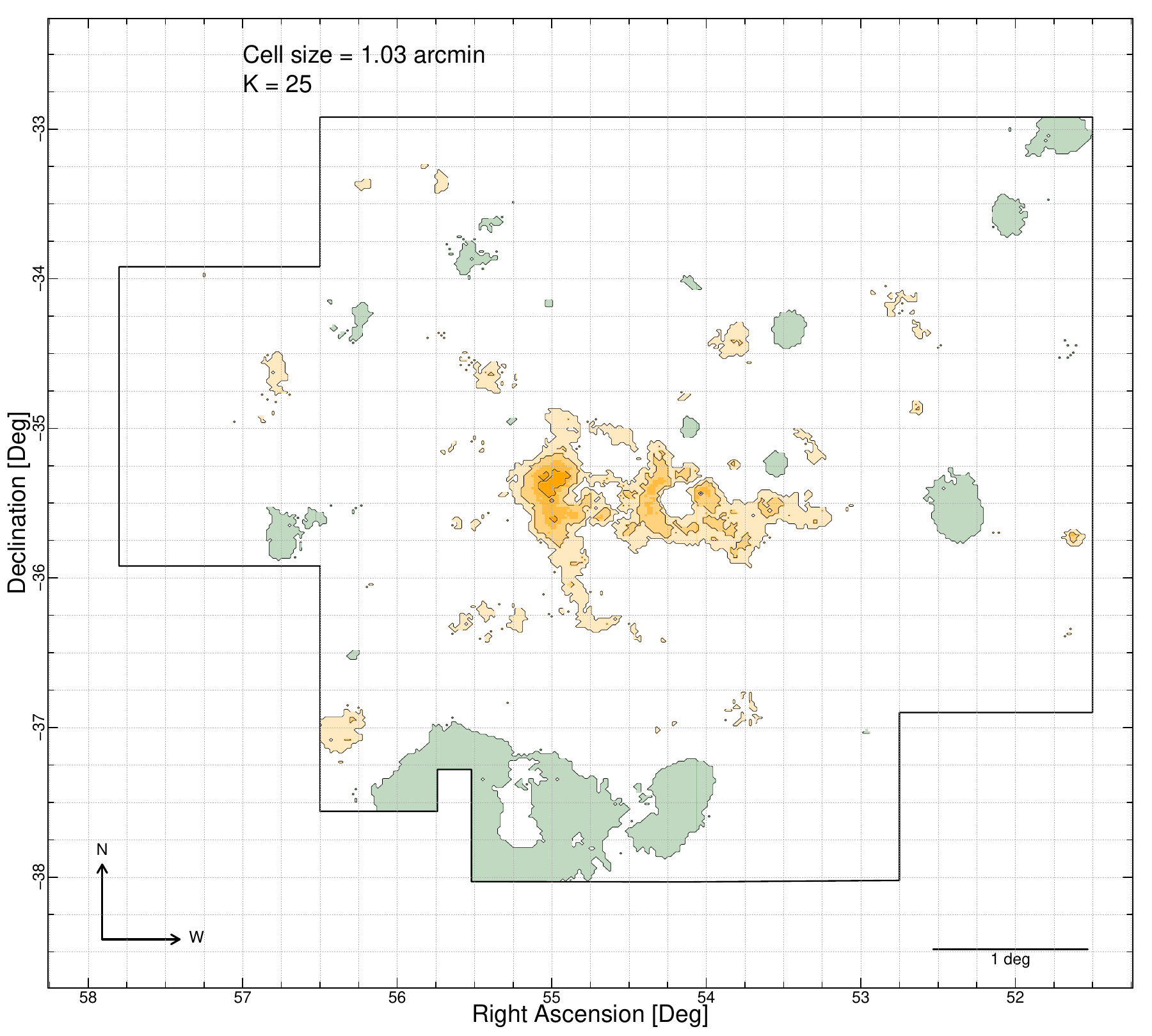}    
    \includegraphics[width=0.195\linewidth]{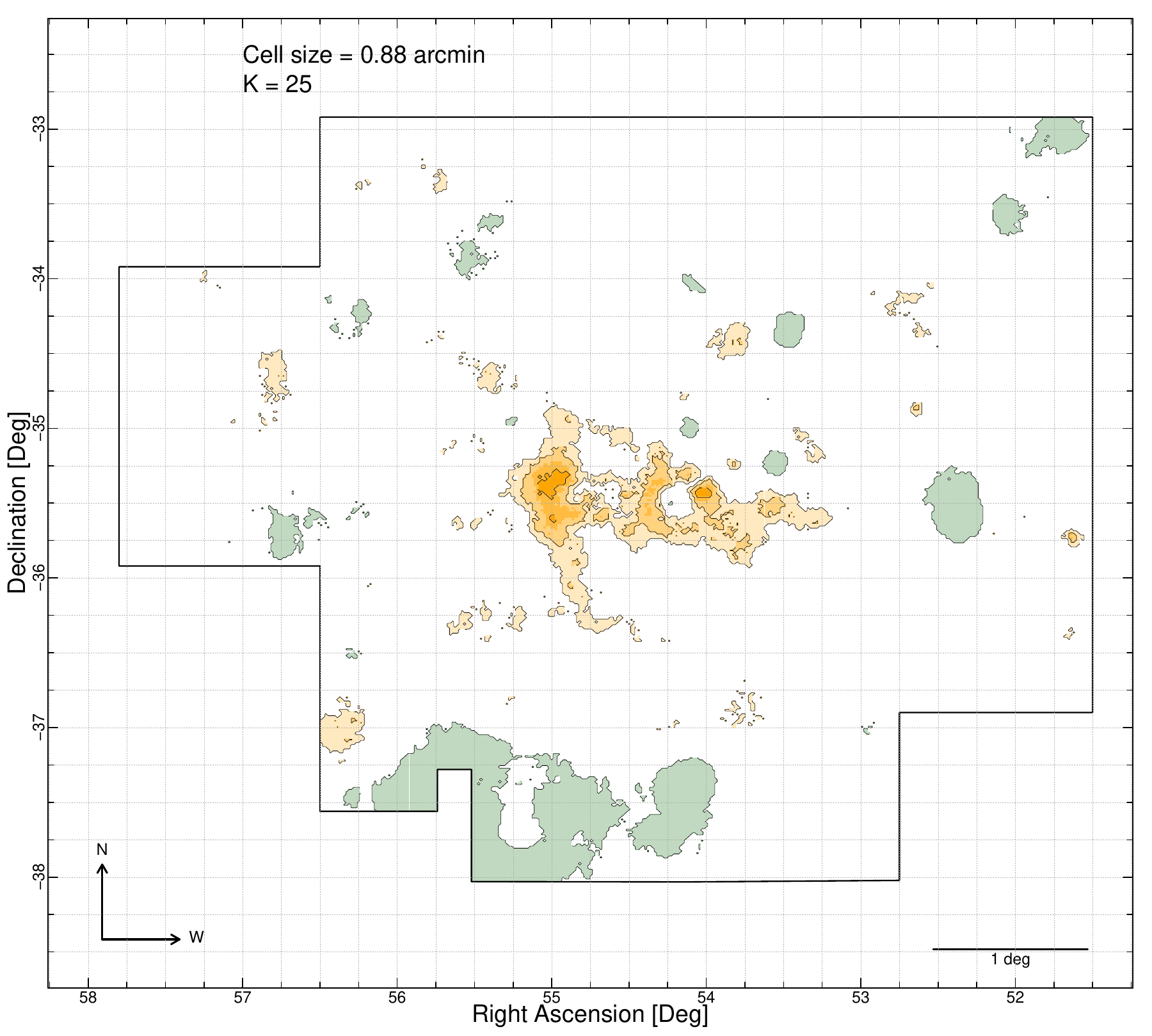}\\  
    \includegraphics[width=0.195\linewidth]{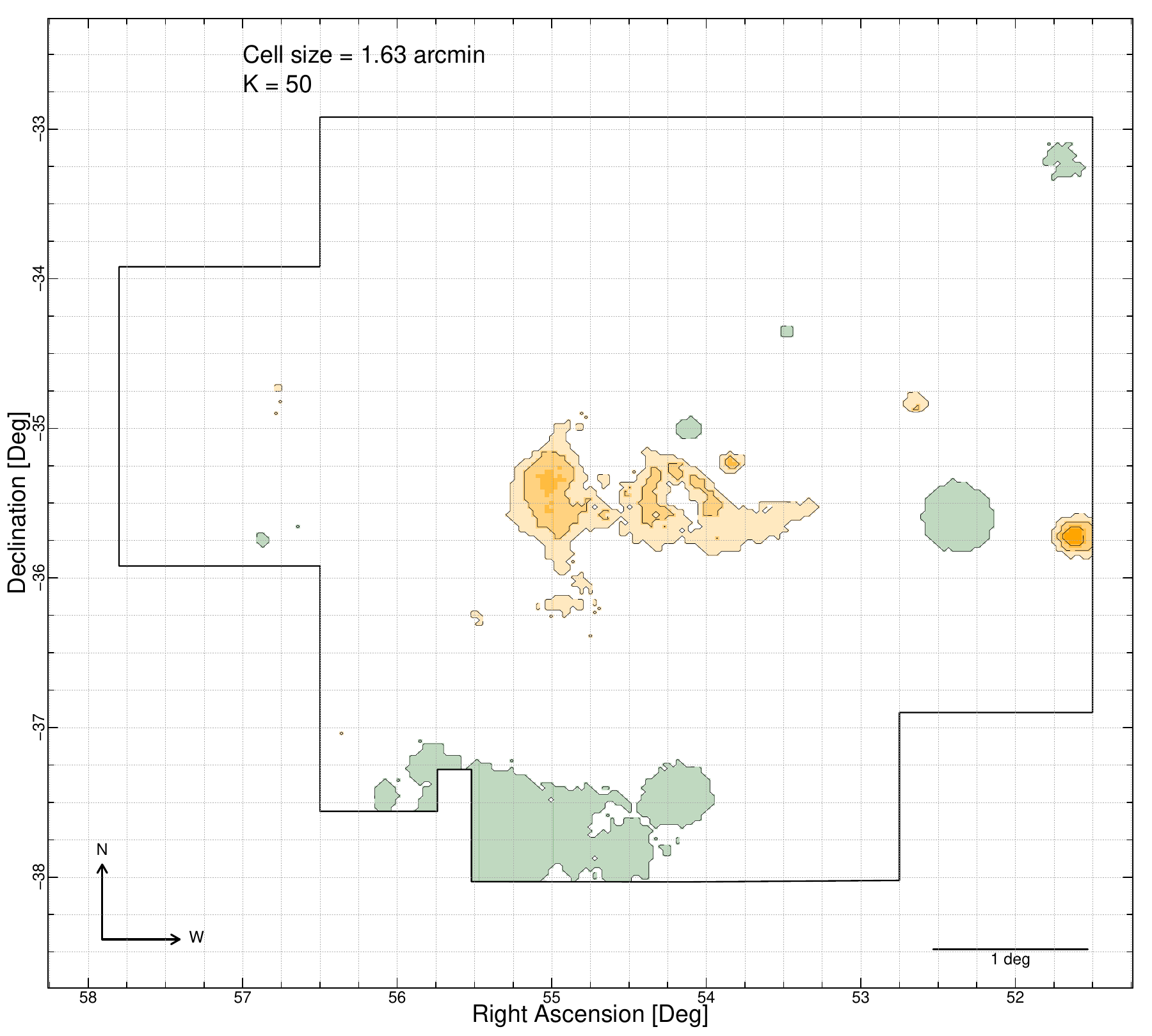}
    \includegraphics[width=0.195\linewidth]{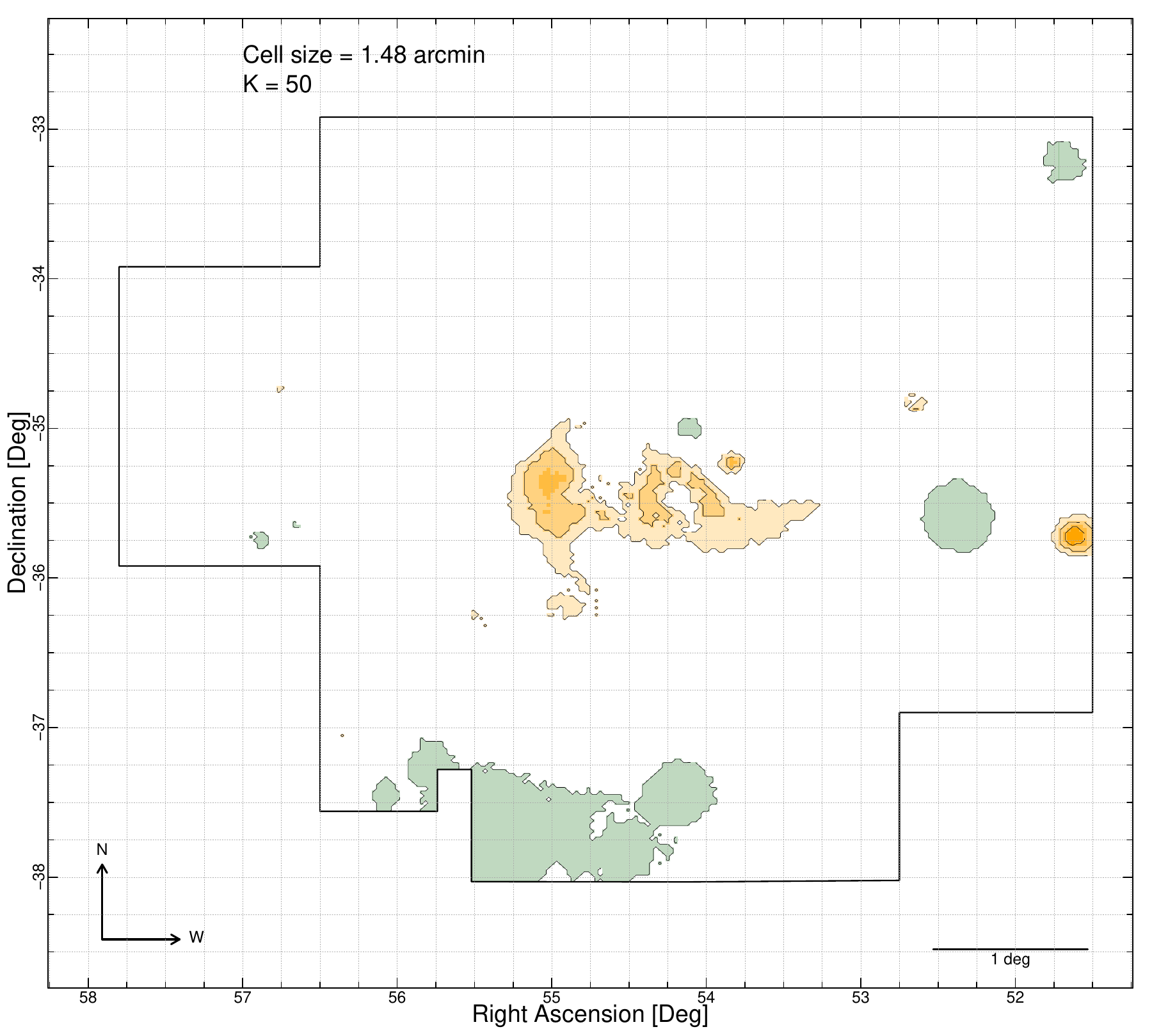}
    \includegraphics[width=0.195\linewidth]{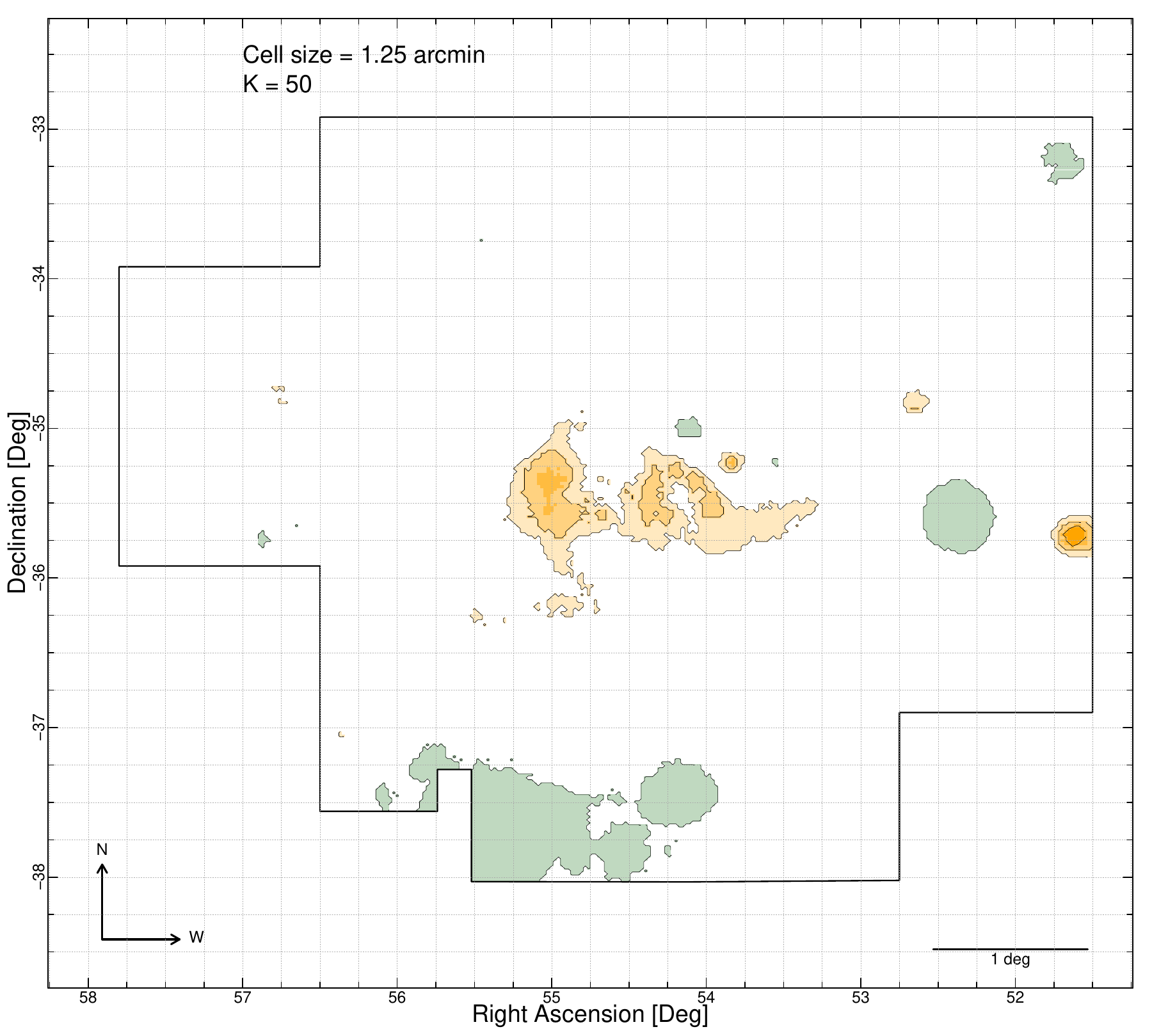}
    \includegraphics[width=0.195\linewidth]{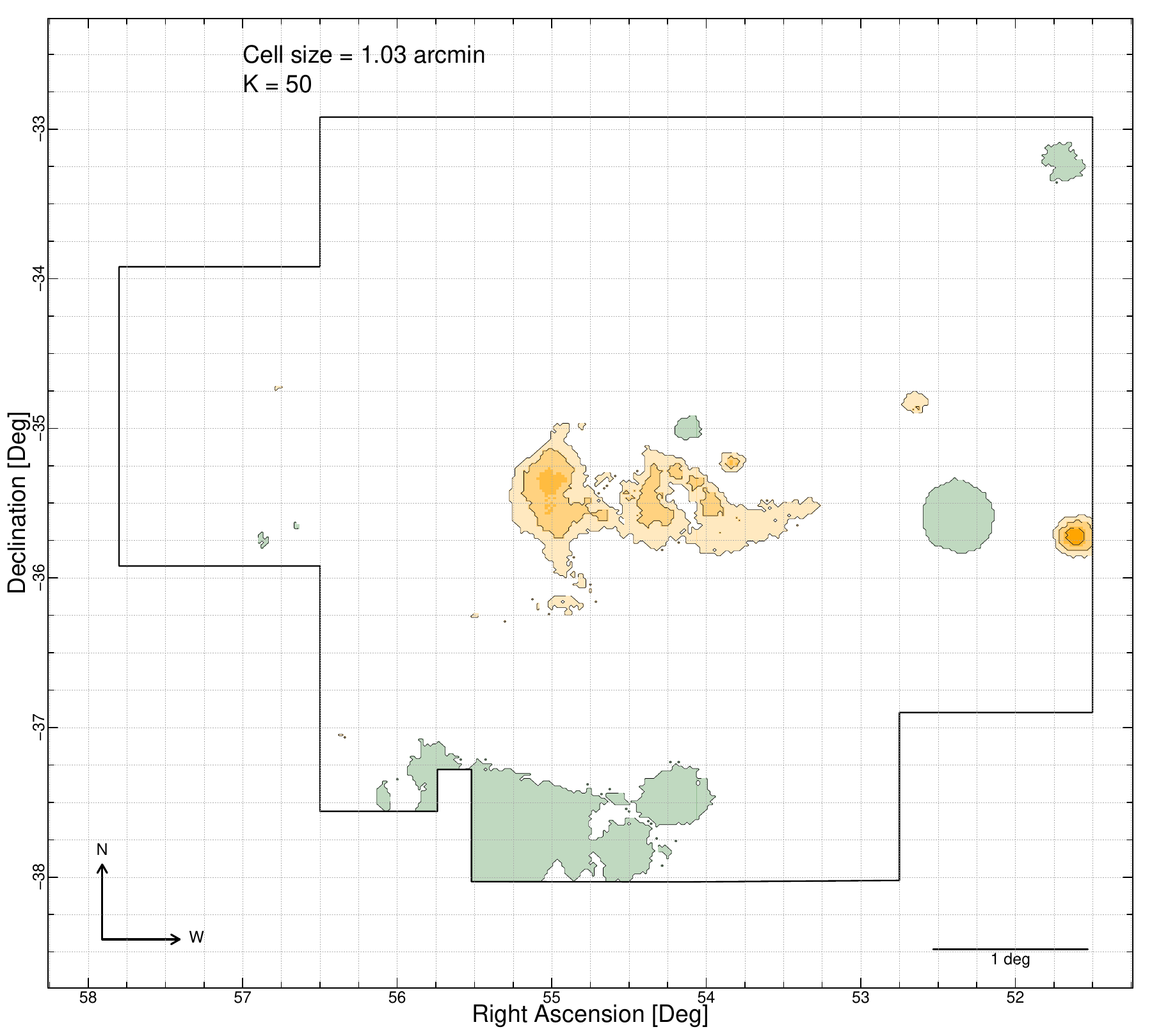}    
    \includegraphics[width=0.195\linewidth]{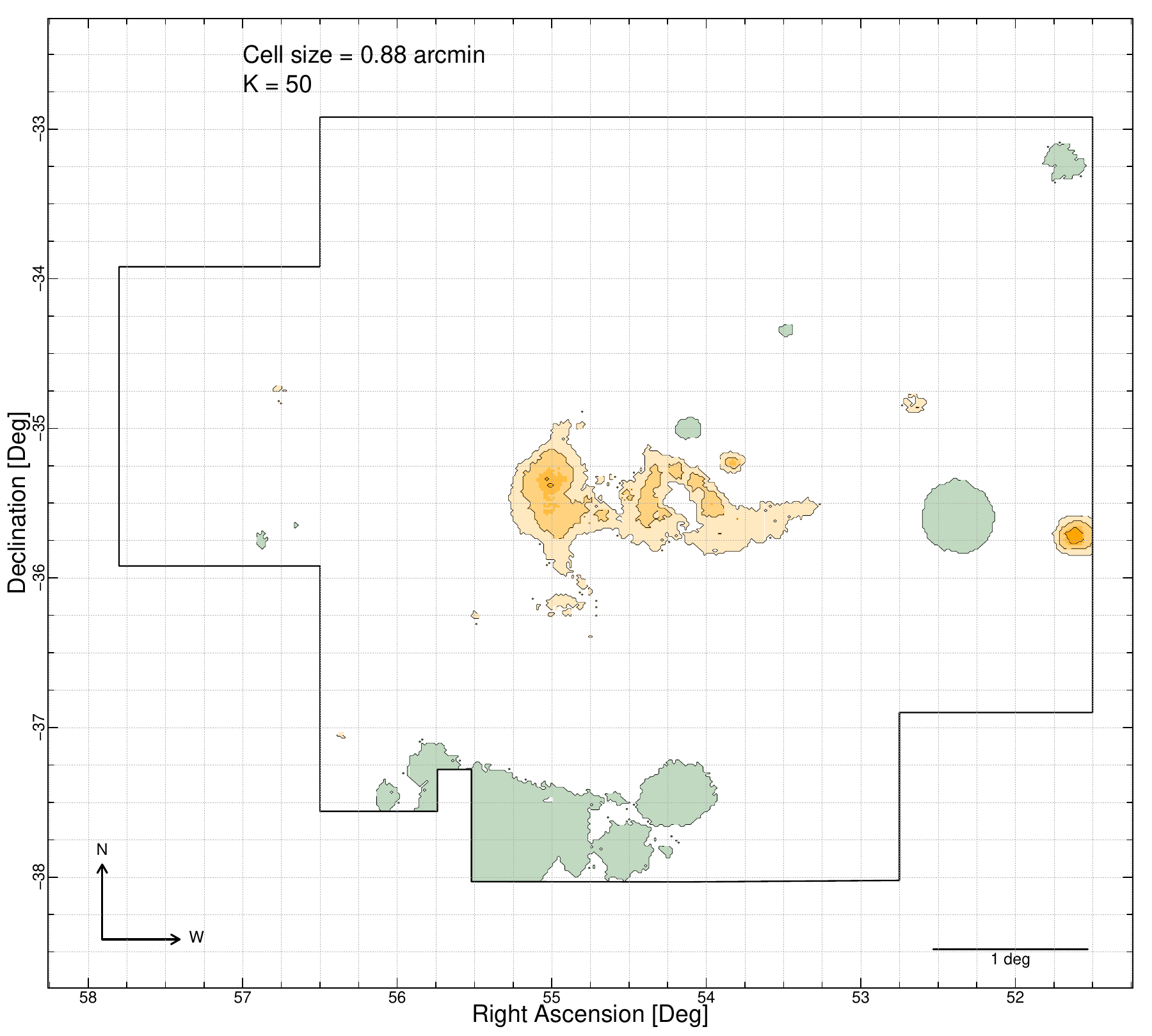}\\  
    \includegraphics[width=0.195\linewidth]{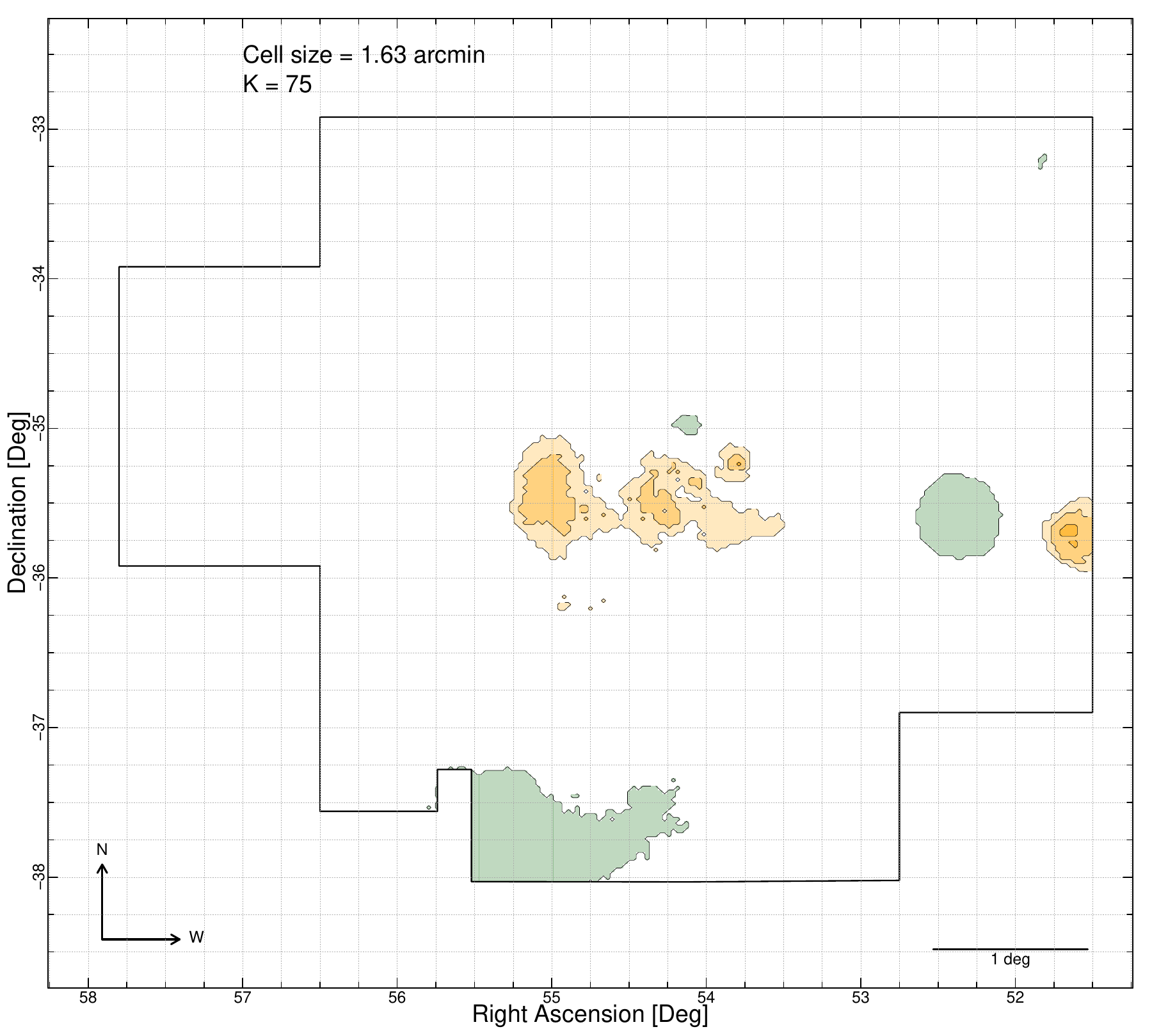}
    \includegraphics[width=0.195\linewidth]{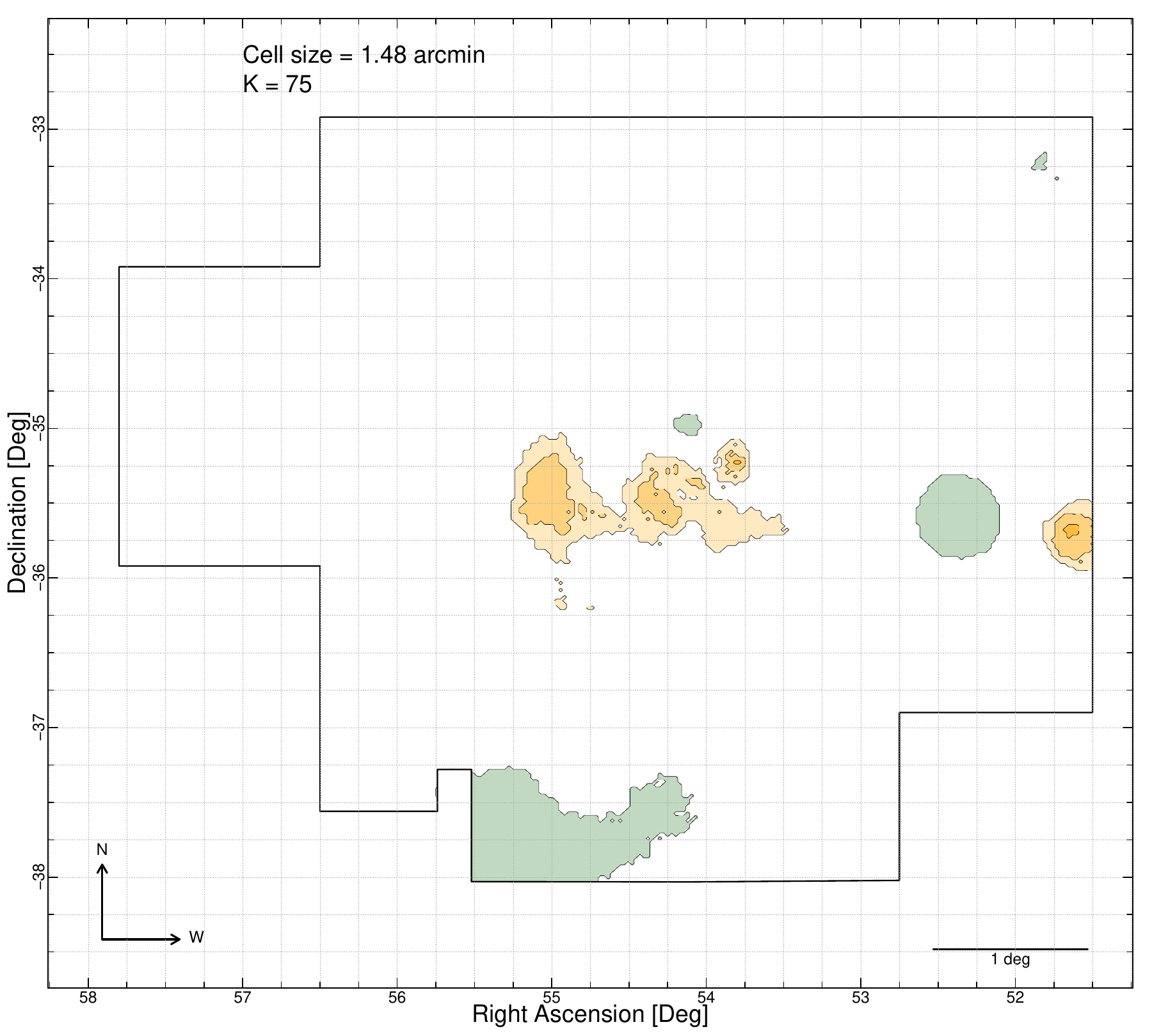}
    \includegraphics[width=0.195\linewidth]{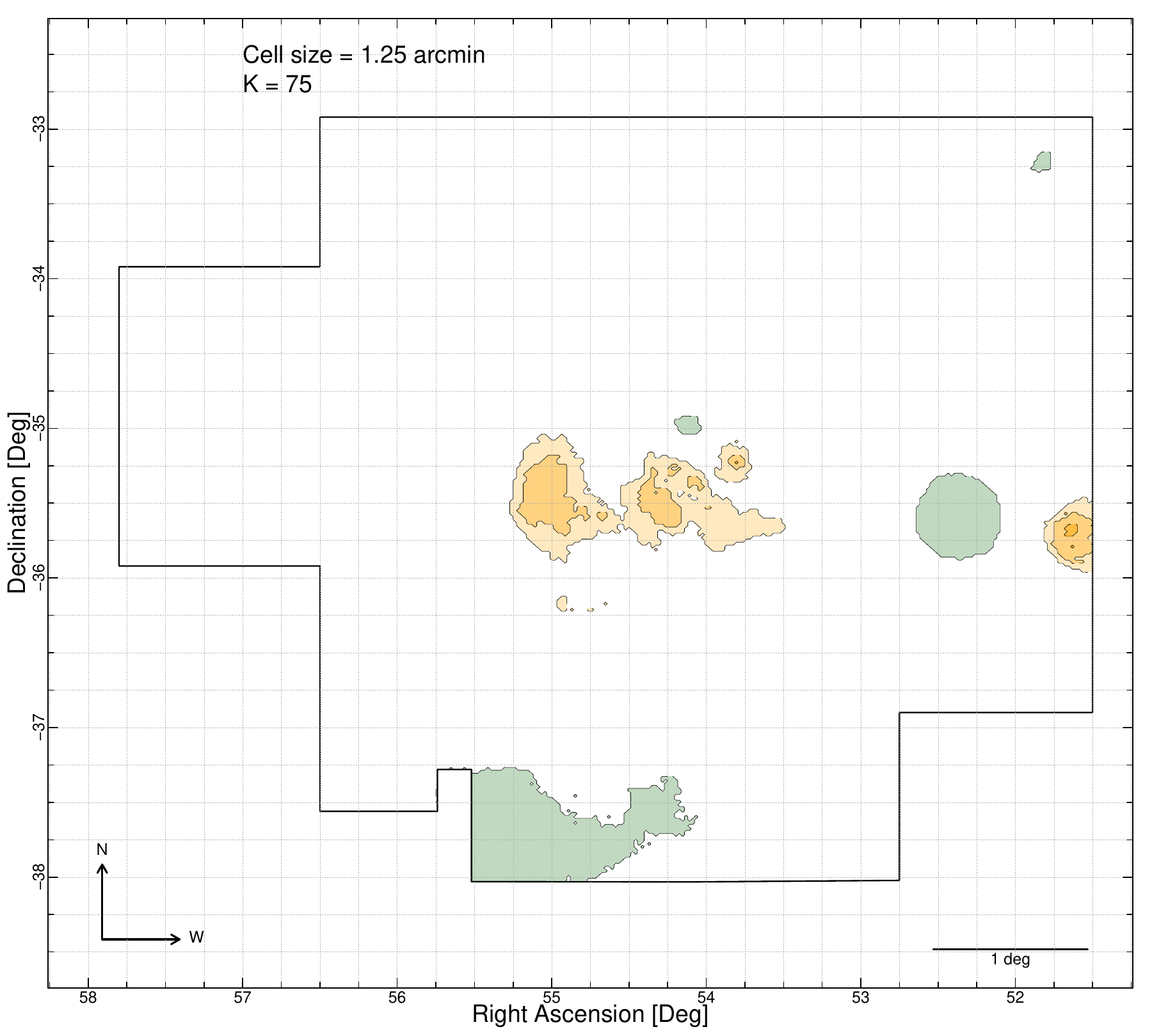}
    \includegraphics[width=0.195\linewidth]{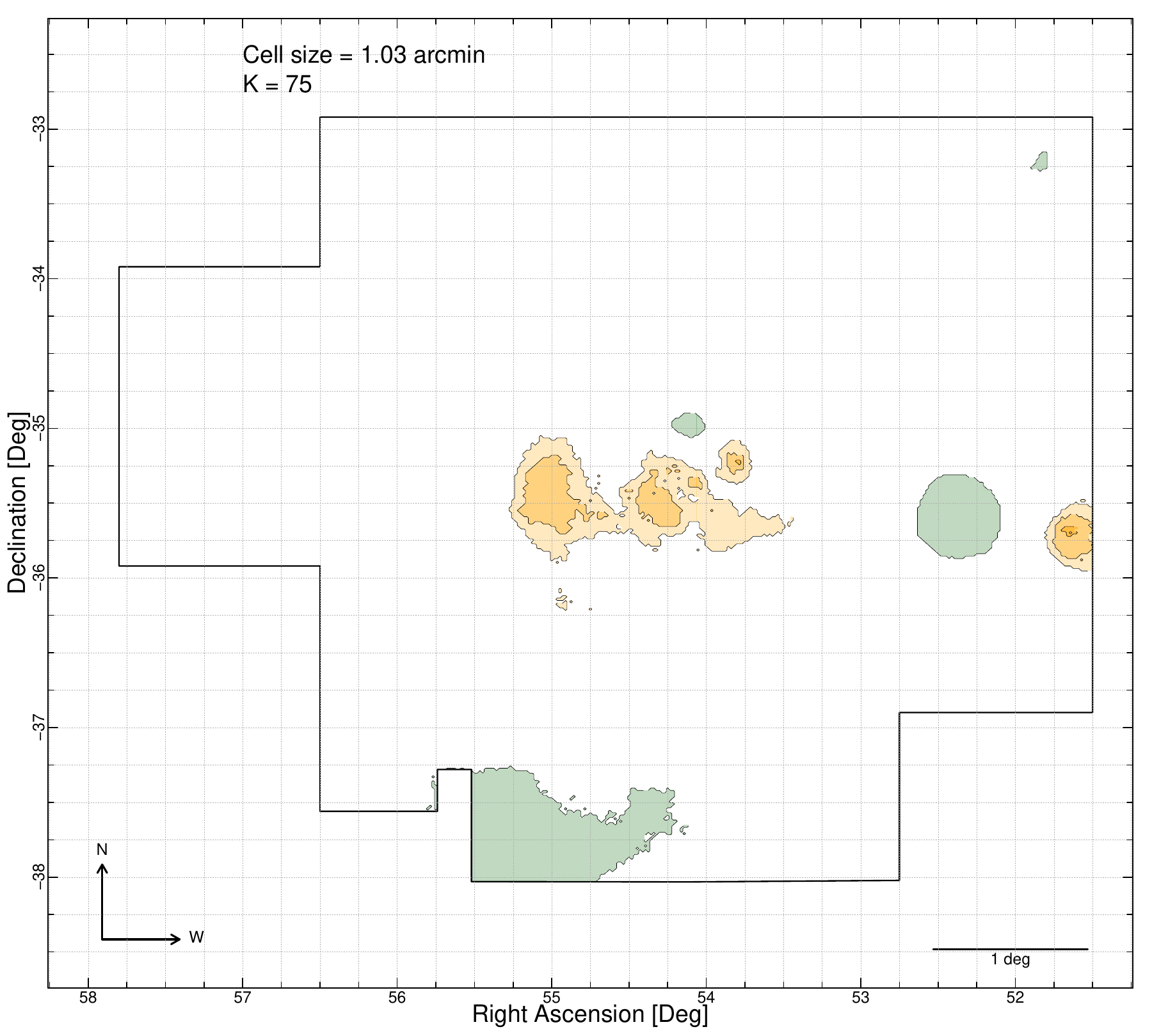}    
    \includegraphics[width=0.195\linewidth]{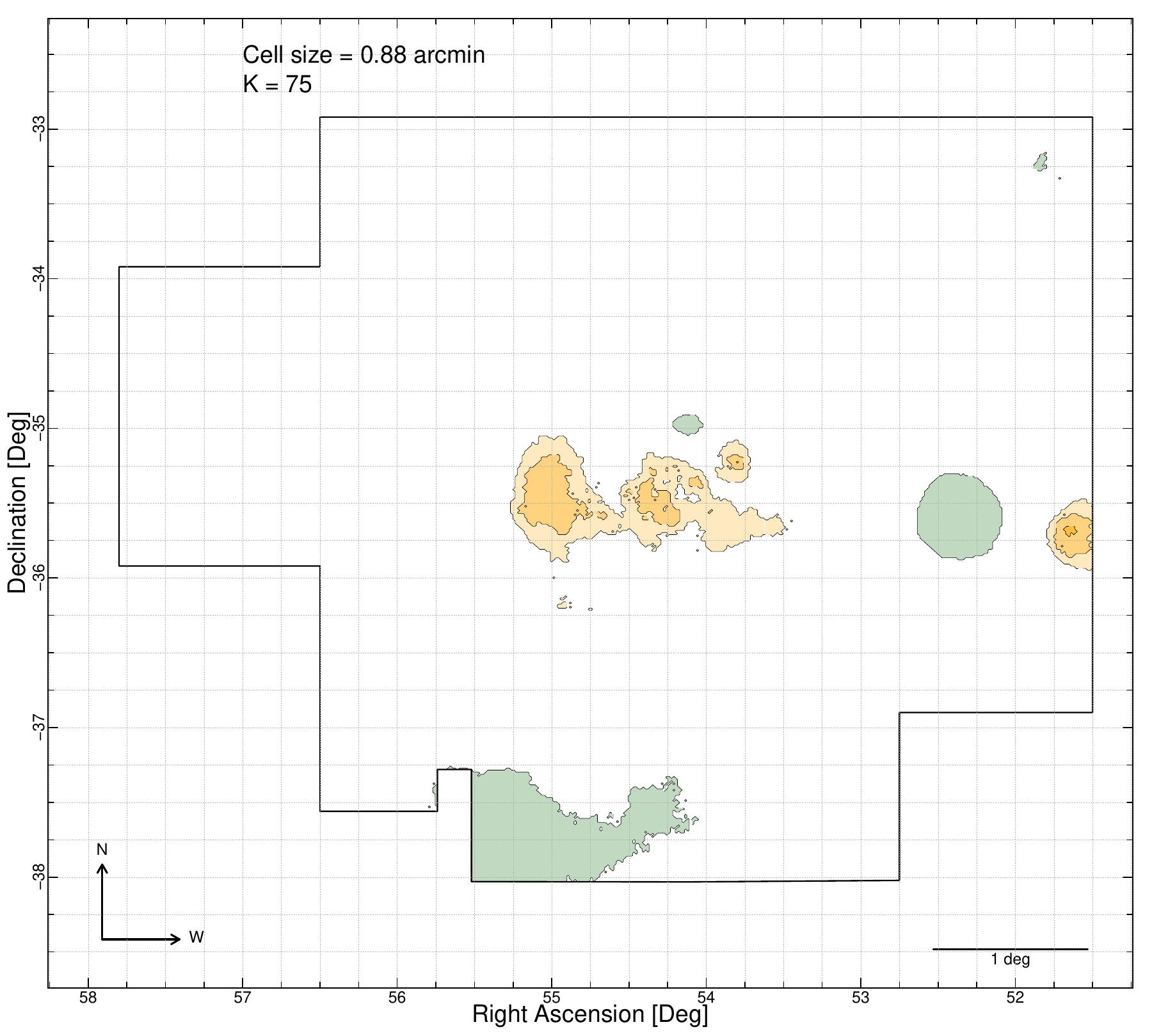}\\     
    \includegraphics[width=0.195\linewidth]{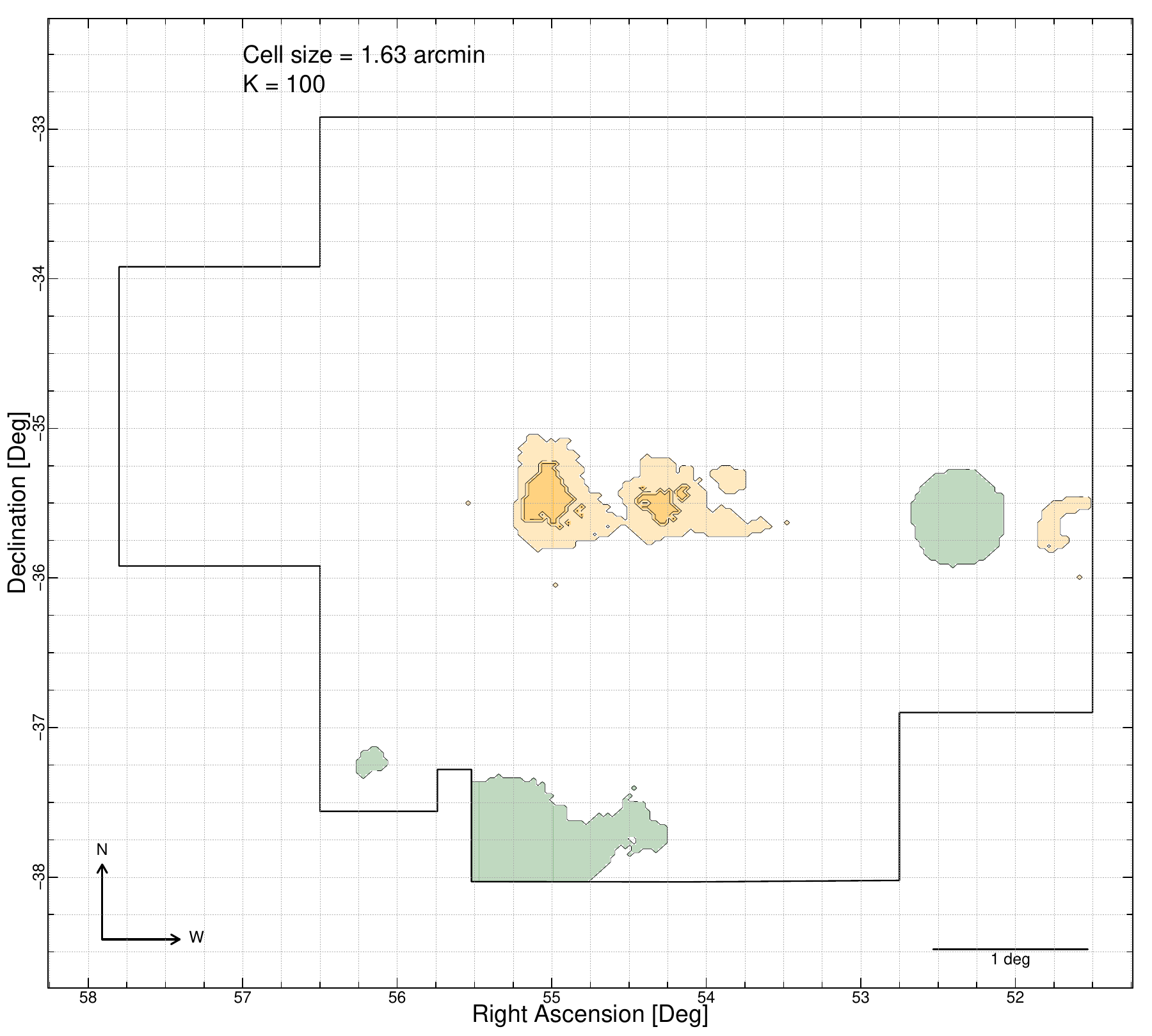}
    \includegraphics[width=0.195\linewidth]{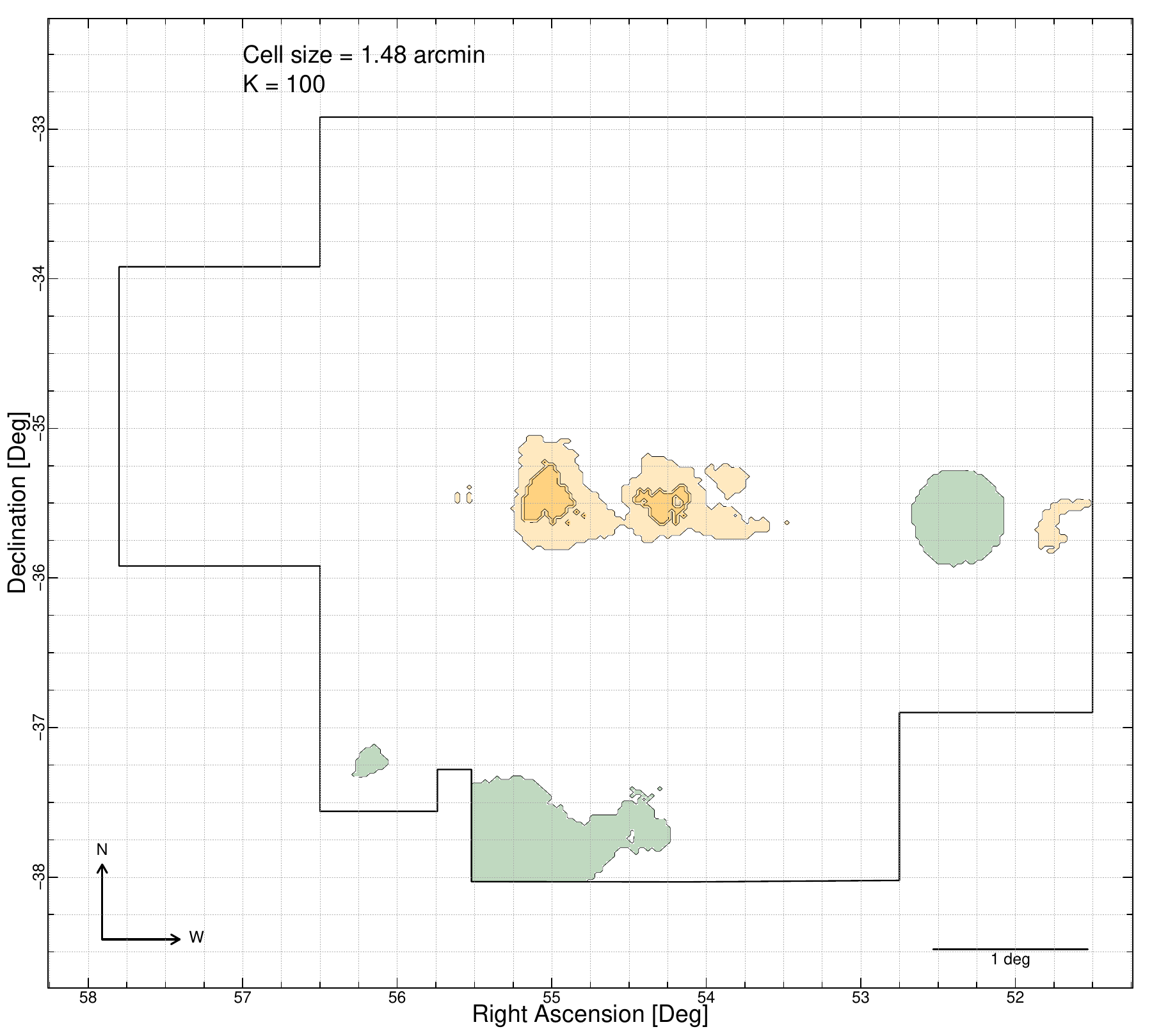}
    \includegraphics[width=0.195\linewidth]{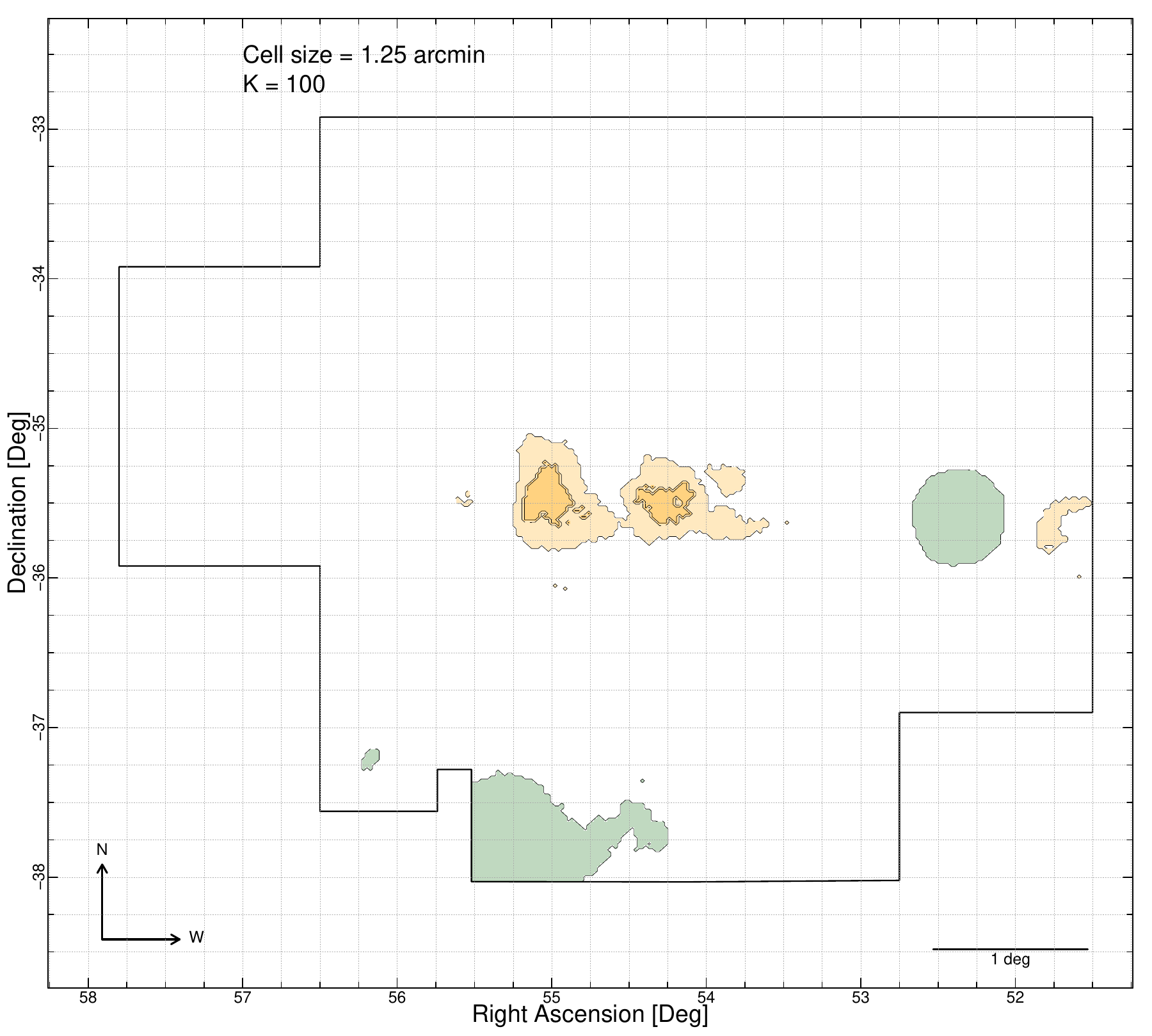}
    \includegraphics[width=0.195\linewidth]{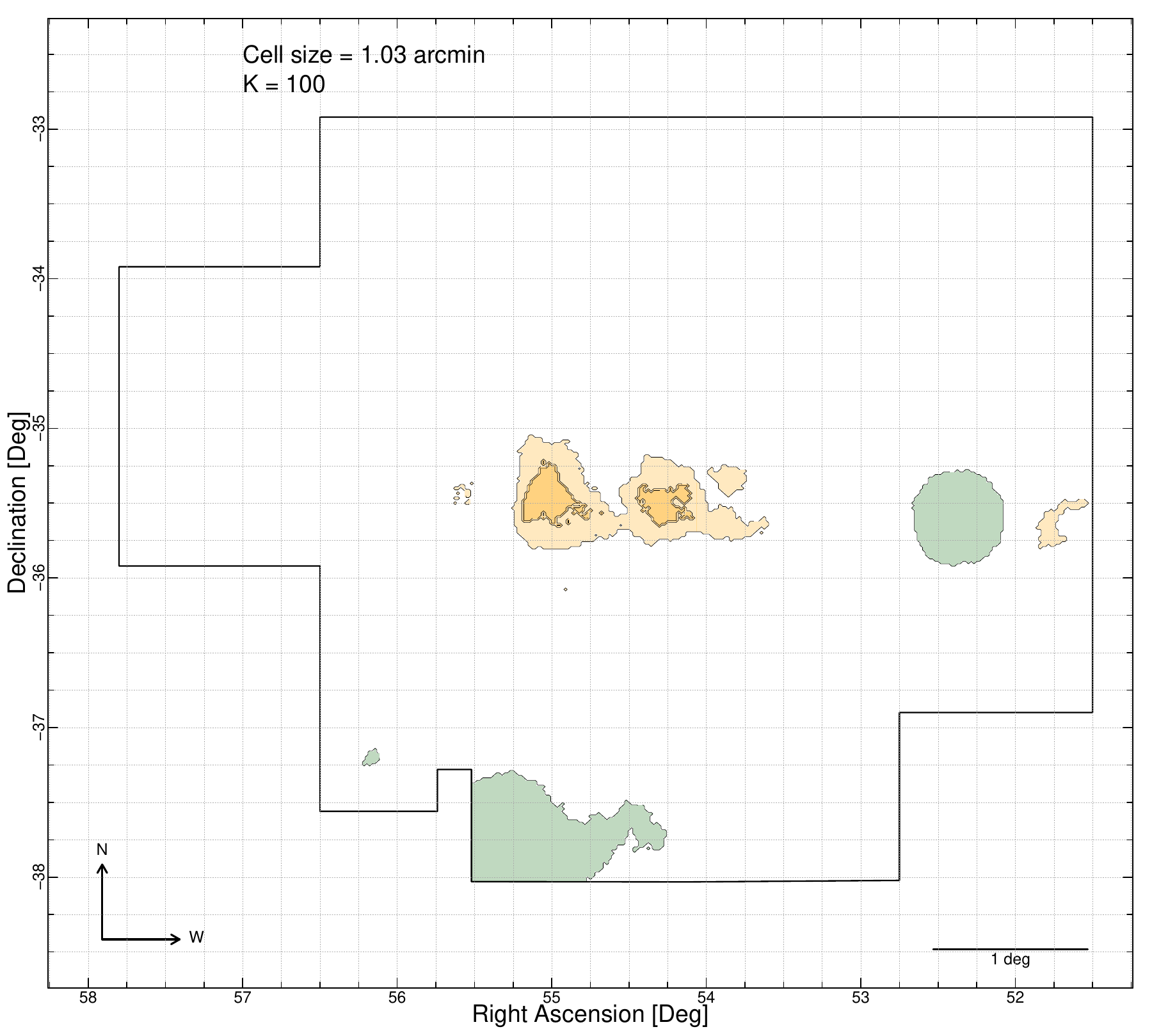}    
    \includegraphics[width=0.195\linewidth]{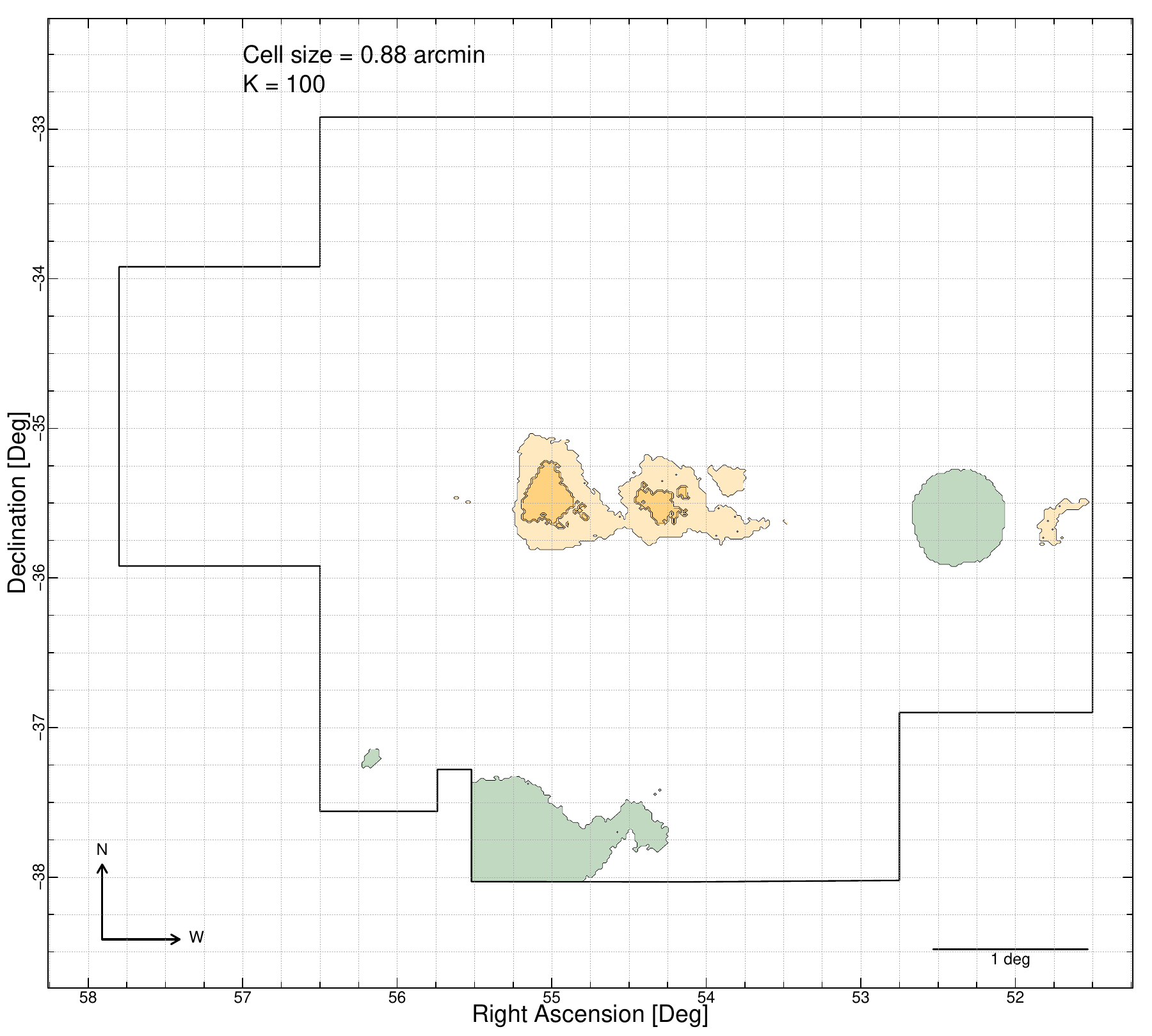}\\      
    \caption{Residual maps as in Figure~\ref{fig:residual_k10} from KNN density maps obtained 
    (from upper to lower) with $K\!=\!(5,10,25,50,75,100)$ and cell sizes 
    $(1.63,1.48,1.25,1.03,0.88)$ arc-minutes (from the left to right) and other standard parameters.}
    \label{fig:residuals_cell_sizes}
\end{figure*}

Figure~\ref{fig:residuals_cell_sizes} shows the residual maps of the spatial distribution of the observed FDS
GC candidates obtained with the standard set of parameters (see above) for
$K\!=\!(5,10,25,50,75,100)$ (from upper to lower) and cell sizes $(1.63,1.48,1.25,1.03,0.88)$ arc-minutes 
(from left to right). The five different cell sizes correspond to the 
default value employed for the experiments described in the paper ($\approx1.25\arcmin$) $\pm15\%,30\%$ of the
central value. As discussed in 
Section~\ref{sec:results}, the default value $\approx1.25\arcmin$ was chosen because it returns very similar 
average cell occupancy within the Fornax cluster 
virial radius and in its outskirt. The differences between residual maps obtained with different cell sizes
in this interval of values are negligible for the same $K$ value. 

\section{Effect of the homogeneous component density estimation and different azimuthal distributions for MGS simulated GCs}
\label{sec:appendix3}

\begin{figure*}[ht]
    \centering
    \includegraphics[width=0.495\linewidth]{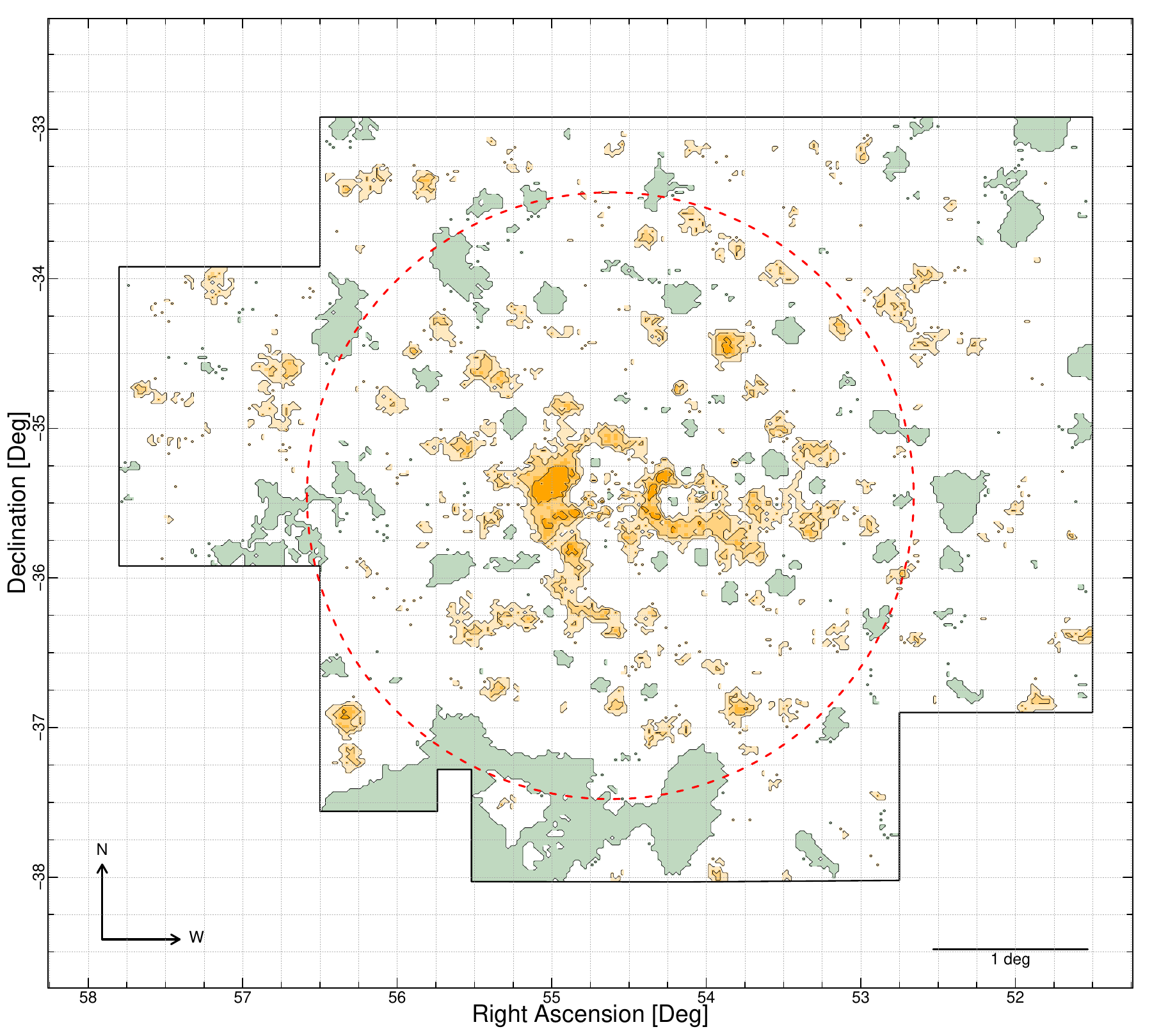}
    \includegraphics[width=0.495\linewidth]{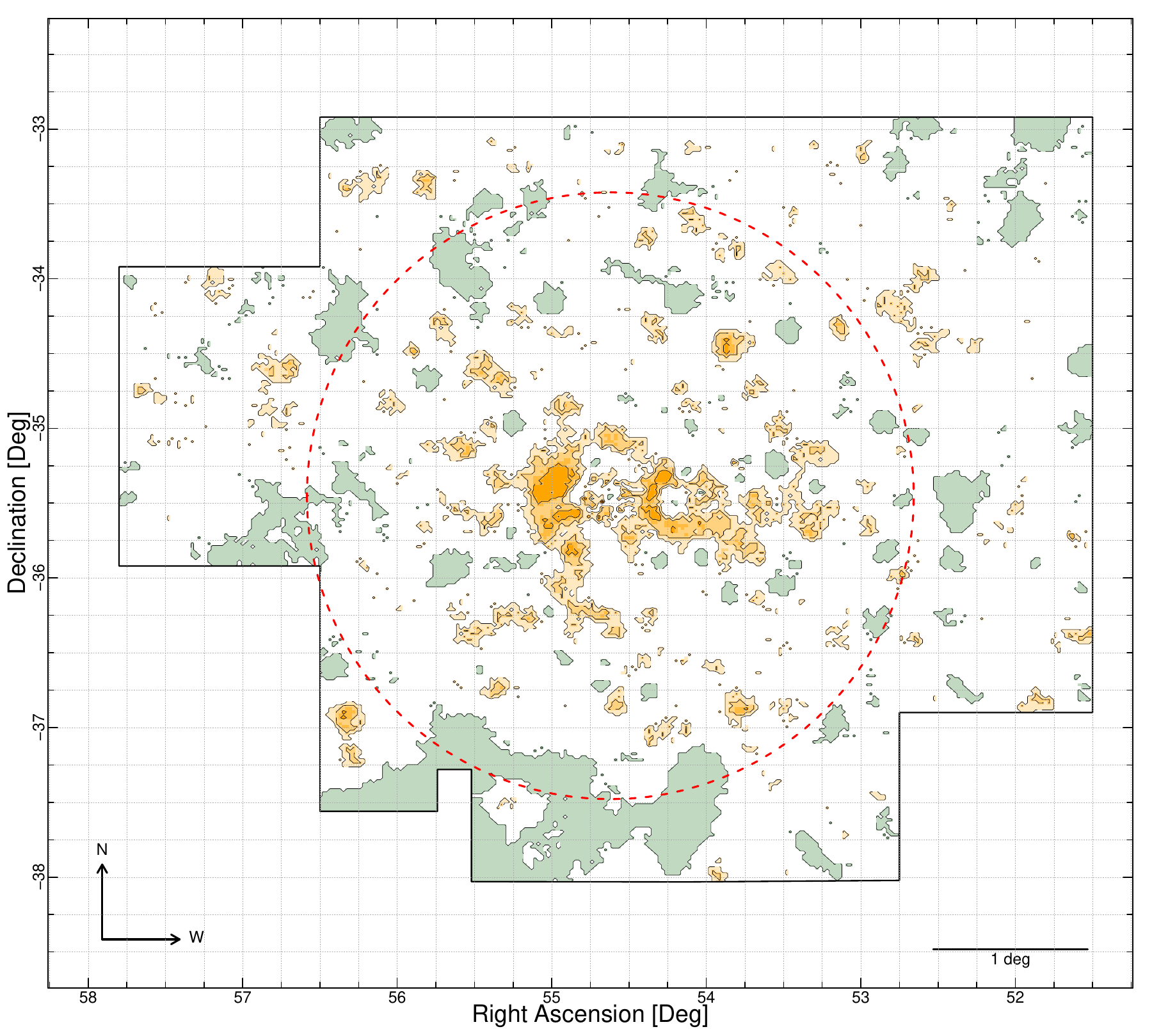}\\ 
    \includegraphics[width=0.495\linewidth]{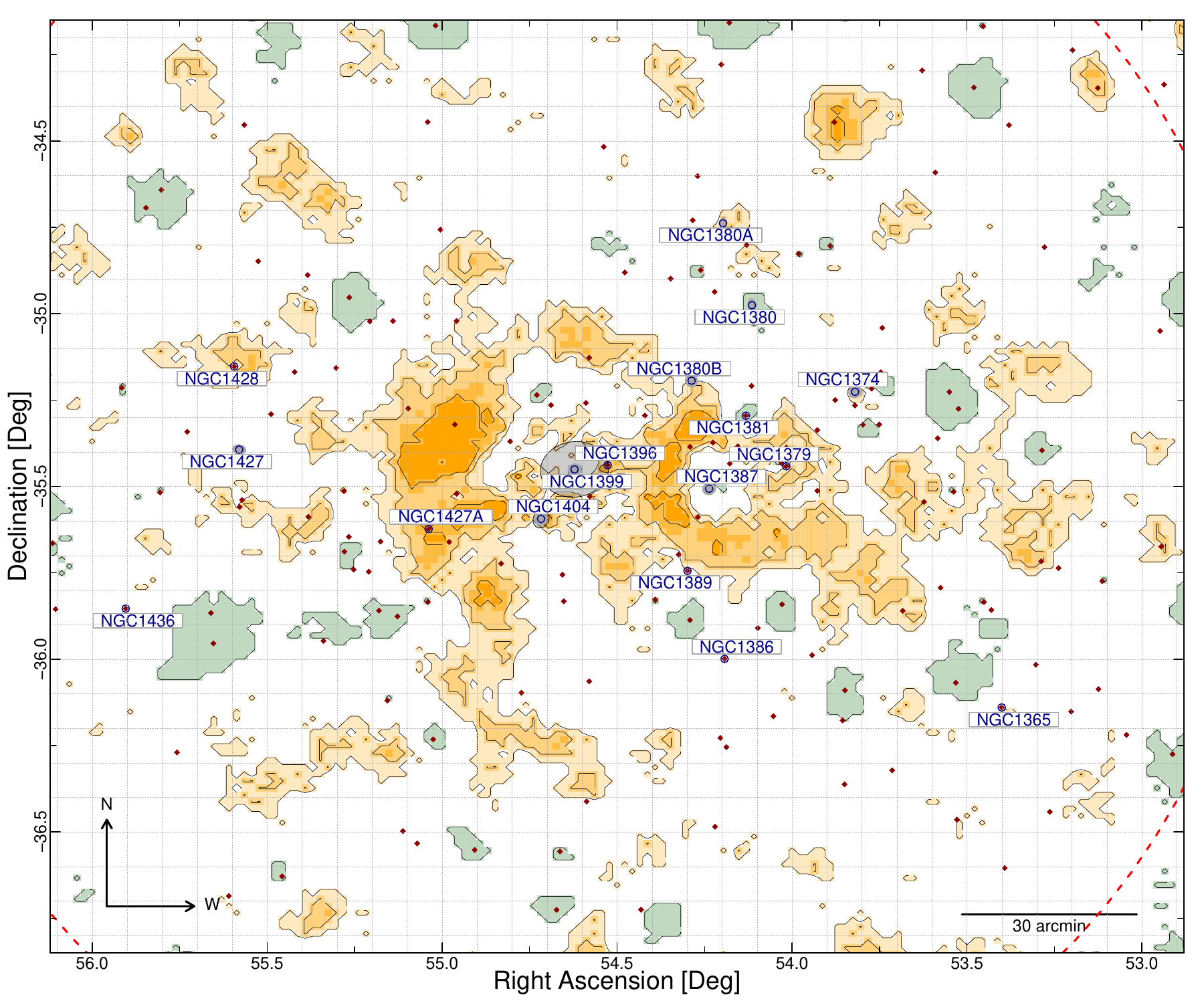}
    \includegraphics[width=0.495\linewidth]{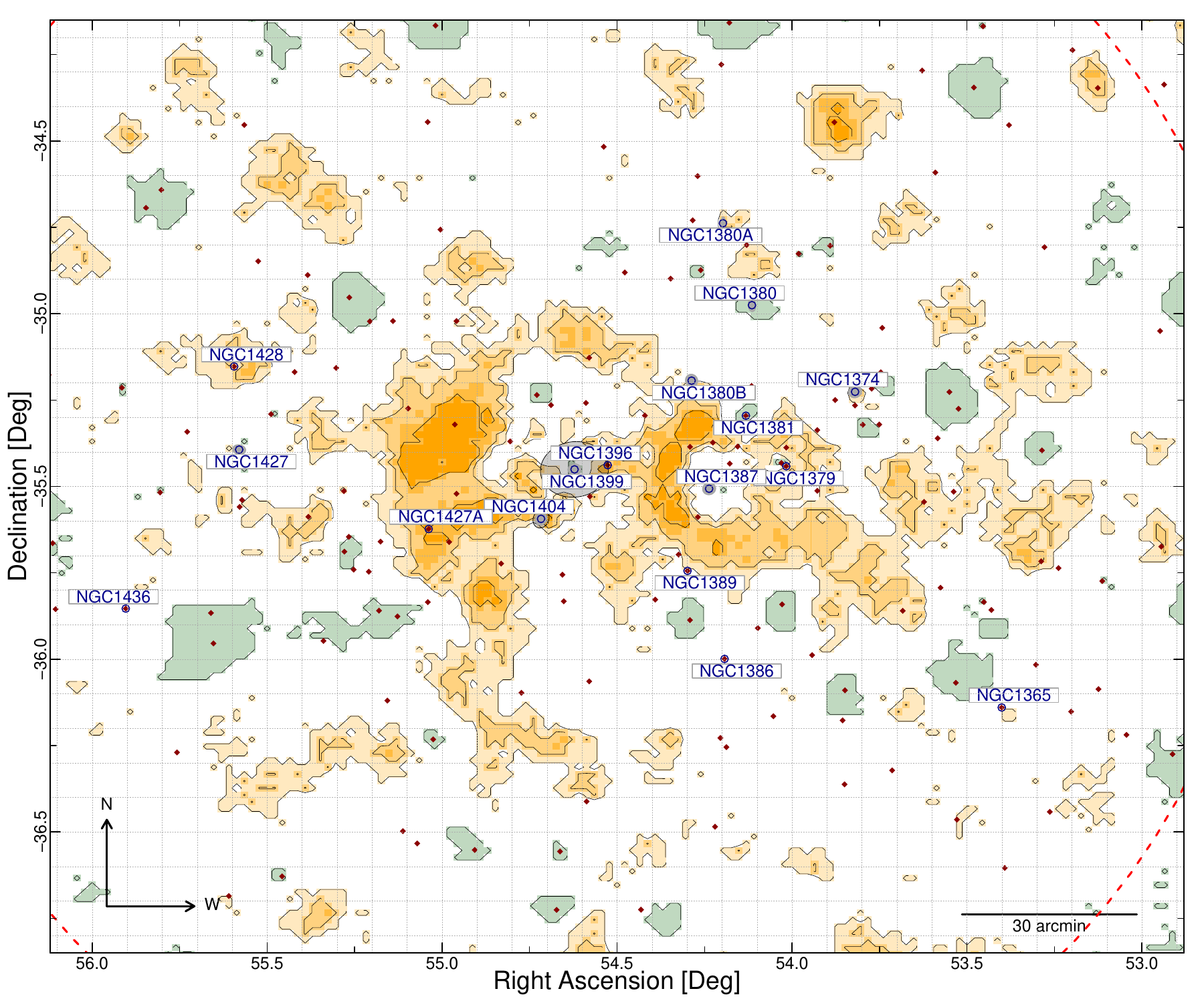}  
    \caption{Upper: residual maps of the full FDS field-of-view obtained with the two different 
    prescriptions for the estimation
    of the density of the homogeneous component (left: control field; right: outside of cluster
    virial radius) described in Section~\ref{subsubsec:homocomponent}. Lower: residual maps of the 
    core of the Fornax cluster obtained assuming circular symmetry in the azimuthal distribution
    of simulated MGS GCs (left), and same elliptical geometry of the host galaxy (right).}
    \label{fig:residuals_backgrounds_ellipticity}
\end{figure*}

As described in Section~\ref{subsubsec:homocomponent}, the density of the
spatially homogeneous component of the GC candidates spatial distribution 
has been estimated with two different methods: A) using the population of
observed FDS sources within four control fields located in the outskirts of the FDS
field-of-view, or B) using all candidate GCs located outside the virial radius of the Fornax
cluster and whose positions are 5 effective radii away from all galaxies in the FCC~\citep{ferguson1989}.
In this paper, we adopted method B) for the characterization of the over-density of ICGCs observed 
in the core of the Fornax cluster. The uncertainty on the density is calculated as the 
standard deviation of the densities in the four control fields for method A, and the standard
derivation of the distribution of densities derived twelve circular regions mostly located outside
of the cluster virial radius for method B. The upper plots in Figure~\ref{fig:residuals_backgrounds_ellipticity}
show the residual maps obtained using the two different methods for the estimation of the homogeneous component
density (method A on the left, and method B on the right) with the same set of parameters
(10,000 simulations, $K\!=\!10$, cell size $\approx\!1.25\arcmin$ and modified Hubble model for the radial profile 
and host galaxy ellipticity for the azimuthal distribution of MGS GCs component). 
The two plots show no significant differences: while the overall 
morphology of the residual structures remains the same, a small decrease of the size of some structures 
can be noticed in the residual map obtained with approach B, because of the slightly larger density value 
in this case that reduces the total number of adjacent cells with positive residuals defining the spatial structures
in the GC candidates spatial distribution.

The lower plots in Figure~\ref{fig:residuals_backgrounds_ellipticity} show
the residual maps of the distribution of FDS GC candidates in the core of the Fornax cluster 
obtained using the standard set of parameters and
assuming circular symmetry for the azimuthal distribution of MGS simulated GCs 
(left), and same elliptical geometry as the host galaxy (right), as described in Section~\ref{subsubsec:galcomponent}. 
Small differences in the shape of the residual structures can be observed in the vicinity of some isolated
galaxies (like NGC\,1374 and NGC\,1380) and N of NGC\,1399, caused by the combined effect of the host galaxy 
position angle and ellipticity. The morphologies of the larger-scale, over-density spatial structures 
observed in the residual maps obtained with the two methods do not change. 

\section{Comparison between the distribution of observed and simulated GC candidates}
\label{sec:appendix5}

\begin{sidewaysfigure}[ht]
    \centering
    \includegraphics[width=\linewidth]{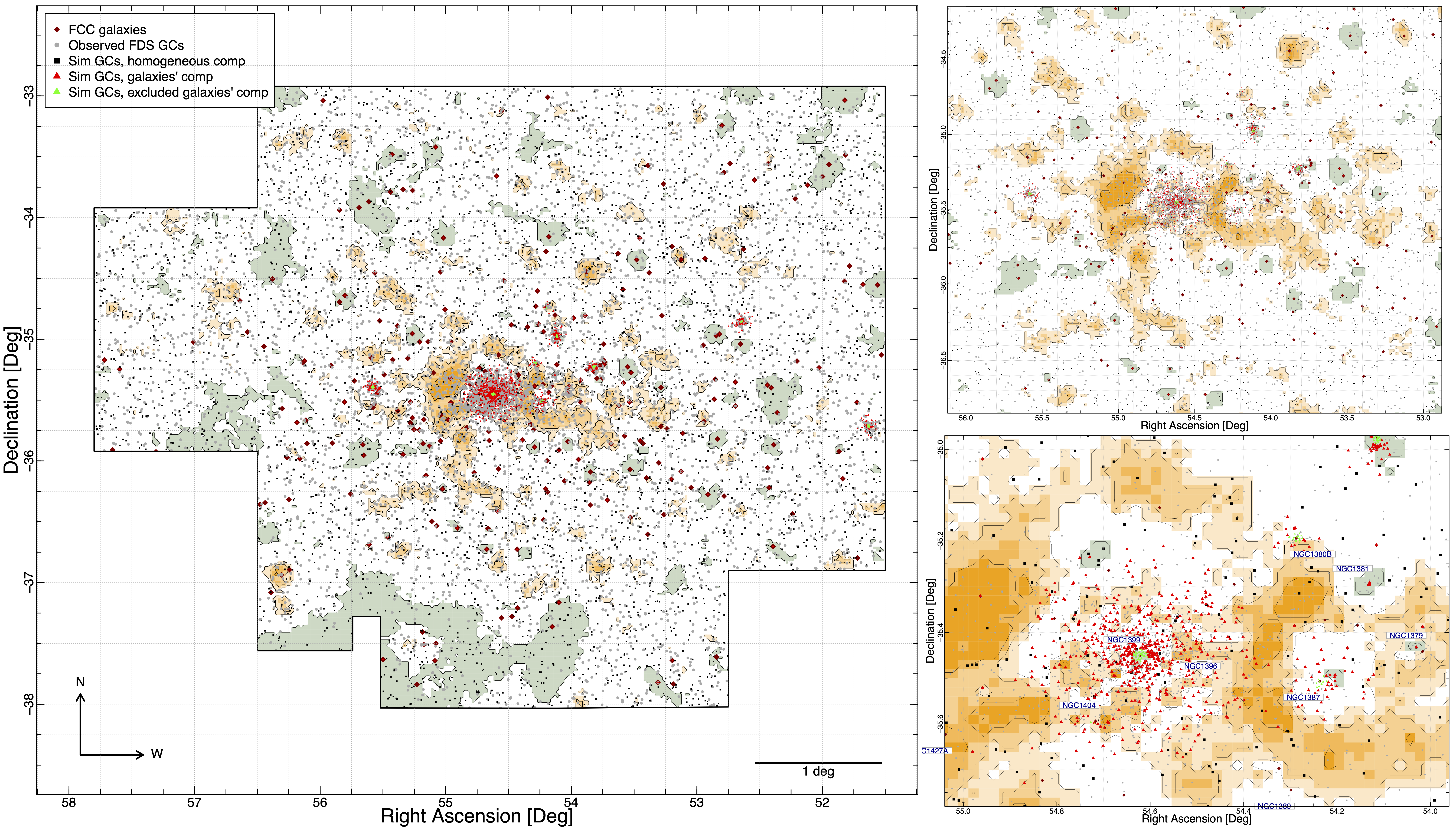}
    \caption{Residual map of the spatial distribution of GC candidates (full field on the left, 
    two different zoomed-in regions of the core of the cluster in the upper and lower right side) 
    obtained with the standard set of parameters (10,000 simulations, $K\!=\!10$, 
    density of the homogeneous component estimated outside of the cluster virial radius, 
    cell size $\approx1.25\arcmin$, modified Hubble model for the radial profile and host 
    galaxy ellipticity for the azimuthal distribution of MGS GCs component). FCC galaxies, the 
    observed GC candidates (gray circles), and
    simulated GCs from a randomly picked simulation used to determine the residual maps are 
    overplotted to the residual map. Simulated GCs from the  
    the homogeneous and galaxy components are displayed with black squares and 
    red triangles, respectively. Green triangles represent simulated GCs in the galaxies
    components located within the avoidance area and not used to determine the residual
    map~(see Section~\ref{subsubsec:galcomponent} for details).}
    \label{fig:mosaic_residuals_sims}
\end{sidewaysfigure}

Figure~\ref{fig:mosaic_residuals_sims} shows the observed distribution of FDS GC candidates
(gray circles) and one, randomly chosen, set of simulated GCs (triangles) overlayed to the 
residual maps obtained with the standard set of parameters (10,000 simulations, $K\!=\!10$, 
density of the homogeneous component estimated outside of the cluster virial radius, 
cell size $\approx1.25\arcmin$, modified Hubble model for the radial profile and host 
galaxy ellipticity for the azimuthal distribution of MGS GCs component) over the whole 
f.o.v. covered by the FDS data (left) and two zoomed-in regions in the core of the 
cluster (upper and lower right). The simulated GCs belonging to the galaxies component 
are associated with the GC systems of both MGC or SGS galaxies. The green triangles show 
simulated GC candidates at galacto-centric distances smaller than the avoidance radii of
MGS hosts~(Table~\ref{tab:main_galaxies_radial}), which have not
been used to determinate the residual maps, in order to partially account for the reduced 
GC detection efficiency in the ground-based, FDS data in the core of bright galaxies~ 
(Section~\ref{subsubsec:galcomponent} for details). 

\end{document}